\documentclass[acmsmall,screen]{acmart}

\setcopyright{rightsretained}
\acmPrice{}
\acmDOI{10.1145/3586038}
\acmYear{2023}
\copyrightyear{2023}
\acmSubmissionID{oopslaa23main-p64-p}
\acmJournal{PACMPL}
\acmVolume{7}
\acmNumber{OOPSLA1}
\acmArticle{86}
\acmMonth{4}

\usepackage{booktabs}   
\usepackage{subcaption} 

\usepackage{listings}
\usepackage{footmisc}  
\usepackage{caption}
\usepackage{float}
\usepackage{multirow}
\usepackage{pifont}

\def\Comment#1{\textbf{\textsl{\color{red}  $\langle\!\langle$#1$\rangle\!\rangle$}} }
\newcommand{\mwh}[1]{\Comment{MWH: {#1}}}
\def\Comment#1{}

\def\CheckedC{Checked~C}
\def\Lstinline{\lstinline[basicstyle=\normalsize\ttfamily,keywordstyle=\color{black}]}
\newcommand{\checkedc}{Checked~C }
\def\struct{\tt struct}
\def\itype{\tt itype}
\newcommand{\singlequote}[1]{\textquotesingle{#1}\textquotesingle}
\newcommand{\var}[1]{\tt{#1}}
\newcommand{\paratitle}[1]{\noindent\textbf{#1}. \ }
\newcommand{\secref}[1]{\mbox{\textsection\ref{#1}}}

\def\SPEC17{SPEC~CPU2017}
\newcommand{\cmark}{\ding{51}}
\newcommand{\xmark}{\ding{55}}
\newcommand{\qmark}{\bf ?}
\newcommand{\ptr}{\Lstinline|ptr|}
\newcommand{\ptrT}{\mbox{\Lstinline|ptr<T>|}}
\newcommand{\arrayptr}{\mbox{\Lstinline|array_ptr|}}
\newcommand{\arrayptrT}{\mbox{\Lstinline|array_ptr<T>|}}

\newcommand{\ntarrayptrT}{\Lstinline|nt_array_ptr<T>|}
\newcommand{\mmptr}{\mbox{\Lstinline|mm_ptr|}}
\newcommand{\mmptrT}{\mbox{\Lstinline|mm_ptr<T>|}}

\newcommand{\mmarrayptr}{\mbox{\Lstinline|mm_array_ptr|}}
\newcommand{\mmarrayptrT}{\mbox{\Lstinline|mm_array_ptr<T>|}}
\newcommand{\mmarrayptrinst}[1]{\mbox{\Lstinline|mm_array_ptr<|{\var{#1}}\Lstinline|>|}}

\newcommand{\checkable}{\tt\_checkable}
\newcommand{\largeptr}{\mbox{\Lstinline|large_ptr|}}
\newcommand{\largeptrT}{\mbox{\Lstinline|large_ptr<T>|}}
\newcommand{\largearrayptr}{\mbox{\Lstinline|large_array_ptr|}}
\newcommand{\largearrayptrT}{\mbox{\Lstinline|large_array_ptr<T>|}}

\def\mmalloc{\tt mm\_alloc}
\def\mmfree{\tt mm\_free}

\begin{document}

\title[Fat Pointers for Temporal Memory Safety of C]
{Fat Pointers for Temporal Memory Safety of C}



\author{Jie Zhou}
\orcid{0000-0001-7493-2212}             
\affiliation{
  \institution{University of Rochester}            
  \country{USA}                    
}
\email{jiezhou@rochester.edu}          

\author{John Criswell}
\orcid{0000-0003-2176-3659}             
\affiliation{
  \institution{University of Rochester}           
  \country{USA}                   
}
\email{criswell@cs.rochester.edu}         

\author{Michael Hicks}
\authornote{Work completed prior to starting at Amazon.}

\orcid{0000-0002-2759-9223}             
\affiliation{
  \institution{Amazon and University of Maryland}           
  \country{USA}                   
}
\email{mwh@cs.umd.edu}         


\begin{abstract}

    Temporal memory safety bugs, especially use-after-free and double free bugs,
    pose a major security threat to C programs. Real-world exploits utilizing
    these bugs enable
    attackers to read and write arbitrary memory locations, causing disastrous
    violations of confidentiality, integrity, and availability.
    Many previous solutions retrofit temporal memory safety to C,
    but they all either incur high performance overhead and/or miss
    detecting certain types of temporal memory safety bugs.

    \Comment{JTC: JZ, the abstract should summarize the performance and
    memory overhead results once that data is available.}

    In this paper, we propose a temporal memory safety solution that
    is both efficient and comprehensive. Specifically, we extend
    {\CheckedC}, a spatially-safe extension to C, with  
    temporally-safe pointers. These are implemented by combining two
    techniques: fat pointers and dynamic key-lock checks.
    We show that the fat-pointer solution significantly improves
    running time and memory overhead compared to the disjoint-metadata approach that provides
    the same level of protection. With empirical program data and hands-on
    experience porting real-world applications, we also show that our
    solution is practical in terms of backward compatibility---one of the major
    complaints about fat pointers.


\end{abstract}



\maketitle

%
%
\section{Introduction}
%
%

A temporal memory safety violation occurs when
a program dereferences a pointer whose referent memory object has already
been freed (\emph{use after free} or \emph{UAF}),
frees a pointer more than once (\emph{double free}), or
frees a pointer that does not point to the start of a
heap object (\emph{invalid free}).
Exploiting a temporal safety-violating bug may allow an attacker to
read or write an arbitrary memory location and thereby to steal
information, corrupt memory, or even execute arbitrary
code~\cite{UAF:CWE416,UAF:BlackHatUSA07,UAFKernel:CCS15,MallocMaleficarum}.
Unfortunately, the past decade has seen an increase in such exploits
used in the real world~\cite{ExploitTrend:MS13,VulTrend:BlueHatIL19,ChromeSecuritySurvey:2020}.


One can enforce temporal memory safety by associating with
each memory object, at allocation time, a distinct \emph{lock},
and with each pointer a \emph{key}. An object's lock is invalidated
when it is freed. At each pointer dereference, a dynamic check
confirms the pointer's key matches the referent object's lock; if not,
it signals a UAF\@.
Prior solutions~\cite{SafeC:PLDI94,Guarding:SPE97,CETS:ISMM10,
XuMemSafe:FSE04} have explored this approach but
incur high memory and/or performance overhead.
For example, CETS~\cite{CETS:ISMM10} incurs 48\% performance overhead
on selected SPEC CPU2006 benchmarks.
PTAuth~\cite{PTAuth:Sec21} and ViK~\cite{ViK:ASPLOS22} lower the
performance overhead on SPEC to 26\% and 9\%, respectively,
but they trade security and scalability for speed.
Both use a small key space (10--16 bits), and they require a predetermined
relatively small maximum object size---for objects larger than the
maximum, PTAuth may raise false alarms while ViK may miss
safety violations (see Section~\ref{section:related:id} for details).


When experimenting with key-lock check approaches, we found that
a crucial reason for their high cost is
the use of \emph{disjoint} data structures, such as look-up tables,
to maintain the association between keys/locks and pointers/objects.
Doing so keeps objects and pointers unchanged from their legacy
representation, which makes it easy for compiled-to-be-safe code to
interoperate with unchanged (e.g., library) code. However, the approach
requires the compiler to add instructions to locate the keys and locks at
pointer propagations and dereferences.

An alternative to making metadata disjoint from a pointer is to store
it \emph{in place}, yielding a kind of \emph{fat pointer}.\footnote{Some works
(e.g., \citet{SoftBoundCETS:SNAPL15})
refer to all techniques that
  associate pointers with metadata as fat pointers.  In this paper, we
  use ``fat pointers'' to only mean pointers with in-place metadata.}
Fat pointers were oft-proposed for enforcing spatial memory safety (i.e.,
bounds checks)~\cite{Cyclone:ATC02,CCured:POPL02} but fell out of
favor because of both high memory- and run-time overhead
and difficulties interoperating with legacy components.
However, we observe that fat pointers do not present the same interoperability
issues when temporal memory safety checks are included
with a new \emph{language feature} rather than added
automatically as part of a \emph{compilation strategy} for unmodified C~code.
This is because a language extension can provide
different pointer types that expose to the programmer the difference
in representation between safe and legacy pointers,
enabling programmers to choose when and how to convert
between the two representations.
However, it is an open question whether fat pointers could be a
practical solution for writing new temporally safe C~code or to retrofit
temporal memory safety to existing C~code.

To that end, we extended the {\CheckedC}
programming language~\cite{CheckedC:SecDev18,CheckedCModel:CSF22,
  CheckedCTR:Microsoft} with new types of temporally safe pointers,
storing keys in place with pointers, and locks in place with pointed-to objects.
{\CheckedC} 
is a new safe extension to C which efficiently enforces spatial memory safety,
providing a solid foundation on which we can build to explore our
in-place key-lock strategy for temporal memory safety.
{\CheckedC}'s type system provides strong security guarantees at
compile time, eliminating common dangerous C idioms such
as arbitrary casts that could break the security benefits of our safe pointers.
As {\CheckedC} is a proper programming language, programmers can
naturally address any compatibility issues caused by the use of fat pointers,
whether initially or during program maintenance, whereas doing so
would be far more challenging when working with an 
automatic compiler transformation~\cite{CCured:TOPLAS05,SoftBound:PLDI09}.


We implemented our new pointer types in the {\CheckedC} compiler,
which is based on Clang and LLVM~\cite{LLVM:CGO04}.
By design, these pointers extend {\CheckedC}'s spatially-safe pointers,
so in a full implementation they would be subject to spatial
safety checks. We have delayed this (conceptually straightforward,
though nontrivial) integration effort in order to first evaluate
the effectiveness of in-place metadata for temporal memory safety checking.
To do so, we ported Olden (a small pointer-intensive benchmark suite),
one pointer-intensive SPEC benchmark, three
real-world applications, and the engine and HTTP protocol part of a
large ubiquitous program {\tt curl} to use temporally safe pointers and measured
the resulting performance and memory overhead.
We compared our {\CheckedC} solution with CETS~\cite{CETS:ISMM10}---the
state-of-the-art key-lock check approach\footnote{As mentioned
earlier, PTAuth~\cite{PTAuth:Sec21} and ViK~\cite{ViK:ASPLOS22} are
more recent but lack the same security and scalability benefits of
CETS and our approach; see Section~\ref{section:related:id}.} using disjoint metadata---and
show that our in-place mechanism significantly reduces
performance overhead (29\% vs. 92\%) and
memory overhead (72\% vs. 202\%) on Olden.

While {\CheckedC} can be used for writing new code,
we also recognize that programmers will want to
port existing C~code to gain memory safety.
We therefore report on our experience porting the
benchmarks and applications.
On average, we can port 1--2 K lines of code per person-day.
We believe that porting can be facilitated by adapting
3C~\cite{3C:OOPSLA22}, a semi-automated porting tool from C to
spatially-safe {\CheckedC}, to work with our temporally safe pointers.

In summary, we make the following contributions:

\begin{itemize}
    \item We explore the benefits and costs of using fat pointers to
        retrofit temporal memory safety to C. We add four new types of
        safe pointers to {\CheckedC} to provide \emph{full} temporal memory
        safety.  We implemented the two most common types for 64-bit systems.

    \item We evaluated the new safe pointers by measuring their performance
        and memory overhead. We show that our fat pointer solution
        is \emph{efficient} in that it incurs significantly lower performance
        and memory overheads compared with a disjoint metadata mechanism.

    \item We show that our solution is \emph{practical} in terms of backward
        compatibility with legacy C code. We support this claim with
        empirical program data and our experience of porting real-world
        applications.
\end{itemize}

\paratitle{Roadmap}
We first briefly review {\CheckedC} in
Section~\ref{section:bg}.
We then describe the motivations for our design choices and details of the new
safe pointers in Section~\ref{section:design}.
We discuss the issue of backward compatibility with legacy C libraries in
Section~\ref{section:compatibility}.
Next, Section~\ref{section:impl} covers the implementation.
We describe the performance and memory consumption evaluation in
Section~\ref{section:eval}.
After that, we report our experience of porting the benchmarks in
Section~\ref{section:port}.
In Section~\ref{section:related}, we compare our work with other key-lock
check works in details, and we also discuss and compare with two other major
types of temporal memory safety solutions.
We then briefly summarize the most important directions for future work
and conclude the paper in Section~\ref{section:concl}.

\section{Background on {\CheckedC}}
\label{section:bg}

{\CheckedC}~\cite{CheckedC:SecDev18} is an extension to C that
enforces spatial memory safety.
It takes inspiration from prior work on
safe-C dialects~\cite{Cyclone:ATC02,ControlC:CASES02, Deputy:ESOP07}
but differs in that it favors \emph{easy incremental porting of and
interoperation with legacy code}.
This section briefly introduces three key concepts of {\CheckedC}.
Design details can be found in the language
specifications~\cite{CheckedCTR:Microsoft} while
\citet{CheckedCModel:CSF22} provide a formal model of {\CheckedC}.


\paragraph{Checked Pointers}
\checkedc extends C with three new types of spatially-safe
checked pointers~\cite{CheckedC:SecDev18}.
{\ptrT} is a pointer to a single type-\texttt{T} object and thus
cannot be used in pointer arithmetic;
{\arrayptrT} and {\ntarrayptrT} (``null-terminated'' {\arrayptr}) are array
types and can be used in pointer arithmetic expressions.
A checked array pointer is associated with a {\it bounds expression}
that delineates the pointed-to array's size. Bounds expressions are normal
program expressions and serve as an \emph{invariant} for checking the validity
of pointer-related operations: the compiler rejects the program if it
detects a violation (e.g., an out-of-bounds access) statically
and inserts dynamic bounds checks when it cannot prove safety statically.
Bounds expressions are not stored \emph{in place}, i.e., as ``fat''
pointers~\cite{Cyclone:ATC02,CCured:TOPLAS05}.
This improves backward compatibility with legacy C code as the
run-time pointer representation remains unchanged~\cite{CheckedCTR:Microsoft}.
{\CheckedC} strictly prohibits casting or assigning a raw C
pointer to a checked pointer, which prevents forging checked pointers
from unsafe sources.

\paragraph{Checked Region}
{\CheckedC} allows mixing uses of checked and raw C pointers, which helps
incremental conversion of legacy C programs. Programmers can use
the  {\tt \_Checked} keyword to explicitly put a block of code
(from a single statement to a whole source file) into a
checked region within which uses of legacy C pointer types are disallowed and
spatial safety is provably assured~\cite{CheckedCModel:CSF22}.
Similarly, programmers can use the {\tt \_Unchecked} keyword to enclose
code in an {\it unchecked region} that disables the compiler's
memory safety checks.

\begin{figure}[tb]
    \centering
    \includegraphics[scale=0.30]{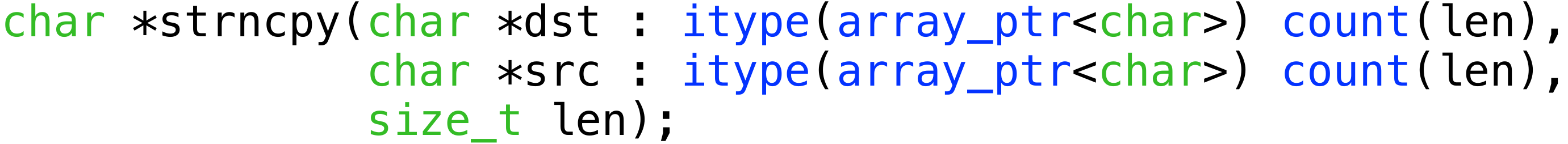}
    \caption{Bounds-safe Interface for {\tt strncpy}}
    \label{fig:bsi}
\end{figure}

\paragraph{Bounds-Safe Interface}
{\CheckedC} provides bounds-safe interfaces (BSI) for better interaction
between checked and unchecked code (legacy libraries and unported source code).
Programmers can declare a BSI for an unchecked function
with {\tt itype} parameters so the function can be called by both checked
and unchecked code.
{\itype} (short for \emph{inter-op type}) is a new keyword introduced by
{\CheckedC} to annotate pointers to be used in both checked and
unchecked contexts.  Notably, the compiler treats an {\itype} pointer
as a raw C pointer when it is passed to or used in unchecked C code,
and treats it as a checked pointer (enforcing necessary memory safety checks)
in checked regions.
Figure~\ref{fig:bsi} shows an example of the BSI for {\tt strncpy};
checked code can pass an {\arrayptr} to {\tt strncpy} via the BSI,
and unchecked code can pass in a raw C pointer.

\section{Temporally Memory Safe Pointers}
\label{section:design}



This section presents our new approach to integrating temporally
memory-safe pointers in {\CheckedC}. We begin
with the motivations and overview of our design and then describe the details
of our various pointer types and their metadata, including how we
manage locks and keys. 
We discuss how we handle backward compatibility with legacy libraries
in Section~\ref{section:compatibility}.

\subsection{Design Overview}
\label{section:design:overview}

\subsubsection{Goal}
\label{section:overview:goal}
We aim to prevent \emph{all} temporal safety
violations, of which there are three varieties: \emph{use after free} (UAF),
\emph{double free}, and \emph{invalid free}. While UAFs can happen with
pointers to stack objects, they are arguably challenging to
exploit~\cite{ExploitTrend:MS13,DANGNULL:NDSS15}. As a result, they
are ignored by most related
work~\cite{PTAuth:Sec21,ViK:ASPLOS22,DangSan:EuroSys17,pSweeper:CCS18,
  CRCount:NDSS19,HeapExpo:ACSAC20,MarkUs:Oakland20,FFmalloc:Sec21}.
Our design aims to detect them nevertheless to ensure comprehensive
temporal memory safety enforcement. At present, we do not check for
spatial safety violations via the new pointers; our focus is to
prototype temporal safety enforcement and its performance. Though the
engineering effort of adding spatial checks is nontrivial (estimated to be
at least one person-year; more details in \secref{section:impl:unimpl}),
they are essentially orthogonal to temporal safety support (e.g., they do not
affect pointer/object representation).

\subsubsection{In-place metadata}
%
Prior work has enforced temporal safety by checking that a pointer's
\emph{key} matches its object's \emph{lock} on a dereference. Key-lock
metadata is typically kept in data structures disjoint from the
pointer and object~\cite{Guarding:SPE97,
XuMemSafe:FSE04,CETS:ISMM10,UAFSan:ISSTA21}.
For example, CETS~\cite{CETS:ISMM10} uses a two-level lookup table to locate
its keys and locks.
Disjoint metadata approaches afford good backward compatibility with
legacy code and lowers the possibility of metadata corruption when
spatial memory safety is not assured.  However, to query and update the
metadata, a program must first \emph{dynamically} locate it,
and the lookup procedure can be very slow.
Inspired by this observation, we decided to see whether
pointers with in-place keys (i.e., \emph{fat}
pointers~\cite{Cyclone:ATC02,CCured:TOPLAS05}) and objects with
in-place locks could make metadata accesses significantly faster.

%
Similar to previous key-lock approaches~\cite{SafeC:PLDI94,Guarding:SPE97,
CETS:ISMM10,PTAuth:Sec21,ViK:ASPLOS22}, we associate each pointer with a key and
its referent with a lock.
Memory allocation sets the lock to a unique value and returns a
pointer with a key set to the same value as the lock;
memory deallocation invalidates the lock to a value that will
never be used for any key.
On a dereference, the compiler inserts a check (omitted if proven safe)
to confirm that the pointer's key matches its referent's lock.
A failed key check signals a temporal memory safety violation.
Different from disjoint key-lock
methods~\cite{CETS:ISMM10,Guarding:SPE97,UAFSan:ISSTA21},
our solution locates the key as part of a fat pointer, and locates
the lock just before the referent object.
Thus, the location of a key is \emph{statically}
known, and the location of a lock is either statically known or can be
computed by a few simple arithmetic and bitwise instructions
(\secref{section:design:mmptr}).
Consequently, metadata propagation, updates, and validity checks are
much faster.


\subsubsection{{\CheckedC}}
There are two main benefits offered by {\CheckedC} as a host language for
temporally safe pointers with in-place metadata. First,
{\CheckedC} ensures that metadata will not be corrupted or fabricated.
{\CheckedC}'s checked pointers (\secref{section:bg}) can only originate
from a heap allocation, the address of a stack/global/thread-local object,
or from another checked pointer. {\CheckedC}'s type system disallows
casting a raw C pointer to a checked pointer.
This eliminates the possibility of forging checked pointer
metadata from untrusted sources at compile time.
Second, fat pointers integrate well with {\CheckedC}'s approach for
enforcing spatial memory safety.
Fat pointer approaches for spatial safety usually have at least two
fields added per pointer,
(base and upper bound or size)~\cite{Cyclone:ATC02,CCured:TOPLAS05}.
Additional metadata for temporal memory safety would make fat pointers
heavier, thus causing slow metadata propagation and updates.
However, {\CheckedC} achieves spatial memory safety with
low performance overhead~\cite{CheckedC:SecDev18,FreeBSDCheckedC:SecDev20}
\emph{without} using fat pointers. Therefore, 
we may afford to combine its current mechanism with fat pointers to
realize full memory safety efficiently.
Additionally, {\CheckedC} gives programmers fine-grained control over the
source code, permitting manual handling of backward compatibility
issues---a major concern for fat pointer approaches.

\subsection{Pointer to Singleton Memory Objects}
\label{section:design:mmptr}

\newlength{\microfigwidth}
\begin{figure*}
    \centering
    \microfigwidth .4\textwidth
    \strut\hfill
    \subfloat[Two {\tt mm\_ptr}s to a {\struct}. One points to the beginning
    and another points to the middle of it.
    \label{fig:mmptr}]%
        {\includegraphics[width=1.15\microfigwidth]%
            {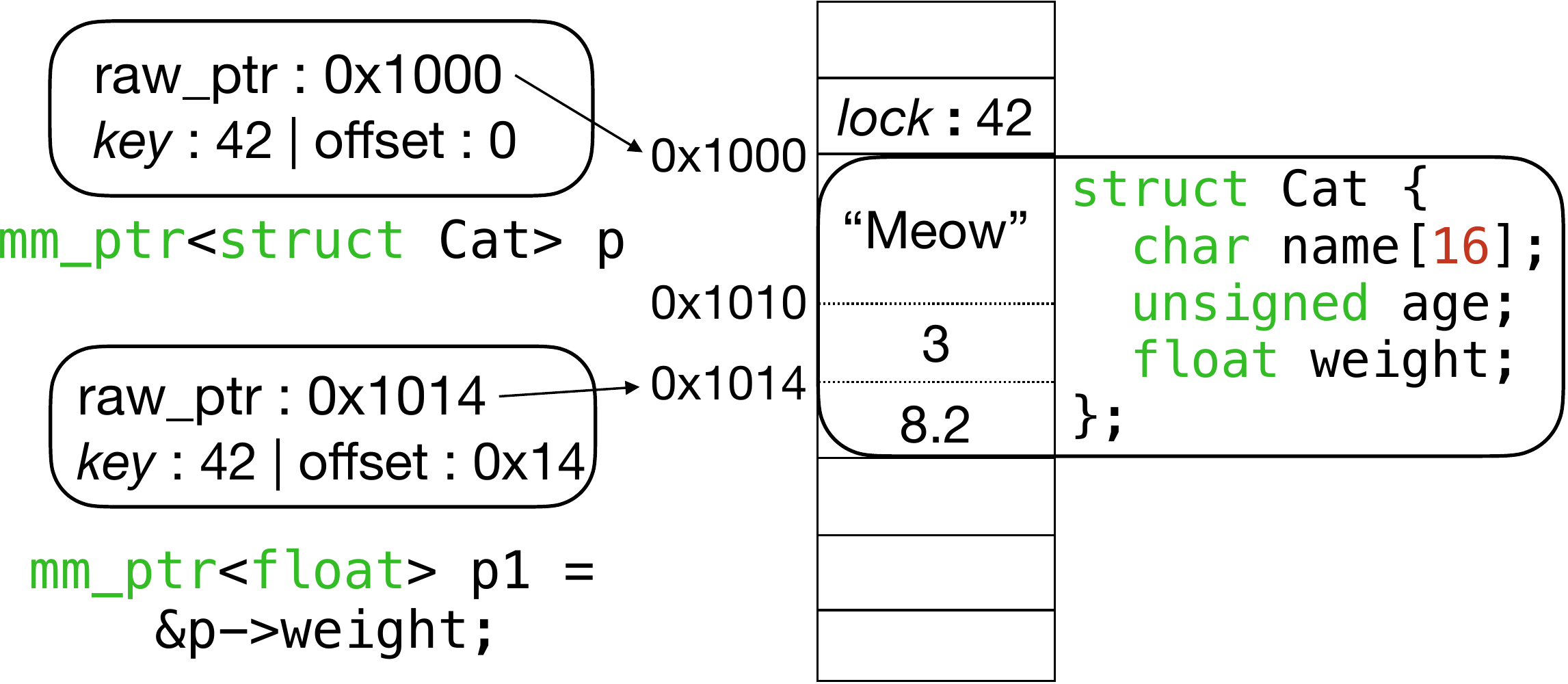}%
        }
    \hfill
    \subfloat[Two {\tt mm\_ptr}s to a large array of integers greater than
    4~GB.
    \label{fig:largearrayptr}]%
        {\includegraphics[width=1.12\microfigwidth]%
            {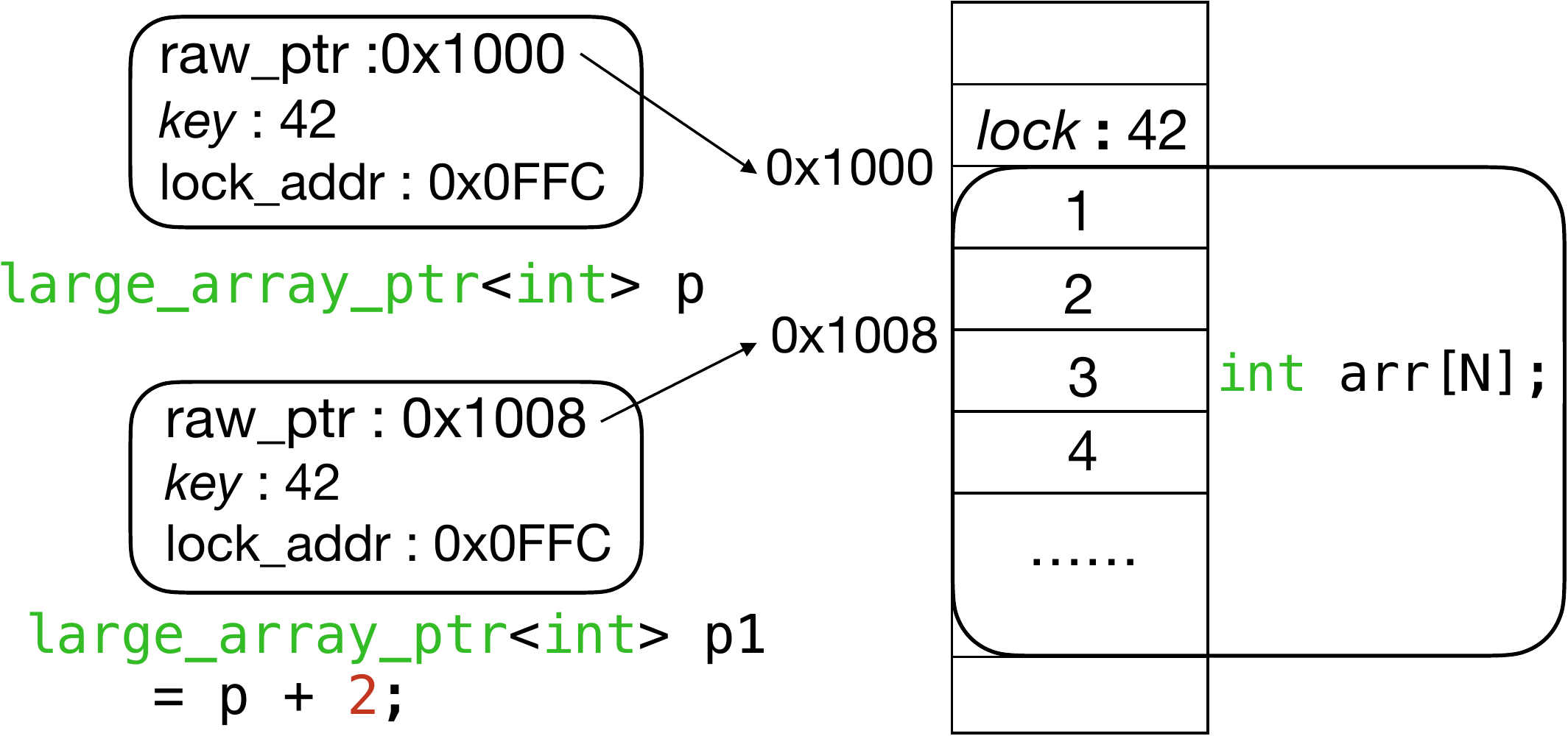}%
        }
    \hfill\strut
    \caption{Structure of Temporally Safe Fat Pointers}
    \label{fig:checkedc_perf}
  \end{figure*}

We add four new types of checked pointers to {\CheckedC}\@.
{\mmptrT}\footnote{``mm'' stands for ``memory management''.}
extends {\CheckedC}'s {\ptrT} (\secref{section:bg}) and thus
types a pointer to a single memory object of type {\tt T}, while
{\mmarrayptrT} extends {\arrayptrT} and types a pointer to an array
of objects of type {\tt T};
the latter allows pointer arithmetic and array subscripts but the
former does not. {\largeptrT} and {\largearrayptrT} play similar
roles, but may point to exceptionally large objects. We cover each in
the coming subsections,  starting in this subsection with {\mmptrT}.

An {\mmptrT} consists of three logical components: a raw C pointer
to its referent memory object of type {\tt T}, a key, and an offset used
to compute the location of the lock.
{\tt T} can be any singleton data type:
a primitive type, a {\tt struct}, or a {\tt union}.
Although a lock is always located right before its object, and
pointer arithmetic is disallowed on {\mmptr}, it is common for
programs to use the address-of operator \singlequote{\&} to
compute the address of an inner field of a {\struct} and to assign
the result to a pointer.
When a checked pointer points to the beginning of a {\tt struct}, it
knows the location of the lock at compile time,
but when the address-taken field is not the first field of the
{\struct}, a checked pointer would lose track of the lock's location.
We solve this problem by adding a second piece of metadata which
contains the offset of the pointer from the referent's start address.
The key and the offset share one single 64-bit integer.\footnote{Our
  current prototype only supports 64-bit systems;
  Section~\ref{section:design:largeptr} describes a straightforward
  design for 32-bit systems.}
We describe the bits allotted for the key and offset in
Section~\ref{section:design:metadata}.

Figure~\ref{fig:mmptr} shows an example of two {\mmptr}s pointing to
a {\tt struct Cat}.
Eight extra bytes for the lock are allocated at the beginning of the
memory object (it may also need to allocate eight more bytes of padding
to properly align the first byte of the memory object), and
the lock is set to a unique number ($42$ in Figure~\ref{fig:mmptr}).
In Figure~\ref{fig:mmptr}, both {\mmptr}s have the same key value and
share the same lock.
Pointer {\tt p} points to the beginning of the {\struct}
and thus has offset $0$ while
{\tt p1} is created by an address-of expression and has offset {\tt 0x}$14$.

The dynamic temporal memory safety check for a pointer dereference
is straightforward: the compiler inserts instructions to extract
the key from the {\mmptr}, compute the lock's address by simply
subtracting the offset from the {\mmptr}'s raw C pointer, load the lock,
and check if the key matches the lock.
At memory deallocation, both the referent's memory and the lock and any
added padding are released, and the lock is set to a reserved value that
is never used for any key. As a result,
there is no need to invalidate any {\mmptr} to the freed memory because
the key of a dangling pointer will never match the lock.
Additionally, although not a pointer dereference,
using the {\tt ->} and {\tt []} operators to
compute the address of an inner object of a {\struct} or an array
pointed to by
a dangling checked pointer (e.g., {\tt \&p->obj} and {\tt \&p[i]})
will be checked and caught as a runtime error.

\subsection{Pointer to Arrays}
\label{section:design:mmarrayptr}

\Comment{JTC: JZ, T can be any type?  Can it be a standard C array?
What's the difference between an {\mmarrayptr} of an {\mmarrayptr} and
an {\mmarrayptr} of a regular C array?}

\Comment{JTC: JZ, I think you're missing my point: it isn't clear what
the restrictions are on type T in the text below.  When reading the
text, you stated that we can have a checked array of a checked array
to create a multi-dimensional checked array.  That sound good, but it
raises the question of what happens when T is just a regular C array
instead of a checked array.  Is that a type error, or is that supported?}

\Comment{JZ: T can be any type, including a C array.}

Similar to singleton memory objects, the lock of an array is also located
right before the object.
Since array pointers allow pointer arithmetic, they
must track the location of the lock when they do not
point to the start of an array.
{\mmarrayptr} has the same inner structure as {\mmptr}, and
the {\tt offset} subfield is the distance between the current raw C
pointer and beginning of the array's first element.
{\mmarrayptrT} extends {\CheckedC}'s {\arrayptrT} and can point to an
array of any data type, including an array of pointers.\footnote{More about
{\CheckedC}'s type system is in \citet{CheckedCModel:CSF22} and
Chapter~2 and Chapter~5 of \citet{CheckedCTR:Microsoft}.}
Pointer dereference checking is the same as that
done for {\mmptr}s (\secref{section:design:mmptr}).

For pointer arithmetic, we follow {\CheckedC}'s
design~\cite{CheckedCTR:Microsoft}: the result of
a pointer arithmetic operation on {\mmarrayptr} can be either an {\mmptr} or an
{\mmarrayptr}, depending on the type of pointer to which the result is assigned.
In both cases, the resulting pointer shares the same key as the source pointer,
and the offset is updated based on the arithmetic.
C programs also use the address-of operator \singlequote{\&}
to create a pointer to an element of an array.
In standard C, an address-of expression with index {\tt i} of an
array pointer {\tt p} is semantically equivalent to a pointer addition
operation of {\tt p} and {\tt i}, i.e., {\Lstinline |&p[i] == p + i|}.
However, {\CheckedC} differentiates these two types of operations.
By default, an address-of expression of an {\arrayptr} generates
a {\ptr} or an {\arrayptr} with its bounds set to zero
(disallowing pointer dereferences).
Programmers must manually do a dynamic bounds cast~\cite{CheckedCTR:Microsoft}
if they want to use the result as an array pointer.
Like pointer arithmetic, we also follow {\CheckedC}'s design
on address-of expressions on an element of an array: by default, the result of
an address-of expression on an {\mmarrayptr} is an {\mmptr}.
Programmers should use a pointer arithmetic operation if they want to use
the result as an {\mmarrayptr}.

\subsection{Key and Offset Allotment}
\label{section:design:metadata}
\begin{table*}[t]
\caption{Program Statistics.  For {\bf Shared Array of Ptr},
    {\bf Call} is the number of call sites to library functions
    with double-pointer argument(s). {\bf Largest} and {\bf Total} are the
    size of the largest shared array and total size of shared arrays of
    pointers, respectively.}
\label{table:stats}
\centering
{\sffamily
\footnotesize{
\resizebox{\textwidth}{!}{
\begin{tabular}{@{}lrrrrrr|lrrrrrr@{}}
    \toprule
    \multirow{2}{*}{{\bf Program}} & \multirow{2}{*}{{\bf LOC}} &
    {\bf Largest} & {\bf Largest } & \multicolumn{3}{c|}{{\bf Shared Array of Ptr}} &
    \multirow{2}{*}{{\bf Program}} & \multirow{2}{*}{{\bf LOC}} &
    {\bf Largest} & {\bf Largest } & \multicolumn{3}{c}{{\bf Shared Array of Ptr}} \\
     & & {\bf struct} & {\bf Heap Obj } & {\bf Call} & {\bf Largest} & {\bf Total} &
     & & {\bf struct} & {\bf Heap Obj } & {\bf Call} & {\bf Largest} & {\bf Total} \\
    \midrule
    500.perlbench & 291~K & 8~KB & 25~MB & 5 & 0 & 0 &
    curl-7.79.1 & 122~K & 8~KB & 10~MB & 29 & 200~B & 200~B \\
    502.gcc & 972~K & 30~KB & 5~MB & 0 & 0 & 0 &
    ffmpeg-n4.1.7 & 1.1~M & 64~MB & 48~MB & 46 & 0 & 0 \\
    505.mcf & 3~K & 648~B & 289~MB & 0 & 0 & 0 &
    httpd-2.4.46 & 204~K & 8~KB & 2~KB & 383 & 200~B & 3~KB \\
    519.lbm & 1~K & 216~B & 204~MB & 0 & 0 & 0 &
    nginx-1.21.1 & 145~K & 32~KB & 224~KB & 2 & 0 & 0 \\
    525.x264 & 73~K & 33~KB & 4~MB & 2 & 0 & 0 &
    openssl-3.0.0 & 403~K & 15~KB & 16~MB & 181 & 280~B & 1~KB \\
    557.xz & 20~K & 64~KB & 320~MB & 0 & 0 & 0 &
    php-7.4.9 & 1.2~M & 1~MB & 9~MB & 92 & 48~B & 48~B \\
    538.imagick & 174~K & 228~KB & 5~MB & 0 & 0 & 0 &
    redis-6.2.6 & 148~K & 10~MB & 3~MB & 26 & 0 & 0 \\
    544.nab & 16~K & 720~B & 1~MB & 0 & 0 & 0 &
    sqlite-3.37.0 & 598~K & 8~KB & 74~MB & 12 & 0 & 0 \\
    \bottomrule
\end{tabular}
}}
}
\end{table*}


The 64-bit metadata field of both {\mmptr} and {\mmarrayptr} is split into
two subfields: a key and an offset, as illustrated in Figure~\ref{fig:mmptr}.
By default, we use the highest 32 bits for the key and the remaining
32~bits for the offset, allowing memory objects of up to 4~GB in size.
This should suffice for most programs.
(Currently the \emph{entire} address space of WebAssembly is
4~GB~\cite{wasm:memory64}.)
We measured the largest {\tt struct} and heap object of
all the C programs in the {\SPEC17} benchmark suite
and eight popular large open-source C programs, totaling $5.5$~million lines
of code.
Table~\ref{table:stats} shows the statistics.
We measured the size of {\tt struct} using an LLVM IR pass.
To collect dynamic data, we used LLVM's test-suite to
run the SPEC benchmarks with the {\tt train} dataset.
For {\tt httpd} and {\tt Nginx}, we used {\tt ab}~\cite{ab:Apache} to fetch
random files ranging from 1~MB to 32~MB from a local server.
The remaining six programs all have extensive built-in test cases.
The largest {\tt struct} is 64~MB from {\tt ffmpeg-n4.1.7},
and the largest dynamic heap object is only 320~MB from {\tt 557.xz}.

We reserve integer $0$ as the invalid key value. Thirty-two bits offer
over 4~billion different keys which should suffice for most programs.
(\citet{FFmalloc:Sec21} counted the \emph{accumulated} number of heap
allocations of the SPEC~CPU2006 benchmarks, and the largest
is 365~million.)
For security-sensitive programs, we provide a compiler option to increase
the number of bits for the key to 48: we leverage the unused highest 16 bits
of a raw C pointer on a 64-bit system and combine it with the 32 bits in the
key metadata subfield. Forty-eight bits provide over 281~trillion keys.
Consequently, the odds of key collision are extremely low.
Alternatively, we can provide compiler options to control the key-offset
allotment within the 64~bits of metadata. For example, if programmers
are certain that a target program never allocates objects larger
than 16~MB, they can configure the compiler to use 40-bit keys
(over 1 trillion keys) and 24-bit offsets.
When compiled with such an option, the compiler will insert
size checks before memory allocation operations to
guarantee that the compiled program will not allocate objects larger
than the maximum offset.

Ideally, the runtime should generate a \emph{unique} and \emph{random}
key for each new allocation, but the performance cost can be prohibitive---e.g.,
x86-64's random number generation
instruction {\tt RDRAND}~\cite{IntelManual:2019} takes around
$100$ to $1,500$ CPU cycles on Intel processors and
up to $2,500$ cycles on AMD processors~\cite{CPUPerf:AgnerFog}.
Additionally, to guarantee the uniqueness of keys, the runtime
would need to maintain a set of keys in use and check whether a newly
generated key is already in use, exacerbating the performance cost.
We describe our key generation for x86-64 in
Section~\ref{section:impl:keycheck}.
Other architectures may take different approaches.

%

\subsection{Pointers to Exceptionally Large Objects}
\label{section:design:largeptr}


As Section~\ref{section:design:metadata} explains,
{\mmptr} and {\mmarrayptr} can point to a memory object up to 4~GB in size.
To support the rare case when a \emph{single} memory object is greater
than 4~GB, we add two types of checked pointers ({\largeptr} and
{\largearrayptr}) for exceptionally large {\struct}s and arrays, respectively.
Similar to a few prior works~\cite{Guarding:SPE97,XuMemSafe:FSE04,CETS:ISMM10},
these two checked pointers both
have two separate 64-bit metadata fields:
one for the key and one for the address of the lock.
Figure~\ref{fig:largearrayptr} shows an example.
A {\largeptr} can also be used to point to a single element of an
exceptionally large array pointed to by {\largearrayptr}
because the distance between the element
and the lock of the array can be as large as the array.


We considered using the 3-field structure of large pointers for
all checked pointers.  It simplifies the language design and implementation,
and it provides tremendously more keys and consequently lowers the
possibility of key collisions compared with the 32-bit or even 48-bit key space.
It is also our choice for 32-bit systems because a single 32-bit field
is insufficient to provide both high entropy for keys and to support
large object sizes.

However, we choose the current design over
the universal structure option mainly due to performance and memory
consumption concerns. Specifically, a 3-field pointer takes 192~bits
on 64-bit systems, and for the AMD64 ABI~\cite{SystemVAMD64ABI},
function parameters and return value of structure types that contain
two eight-byte integers should be passed by registers,
while structures of more than two eight-byte integers should be passed
via the stack memory.
A compiler may opt to break a structure argument into multiple
scalar arguments, but it is not guaranteed.
Considering that DRAM is usually two orders of magnitude slower than
registers~\cite{SysPerf:Gregg}, the universal 3-field pointer design is
potentially prohibitive in terms of performance.


\subsection{Lock Management}
\label{section:design:lock}

Because C uses different memory management mechanisms for
the heap and the stack, our {\CheckedC} extensions manage
the locks for heap (\secref{section:design:lock_heap})
and stack (\secref{section:design:lock_stack}) objects differently.
Additionally, we add locks for certain address-taken global objects
(\secref{section:design:lock_global}).
Next, we describe how our enhanced {\CheckedC} compiler manages locks for
the three types of memory.

\subsubsection{Heap}
\label{section:design:lock_heap}

We add custom memory allocator/deallocator wrappers, dubbed
{\mmalloc}/{\mmfree}, to a runtime library.
In addition to the requested memory for an object, an {\mmalloc}
allocates extra bytes for the lock and padding for the system's
alignment requirement. It generates a new key, sets the lock to the
key, and returns a new checked pointer with the key and an offset of zero.
{\mmfree} frees the requested memory plus the lock
and padding, and it invalidates the lock before it returns.

Additionally, an {\mmfree} also detects invalid free and double free bugs.
Before an {\mmfree} calls the underlying memory deallocator {\tt free},
they first check the offset value and signals an invalid free bug in case of
a non-zero offset.
They then do a key-lock check. If the key does not match
the lock (the lock being either the reserved invalid key value or a new
valid key value), a double free bug is detected.
Finally, in the case when a pointer arithmetic operation
overflows the offset to 0 (incidentally or maliciously), an invalid
free error will be caught because chances are that
the key would not match the ``lock'' (whatever resides before the updated
pointer).

\subsubsection{Stack}
\label{section:design:lock_stack}

%
%
\begin{figure}[tb]
    \centering
    \includegraphics[scale=0.42]{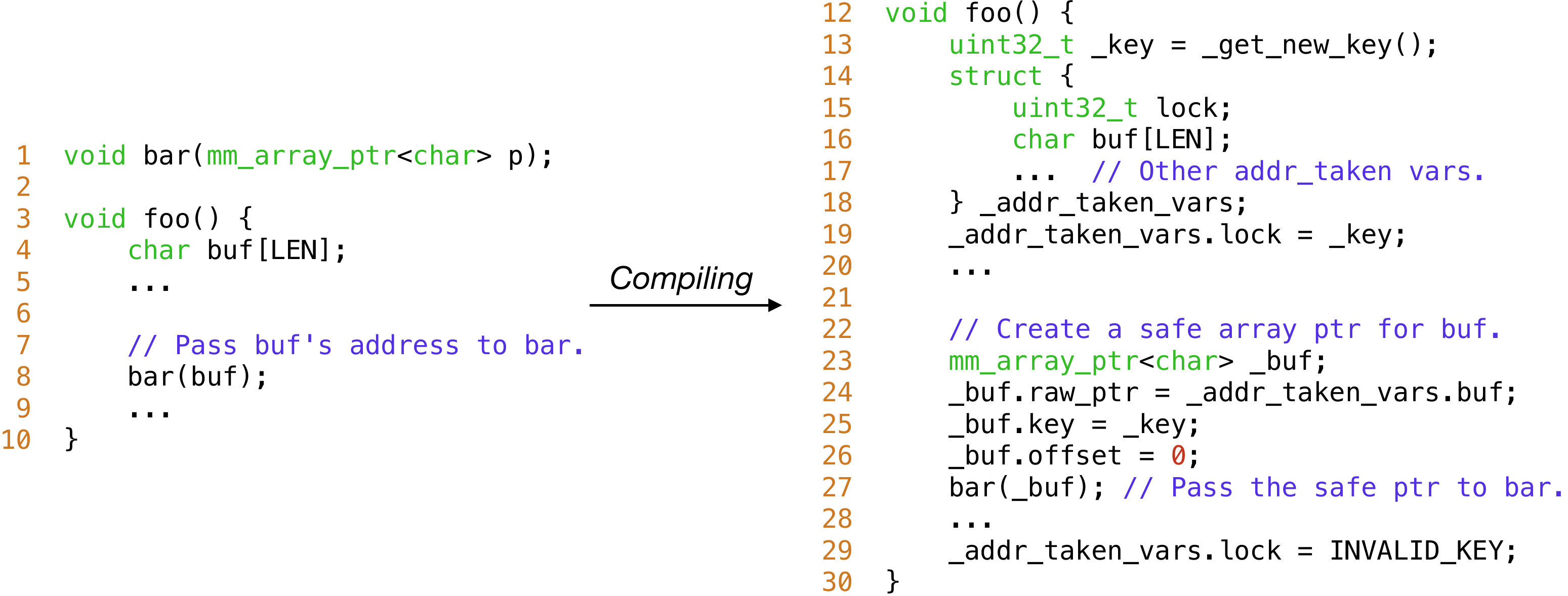}
    \caption{Pair Address-taken Stack Variables with a Lock. The code is only
    for illustration purposes. The real transformation happens during LLVM IR
    code generation.}
    \label{fig:addr_taken_stack}
\end{figure}

Most stack objects have the same lifetime as their enclosing function;
however, compilers may optimize memory by reusing some stack slots for
different objects.  Our compiler guarantees that all
address-taken stack objects allocated in the same function have the same
lifetime
and thus can share the same lock. Specifically, our compiler assembles
all fixed-size address-taken local variables
into one structure and adds a lock (plus necessary padding for alignment)
to the beginning of the structure.
Because each variable's offset within the structure is known statically,
when the variable's address (including its sub-object if it is an aggregate)
is taken and assigned to a checked pointer, the compiler can
compute the pointer's offset metadata.
For each variable-sized array, the compiler creates a lock for it individually.
In the function prologue, the compiler inserts a call to the function in the
runtime library that returns a new key; the compiler then sets all the locks
in the frame to the key and sets the key metadata of every checked pointer
that points to address-taken stack objects.
In the function epilogue, the compiler revokes the lock(s) by setting them to
the reserved invalid key value.
Since a lock is a field of the {\struct} that contains address-taken stack
objects, the lock invalidation guarantees that the lifetime of all such
{\struct}s lasts to the end of the function, thus preventing
the compiler from reusing stack slots for different objects.

Figure~\ref{fig:addr_taken_stack} shows an example. The start address of
a local array {\tt buf} is passed to a function {\tt bar} (line~8)
that takes a checked array pointer argument.
After compilation, {\tt buf} is associated with a lock (line~15), and
a temporary {\mmarrayptrinst{char}} with its key set to the lock and
offset set to 0 (line~23--26) is generated for the call to {\tt bar} (line~27).

\subsubsection{Global}
\label{section:design:lock_global}

Global objects are never freed, and therefore, pointers to them do not suffer
from use-after-free bugs (invalid frees are still possible).
However, a single pointer may point to a global, stack, or heap object,
depending on the execution context.
For example, in Figure~\ref{fig:addr_taken_stack}, {\tt bar} takes
a pointer to a stack object from {\tt foo} (line~8), but another function may
pass a pointer to a global object to {\tt bar}.
The compiler needs to know whether to insert a key check for such pointers.
However, it is an undecidable problem to statically infer to which
variables a pointer points~\cite{IsPASolved:PASTE01}.
Our solution is to have the compiler put a lock right before each
global variable whose address is taken \emph{and} assigned to a checked pointer.
We reserve one value as the lock for all global variables because
global variables are never freed.
We also support directly assigning a string constant to a
{\mmarrayptrinst{char}} because string constants are essentially
static constant global variables.

We acknowledge that paring a global object with a lock and checking the
validity of pointers to global objects incurs unnecessary overhead.
However, this design provides a universal interface for programmers to use:
a checked pointer will always be checked, and programmers do not need to
consider to which type of memory it points.
Our empirical experience on porting the C programs indicates that address-taken
global variables are infrequent.
We therefore believe our design's benefits greatly outweigh
its performance and memory overheads.

\paragraph{External Global Objects}
Address-taken {\tt extern} global variables pose a subtle challenge.
If one translation unit \emph{declares} an {\tt extern}
global variable and assigns its address to a checked pointer
while the translation unit that \emph{defines} the variable does not
assign its address to a checked pointer,
the compiler will not know whether to allocate a lock for the variable.
We solve this problem with a new qualifier {\checkable} and
the name mangling technique that is commonly used in object-oriented
programming languages such as {C{\tt ++}}.
Specifically, programmers should use the {\checkable} type qualifier to label
the definition (and optionally, the {\tt extern} declaration) of
an address-taken global variable.
The compiler will automatically mangle the names of
address-taken {\tt extern} global declarations and
the names of {\checkable} variables, and the compiler will add a lock
for {\checkable} variables.
An ``undefined symbol'' error will be raised during linking if
the definition of an address-taken global variable is missing
the {\checkable} keyword.

\section{Compatibility with Legacy C Libraries}
\label{section:compatibility}

A {\CheckedC} program is rarely self-contained; it relies on legacy
libraries. Such libraries' APIs (in header files) are expressed using
raw pointers or bounds-safe interfaces (\secref{section:bg}). As
such, they make plain their expectations about representation: legacy
and {\tt itype}-annotated pointers have no in-place metadata. The
{\CheckedC} compiler will ensure that only compatible pointers are
passed to/from libraries. We provide library routines to
\emph{(un)marshal} pointers that programmers can use to satisfy
compatibility requirements. We focus on the interoperability between
checked code and legacy libraries in this section. We describe the
challenges and our solutions specific to the interaction between checked
and unchecked parts in a partially ported program in Section~\ref{section:port}.

\subsection{Type Compatibility}
\label{section:compatibility:type}

Recall Figure~\ref{fig:bsi} from \secref{section:bg} which gives
the bounds-safe interface for {\tt strncpy}. Its arguments are
declared with an \emph{inter-op} type
{\Lstinline|char *p : itype(array_ptr<char>)|}.
With such a defintion, the {\CheckedC} compiler will force checked code
to pass an
{\Lstinline|array_ptr<char>|}, but unchecked code may pass a %
{\Lstinline|char *|}. In checked code, we can also safely call
{\tt strncpy} with an {\Lstinline|mm_array_ptr<char>|}. This is
because {\Lstinline|mm_array_ptr<char>|} is \emph{compatible} with
{\Lstinline|array_ptr<char>|}.\footnote{Recall that, in the full design, all
  {\tt mm}$*$ pointers are bounds-checked.} That is, doing something like %
{\Lstinline|x = (ptr<int>)y|} where {\tt y} is an
{\Lstinline|mm_ptr<int>|} assigns only the raw pointer of {\tt y}
into {\tt x}, skipping the key.  The pointer {\tt x} will
not be temporally safe---the lock will be present in
the pointed-to data, but cannot be checked; of course, 
legacy library code will not be doing temporal checks anyway.

Now suppose we had
\begin{lstlisting}[morekeywords={array_ptr,ptr}]
  void foo(int **p : itype(array_ptr<ptr<int>>));
\end{lstlisting}
We could \emph{not} safely pass {\tt foo} a pointer {\tt q} of checked
pointer type
{\Lstinline|mm_array_ptr<mm_ptr<int>>|}. While we can strip off
the key for {\tt q}, that is not enough: the
function {\tt foo} will assume the array points to
single-word pointers, but the caller would be passing in an array of
fat pointers instead. The compiler will prevent this mistake because
it will deem types {\Lstinline|array_ptr<ptr<int>>|} and
{\Lstinline|mm_array_ptr<mm_ptr<int>>|} incompatible.

A similar problem can arise when passing a {\tt struct} with an
{\mmptr}/{\mmarrayptr} field to a library function. However, as a
practical matter, this will not happen because
{\tt struct}s defined in legacy libraries will not declare
the use of fat pointers, even in bounds-safe interfaces; conversely,
{\struct}s defined in applications will not be seen by libraries.
We assume that programmers cannot or will not change the source code of
legacy libraries, and as a result,
we just need to focus on what to do in situations like the one
involving {\tt foo} above (i.e., shared arrays of fat pointers).

\subsection{Marshalling}
\label{section:compatibility:marshal}

%
%
\begin{figure}[tb]
    \centering
    \includegraphics[scale=0.32]{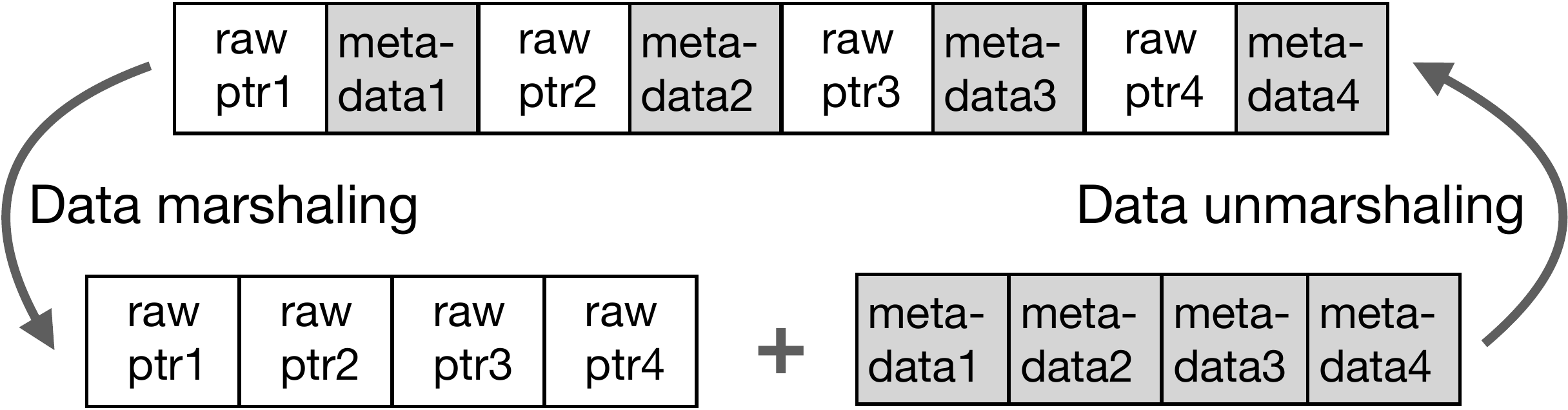}
    \caption{Marshaling an Array of Four Fat Pointers}
    \label{fig:marshal}
\end{figure}



The incompatibility shown above means that,
unfortunately and inevitably, metadata must be separated from
some fat pointers passed to libraries.  We therefore
employ data marshaling~\cite{PtrSplit:CCS17}:
to pass an array of fat pointers to a library function, the
programmer must insert code that copies raw pointers within the
original array into a new array of raw pointers.  Our system
provides data marshaling utility functions such as the one below to
assist the programmer:
\begin{lstlisting}[morekeywords={mm_ptr,mm_array_ptr}]
  void **_marshal_ptr_array<T>(mm_array_ptr<mm_ptr<T>> p, unsigned size);
\end{lstlisting}
The function allocates a new array of raw C pointers, copies {\tt size} raw
C pointers from {\tt p} into the new array, and returns a pointer to the new
array. The
returned {\tt void **} pointer can be cast to a legacy pointer as needed.
Figure~\ref{fig:marshal} illustrates
marshaling an array of four checked pointers.

Programmers can continue to use the original array of checked pointers after
such a library call if the callee only reads the array.
If the library function can write to the array, programmers have two options.
First, they can continue to use the array of raw pointers returned
from the marshaling procedure but at the cost of reduced temporal
safety guarantees; this is possible because {\CheckedC} allows raw
and checked pointers to coexist outside checked regions (\secref{section:bg}).


Second, if programmers want to revive the checked pointers after the call,
they need to write an unmarshalling procedure to reassociate the
array of raw pointers and their corresponding metadata.
Fortunately, this situation is uncommon; {\tt qsort} in {\tt libc}
is the only such library function of which we are aware. We wrote a small
wrapper function (16~lines) for it to call our marshalling procedure and
the original {\tt qsort} and then recover the array of checked pointers.
Note that this inconvenience is shared by disjoint metadata approaches
because essentially such library functions break the
association between a raw pointer and its metadata and there is no easy
way to recover the association both automatically and soundly.
For example, CETS, a compiler-based disjoint key-lock
approach~\cite{CETS:ISMM10}, uses the address of a pointer as an index
to locate its metadata, and therefore a {\tt qsort} call mentioned above
will invalidate a pointer's connection with its metadata.
CETS writes its own version of {\tt qsort} to solve this problem.

\mwh{I realized that the marshal function must be calling malloc, so I
  added the point of calling free. This brings up that marshalling is
  a potential safety problem: You have malloced something you are not
  temporally checking?}


\paragraph{Performance Cost}
Data marshalling could be expensive if it occurs often or must
marshal large arrays. To see whether this might be the case, we analyzed the
eight C programs in {\SPEC17} and eight large open-source
C programs (totalling $5.5$~million lines of code)
to estimate the sizes of shared arrays of pointers.
Specifically, we instrumented programs to record the bounds of each heap object,
and for each call to a library function with double-pointer argument(s),
the runtime library computes the size of a possible array of pointers.
For example, if a double-pointer {\tt 0x1010} is passed to a library function,
and our runtime records an object ranging {\tt 0x1000}--{\tt 0x1020},
then we assume that there is a 16-byte shared array of pointers.


As Table~\ref{table:stats} shows (see \secref{section:design:metadata}
for how we ran the programs),
none of the SPEC benchmarks and only four of the eight applications
use shared arrays of pointers on the heap. The largest single shared array
is only 280~bytes ({\tt openssl});
the largest total size of shared arrays is only 3~KB ({\tt httpd}).

\paragraph{Programmer Effort}
We estimated the effort required from programmers to add calls to
marshalling/unmarshalling code manually by counting the
static call sites with double-pointer argument(s).
Table~\ref{table:stats} shows that, for most programs, there is only a
small number of call sites with double-pointer argument(s).
The largest number is $383$ from {\tt httpd}, which has 200~K SLOC.
However, this is an overestimate for the call sites that require programmers'
manual intervention because many double-pointer arguments do not
point to an array of pointers but only to a single pointer,
e.g., {\tt strtod} in {\tt libc}. Our analysis pass has a whitelist of
several such functions in {\tt libc}, but there could be more.
In all the 51~K lines of code that we ported for evaluation
(\secref{section:eval} and \secref{section:port}), we only need to add a
call to the marshalling procedure once and the {\tt qsort} wrapper twice.
In short, we believe that manual intervention for data marshaling is manageable,
considering the low ratio of the number of call sites to the lines of code.

\section{Implementation}
\label{section:impl}

We extended the {\CheckedC} compiler
\footnote{Originally from Microsoft:
\url{https://github.com/microsoft/checkedc-clang}\label{checkedcclang}.
Now maintained by the Secure Software Development Project:
\url{https://github.com/secure-sw-dev/checkedc-llvm-project}.}
(which is based on Clang and LLVM~\cite{LLVM:CGO04})
to support our new checked pointers.
Our implementation is based on commit {\tt 2eebdf} of its {\tt Clang}
frontend\footref{checkedcclang} and commit
{\tt e5e9ba7} of its {\tt LLVM}
backend\footnote{\url{https://github.com/microsoft/checkedc-llvm}.
The frontend and backend of the {\CheckedC} compiler were initially in separate
repositories. They were later merged.}.
In this section, we explain the implementation of the new checked pointers
(\secref{section:impl:ptr}), the dynamic key-lock checking
(\secref{section:impl:keycheck}), and the runtime library that we added
for safe heap (de-)allocation and compatibility support
(\secref{section:impl:runtime}).
We also describe the current implementation limitations for a fully thread-safe
and memory-safe {\CheckedC} compiler.

\subsection{Checked Pointers}
\label{section:impl:ptr}

We implemented {\mmptrT} and {\mmarrayptrT} using a structure that
consists of two fields: an LLVM pointer~\cite{LLVMPointer} for
the raw C pointer and an {\tt i64} integer for the key-offset metadata.
Pointer propagation operations, e.g., assignments or pointer arithmetic,
will create a new checked pointer with the source pointer's metadata
(offset is updated as needed).
For other regular pointer operations (such as dereferences and
comparison), the raw C pointer is extracted to do the
computation as what is normally done for the original C.
We have not implemented the two large pointer types
(\secref{section:design:largeptr})
because none of our test programs use them.
These two types of pointers can be implemented using a 3-field structure:
an LLVM pointer for the raw pointer, an {\tt i64} for the key, and an
LLVM pointer for the address of the lock.
Additionally, as Section~\ref{section:overview:goal} mentions, we did not
implement spatial memory safety checks for the new checked pointers
(more details in \secref{section:impl:unimpl}).

\subsection{Dynamic Key Checks}
\label{section:impl:keycheck}


We modified Clang's IR generator to create a function that performs
a dynamic key check and to insert a call to this function before
dereferencing a checked pointer (i.e., via operators {\tt *}, {\tt ->}, and {\tt []}).
LLVM's inlining pass will later inline the calls to the key check function to
improve performance. While it would be simpler to implement the
key check function in a runtime library, doing so would require link-time optimization
(LTO) to inline the calls to it. LTO is undesirable in certain
scenarios as it can be both time- and memory-expensive.


\subsubsection{Key Generation}
We reserve integer $0$ as the invalid key and $1$ as the key for
global objects. As Section~\ref{section:design:metadata} describes,
the performance overhead of generating \emph{unique} and \emph{random}
keys for each memory allocation can be prohibitive.
Another option is to omit randomness by selecting an initial key value
and then updating the key predictably on each
memory allocation~\cite{CETS:ISMM10}.
Our current prototype takes the middle ground: it uses Intel's {\tt RDRAND}
instruction~\cite{IntelManual:2019} to generate a 32-bit random integer
as the first key and increases the key by $1$ for all subsequent requests
for new keys.  This creates unique keys with a degree of randomness.
To improve security, our prototype could be easily enhanced to generate a new
random key periodically (e.g., every $1,000$ memory allocations).


\subsubsection{Redundant Key Check Optimization}
\label{section:impl:keycheck:opt}
The initial code generation from Clang AST to LLVM IR creates many
redundant key checks.
A key check on pointer {\tt p} can be safely removed if the compiler
is certain that since the last check on {\tt p},
(1) {\tt p}'s referent is not freed, and (2) {\tt p} is not updated to
point to another object, directly by assignment or indirectly via
a double-pointer to {\tt p}.
We implemented a standard local data-flow analysis to remove redundant
key checks.

Our compiler invokes the redundant key check
optimization pass immediately after LLVM's {\tt mem2reg}
pass~\cite{mem2reg:LLVM} which promotes stack-allocated memory objects
into LLVM IR SSA virtual registers early within LLVM's pass pipeline.
Our current data-flow analysis implementation is very conservative.
It assumes that any function call may free any heap object, and it
assumes that writes through double pointers may change to which
memory object any checked pointer points.

We also add two operators,
{\tt mm\_checked} and {\tt mm\_array\_checked}, for programmers to label
a safe pointer as already-checked so as to assist the compiler for further
key check optimization.  The compiler will directly mark the pointer
as valid in the data-flow analysis, which may result in fewer inserted key checks.
Programmers can use these operators when they are certain that
a checked pointer is valid but the compiler cannot prove the validity.
It is particularly helpful in the case of a loop or recursive function.
However, to ensure temporal memory safety, these operators may only be used
outside safe regions (\secref{section:bg}).
This optimization is inspired by {\CheckedC}'s {\tt dynamic\_check}
operator~\cite{CheckedCTR:Microsoft} which evaluates a
programmer-written boolean expression and informs the compiler that
a condition about a pointer's bounds is true when the compiler is
unable to determine this fact by itself.
Use of our two operators is also similar to when Rust programmers opt to use
{\tt unsafe} blocks to suppress security checks when dereferencing raw pointers
for performance. Prior surveys~\cite{UnsafeRust:OOPSLA20,UnsafeRust:ICSE20}
show that eliding security checks to improve performance is one of the most
important reasons for writing unsafe Rust code, especially for certain types
of crates.

\subsection{Runtime Library}
\label{section:impl:runtime}

We implemented the memory allocator/deallocator
wrappers, i.e., {\mmalloc}/{\mmfree} (\secref{section:design:lock_heap}),
and the data (un)marshalling
procedures (\secref{section:compatibility:marshal}) in a small runtime library.
An {\mmalloc} internally calls one of {\tt libc}'s {\tt malloc}
family of functions and initializes the lock and a checked pointer
(\secref{section:design:lock_heap}).
{\mmfree} invalidates the lock and does necessary
security checks (\secref{section:design:lock_heap}) in addition to
calling the original {\tt free}.

We also implemented safe versions of {\tt libc}'s string duplication
functions {\tt strdup}/{\tt strndup}. These two functions are very
commonly used, and the returned pointer is expected to be passed
to {\tt free}.

Our runtime library also includes several wrappers for
common {\tt libc} functions such as {\tt strchr} and {\tt strtok}. These
library functions take an argument of array pointer, locate a substring/byte,
and then return a pointer to the located target. Our function wrapper
takes a checked array pointer (call it {\tt p}), calls the corresponding
library function with the raw pointer of {\tt p}, creates a new checked array
pointer by adding to {\tt p} the difference between {\tt p}'s raw pointer
and the returned raw pointer of the library function, and then returns
the new checked pointer.
These function wrappers improve the coverage of checked pointers
because, without them, a program using a pointer returned from these
{\tt libc} functions must use the returned raw pointer instead of a
checked pointer.

Our current prototype does not provide safe wrappers for {\tt mmap}
and {\tt munmap}.  Supporting these two system calls is much more complicated.
When {\tt mmap} is only used as a normal memory allocator
(i.e., mapping anonymous, non-shared, read/write memory), programmers can simply
use our {\mmalloc} to replace it; alternatively,
we can add a wrapper to {\tt mmap} to handle this simple case.
If {\tt mmap} is used to allocate read-only memory, the runtime would be
unable to conveniently initialize and invalidate the lock as an {\mmalloc} does.
We can create a wrapper that maps the memory read/write, initializes the lock,
and then uses {\tt mprotect} to make the memory read-only.
Supporting other use cases (e.g., shared memory and memory-mapped files)
may be possible, but their support is not straightforward,
so we leave it to future work.
There is only one use of {\tt mmap}/{\tt munmap} in {\tt thttpd} in our
evaluation's benchmarks (Table~\ref{table:apps}).  We leave its return value
as a raw pointer.

\subsection{Limitations}
\label{section:impl:limitation}

\subsubsection{Multithreading Support}
\label{section:impl:multithread}

Our work is thread-safe for multithreaded programs that are data-race free.
However, our approach suffers two limitations for programs that have certain
data races. First, if there is a data race between a pointer dereference and
a memory deallocation on the same pointer, our approach may miss a UAF bug
if the free happens in the middle of the key checking. This is a common
limitation shared by all key-lock checking approaches.
Second, if there is a data race between reading and updating a checked pointer,
the read may see a half-updated pointer because the update
is not atomic by default. These two problems can be solved by making the
key-checking and pointer update atomic.

We can make key-lock checking atomic using a mutex. For checked pointer updates,
we see three possible solutions for our x86-64 implementation.
First, we can use a 128-bit atomic compare-and-swap instruction
({\tt CMPXCHG16}~\cite{IntelManual:2021}) to update a regular
checked pointer.
Second, we can use XMM registers and SIMD instructions~\cite{IntelManual:2021}
to update a regular
checked pointer, i.e., moving the two 64-bit fields of a checked pointer
to/from an XMM register before/after updating the checked pointer.
The third option is to use a lock, which could potentially
incur severe performance overhead. Atomically updating checked pointers to large
memory objects (which are 192 bits in size) will require a lock.
In addition, we can apply static analysis techniques (such as
LOCKSMITH~\cite{LOCKSMITH:TOPLS11}) to detect data races as an
optimization for performance.

\subsubsection{Unimplemented Functionality}
\label{section:impl:unimpl}

\begin{table}[t]
\caption{Unimplemented Functionality for a Fully Memory-safe {\CheckedC} Compiler}
\label{table:impl}
\centering
{\sffamily
\footnotesize{
\resizebox{\columnwidth}{!}{
\begin{tabular}{@{}ll@{}}
    \toprule
    {\bf Functionality} & {\bf Explanation} \\
    \midrule
    {\it Fully-safe pointers} & The new checked pointers (\secref{section:design})
    do not enforce spatial memory safety checks as {\CheckedC} does.\\
    {\it Large pointers} & The two large pointers (\secref{section:design:largeptr})
    are not implemented. They are not used in the evaluation. \\
    {\tt {\it itype}} & No extention to {\CheckedC}'s {\tt itype} to support the new
    pointers.  Using pointer casts to enable \\
    & calls to unchecked C functions sufficed for porting the
    applications we used in the evaluation. \\
    \bottomrule
\end{tabular}
}}}
\end{table}

Table~\ref{table:impl} summarizes our current prototype's unimplemented features
for a fully memory-safe {\CheckedC} compiler.  Notably, we did not
integrate {\CheckedC}'s spatial memory safety checks with our new checked
pointers. The integration does not introduce many new {\it design} complications
because the spatial and temporal memory safety checks are semantically independent,
but the {\it implementation} is nontrivial.
We therefore opted to first build a prototype to evaluate our temporal-safe
fat pointers. For a full implementation when building upon our
current prototype, we need to modify the AST-to-LLVM-IR code generation to
integrate the original {\CheckedC}'s spatially-safe pointers with our new
checked pointers for both spatial and temporal memory safety checks.
By default, the original {\CheckedC}'s {\tt Clang}
assumes that pointers are lowered to an LLVM~IR singleton type
(an {\tt LLVM::PointerType}, which is a 64-bit integer on 64-bit systems).
However, due to the additional in-place key-offset metadata, we violate that
assumption by lowering pointers to an LLVM~IR struct
(\secref{section:impl:ptr}).
We need to handle the code lowering for \emph{all} types of
expressions/statements that involve pointers, and this
is a labor-intensive task due to C's extraordinarily
freestyle grammar and unrestricted use of pointers.

We believe that the performance and memory overhead evaluation
(\secref{section:eval}) is valid even without a full implementation.
The overhead of spatially-safe {\checkedc} mainly comes from the dynamic
bounds checks and null pointer checks. The overhead of our temporally-safe
pointers comes from two sources: key-lock checking and key-offset metadata
propagation. Because the causes of overhead are independent, we estimate that
the overall overhead would be roughly the sum of the spatial and temporal
memory safety overheads.

\section{Evaluation}
\label{section:eval}

We evaluated the performance and memory consumption overhead incurred
by our new checked pointers for temporal memory safety.
We also compared our approach with CETS~\cite{CETS:ISMM10}, a disjoint
key-lock checking mechanism which we believe is the most relevant
related work.
We describe the benchmarks used for evaluation and the reasons for choosing
them in Section~\ref{section:eval:benchmark},
experimental setup in Section~\ref{section:eval:setup}, and
performance and memory evaluation in Section~\ref{section:eval:perf} and
Section~\ref{section:eval:mem}, respectively.
Finally, we report our porting experience of the benchmarks in
Section~\ref{section:port}.

\subsection{Benchmarks}
\label{section:eval:benchmark}

Our evaluation requires programs that use our new checked pointers.
Consequently, we needed to port existing programs to
use our new pointer types.  Since such modifications must be done
manually and our team had limited engineering person-effort available,
porting large benchmark suites such as SPEC or
large applications like {\tt Nginx} in their entirety was impractical.
Therefore, for our evaluation, we ported
the Olden benchmark suite,
one SPEC benchmark ({\tt 429.mcf}),
three mature real-world applications/libraries,
and
the engine and the HTTP components of {\tt curl}
to use our new checked pointers.
Table~\ref{table:apps} provides brief descriptions of these programs
and their source lines of code (LoC) computed
using {\tt cloc}~\cite{cloc}.



\paragraph{Olden Benchmarks and mcf}
We chose the Olden benchmark
suite~\cite{Olden:TOPLAS95,CompilerPrefetch:ASPLOS96} for three reasons.
First and foremost, Olden is a pointer-intensive benchmark suite
that emphasizes dynamic recursive data structures such as binary trees and
linked lists.
\emph{All} related work that retrofits metadata to pointers
shows much higher performance and memory overhead on their pointer-intensive
benchmarks than other programs~\cite{CETS:ISMM10,DANGNULL:NDSS15,
DangSan:EuroSys17,pSweeper:CCS18,CUP:AsiaCCS18,HeapExpo:ACSAC20,PTAuth:Sec21}.
We believe that such benchmarks will expose the
worst performance and memory overheads that our work can incur.
Second, it is a relatively small benchmark suite amenable to manual
porting.
Third, the original {\CheckedC} paper used it for
evaluation~\cite{CheckedC:SecDev18}.
We similarily chose {\tt 429.mcf} from SPEC CPU2006 as it is
also small yet pointer intensive.

\Comment{JTC: JZ, why is the fact that the original {\CheckedC} paper
used Olden relevant?  Are we comparing our results to theirs?  Upon
further thought, we're not really articulating why the following fact
is relevant.}
\Comment{JZ: To give Olden some credibility.}

\begin{table}[tb]
\caption{Description and Statistics of Programs for Evaluation}
\label{table:apps}
\centering
{\sffamily
\footnotesize{
\resizebox{\columnwidth}{!}{
\begin{tabular}{@{}llrrrrrrl@{}}
    \toprule
    \multirow{2}{*}{\bf Program} & \multirow{2}{*}{\bf Description} &
    \multirow{2}{*}{\bf LoC} & \multirow{2}{*}{\bf Ported} &
    {\bf Porting} & \multicolumn{3}{c}{\bf \# of Safe} & \multirow{2}{*}{\bf Input for Evaluation} \\
    & & & & {\bf Effort} & {\bf Ptr} & {\bf Alloc} & {\bf Free} & \\
    \midrule
    Olden & Data structure benchmark & 5,027 &4,206 & --- & 298 & 28 & 0 & LLVM Test-suite \\
    429.mcf & Vechicle scheduling & 1,574 & Full & 1 day & 129 & 4 & 3 & SPEC ref dataset\\
    thttpd-2.29 & Lightweight HTTP server & 8,360 & Full & --- & 426 & 47 & 36 & 4~KB--32~MB random files \\
    parson-2d7b3dd & JSON parser & 2,783 & Full & --- & 826 & 17 & 31 & 328~KB--232~MB JSON files  \\
    lzfse-1.0 & Data (de-)compressor & 3,383 & Full & 2 days & 148 & 6 & 6 &
    Silesia corpus and enwik9 \\
    curl-7.79.1 & Network data transferer & 122~K & 31~K & 16 days & 1,770 & 267
    & 539 & Built-in tests \\
    \bottomrule
\end{tabular}
}}
    }
\end{table}


\paragraph{Real-world Applications}


We did a survey on recent temporal memory safety violation CVEs and
found that two types of programs are often vulnerable.
The first are long-running programs such as web servers.
This is because long-running programs need to constantly
free allocated objects to prevent memory exhaustion, enabling the
possibility of temporal memory safety bugs.
Examples include CVE-2019-5096 and CVE-2020-1647.
The second type of vulnerable program is one that
parses input from untrusted sources, especially complex files such as PDFs.
This is because the parsing code usually needs to frequently
allocate and deallocate buffers. For example,
CVE-2021-36088 is a double free bug in the JSON parsing component of a
log processing application;
other examples include CVE-2018-1000039 and CVE-2019-17534.
We therefore looked for lightweight long-running programs and
programs that process input files from untrusted sources.
Table~\ref{table:apps} lists the three applications we chose:
the {\tt thttpd}~\cite{thttpd} HTTP server,
the {\tt parson}~\cite{parson} JSON parser,
and
the {\tt lzfse}~\cite{lzfse} data compressor.

We also ported portions of {\tt curl}~\cite{curl}. {\tt curl} is
a command line tool for transferring data with URLs via network.
It consists of an engine and a library that supports over 20 network protocols.
We chose {\tt curl} for three reasons. First, it is a ubiquitous tool that
is ``used daily by virtually every Internet-using human on the
globe''~\cite{curl}. Second, there are nine temporal memory safety
vulnerabilities in its list of 135~published security vulnerabilities as of
February 2023, and
all nine vulnerabilities were reported since 2016.~\cite{curl_security}.
Third, we would like to test our work's interoperability between the
checked code and the original C~code in a partially ported large program.
To that end, we ported {\tt curl}'s engine and its HTTP protocol component
(its most used protocol).

\subsection{Experimental Setup}
\label{section:eval:setup}

\paragraph{Checked Benchmarks}
We manually ported the benchmarks to use our new checked pointers.
Specifically, we replaced all the calls to the {\tt malloc} family
of functions and {\tt free} with their corresponding allocator
and deallocator wrappers, respectively (\secref{section:impl:runtime});
we also replaced all related pointers
with {\mmptrT} and {\mmarrayptrT} except those that cannot be ported
mainly due to backward compatibility issues with original C code, e.g., those
returned from a {\tt libc} function call.

\paragraph{Compilers and System}
We used the original LLVM~8.0.0 compiler, upon which our {\CheckedC} compiler
is built, to compile all the unmodified benchmarks as the baseline.
We compiled the ported programs with our {\CheckedC} compiler.
To compare with a disjoint key-lock checking mechanism,
we used the SoftBoundCETS
compiler~\cite{SoftBoundCETS:Github,SoftBoundCETS:SNAPL15}
(CETS for short) to compile unmodified benchmarks.
The currently open-sourced stable CETS compiler is based on LLVM~3.4;
we ported it to LLVM~8.0.0 and disabled its spatial memory safety component
for a more accurate comparison.
Unfortunately, CETS cannot compile three of the Olden benchmarks
({\tt bh}, {\tt em3d}, and {\tt mst}), {\tt mcf}, and our
real-world applications correctly due to
bugs in its runtime library and assertion failures in its compiler.


We compiled all programs with the standard {\tt -O3} optimizations for
all three settings (baseline, {\CheckedC}, and CETS).
In addition, we used LLVM's linker {\tt LLD} to do
link-time optimization (LTO) on Olden.
This is because CETS' runtime library is compiled separately from
applications, and it contains many functions that can only be inlined
with LTO.  Thus, using LTO considerably improves the performance of
programs compiled by CETS. We did not use LTO for other programs.


We conducted all the experiments on an Ubuntu 20.04.1 OS running on a machine
with an Intel i7-7700 CPU (8 logical cores), 16~GB of DRAM, and 256~GB of SSD.

\subsection{Execution Time Overhead}
\label{section:eval:perf}
We ran the baseline and our checked version of each test program 20~times and
computed their average execution time/transfer throughput/(de)compression rate.

\subsubsection{Olden and mcf}
\label{section:eval:perf:olden}




We used the LLVM test suite~\cite{LLVMTestSuite} to evaluate Olden's
execution time.
We excluded the {\tt voronoi} benchmark because it frequently uses
an unsafe code pattern (casting an integer to a pointer) that was not
supported by the original (and hence our) {\CheckedC}.
We used the outputs from running the baseline programs as the expected
results for our {\CheckedC} and the CETS-compiled benchmarks.  The LLVM
test-suite verified that both versions have the same output as the baseline.
We modified the inputs to increase the input size so that all benchmarks
(except for {\tt perimeter}, {\tt power}, {\tt treeadd}) run for at least
five seconds.
The {\tt perimeter} program runs for less than 2~seconds regardless of its
input size, and {\tt power} takes no user inputs.
For {\tt treeadd}, the input is the depth of a binary tree to build;
the CETS-compiled {\tt treeadd} crashes with a segmentation fault with an input
greater than 26. We therefore chose the largest input ($26$) that
permits the CETS-compiled {\tt treeadd} to execute.



\begin{figure*}
    \centering
    \microfigwidth .4\textwidth
    \strut\hfill
    \subfloat[Olden Execution Time Normalized to Baseline\label{fig:olden_perf}]%
        {\includegraphics[width=1.17\microfigwidth]%
            {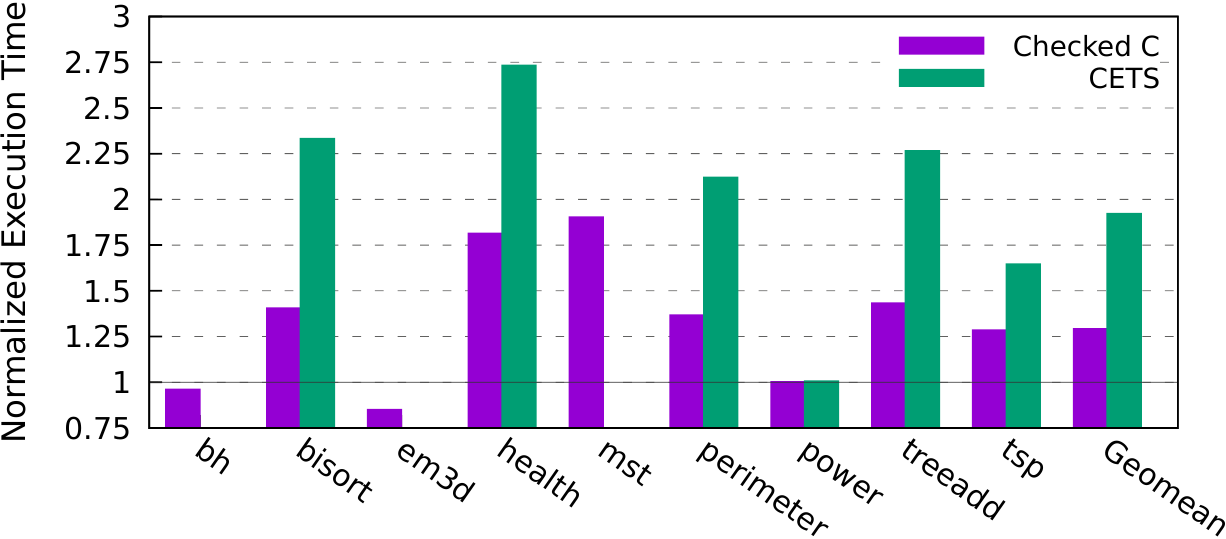}%
        }
    \hfill
    \subfloat[Transfer Throughput of a Local {\tt thttpd}
    Server\label{fig:thttpd_perf}]%
        {\includegraphics[width=1.17\microfigwidth]%
            {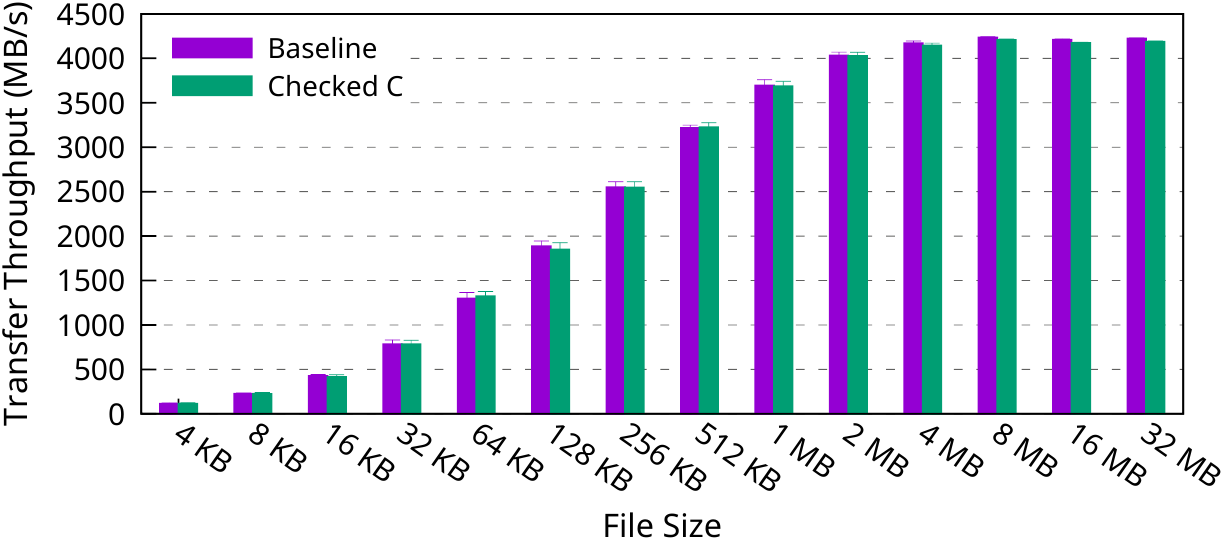}%
        }
    \hfill\strut\\
    \vspace{1.5ex}
    \strut\hfill
    \subfloat[{\CheckedC} Performance on {\tt parson}\label{fig:parson_perf}]%
        {\includegraphics[width=1.17\microfigwidth]%
            {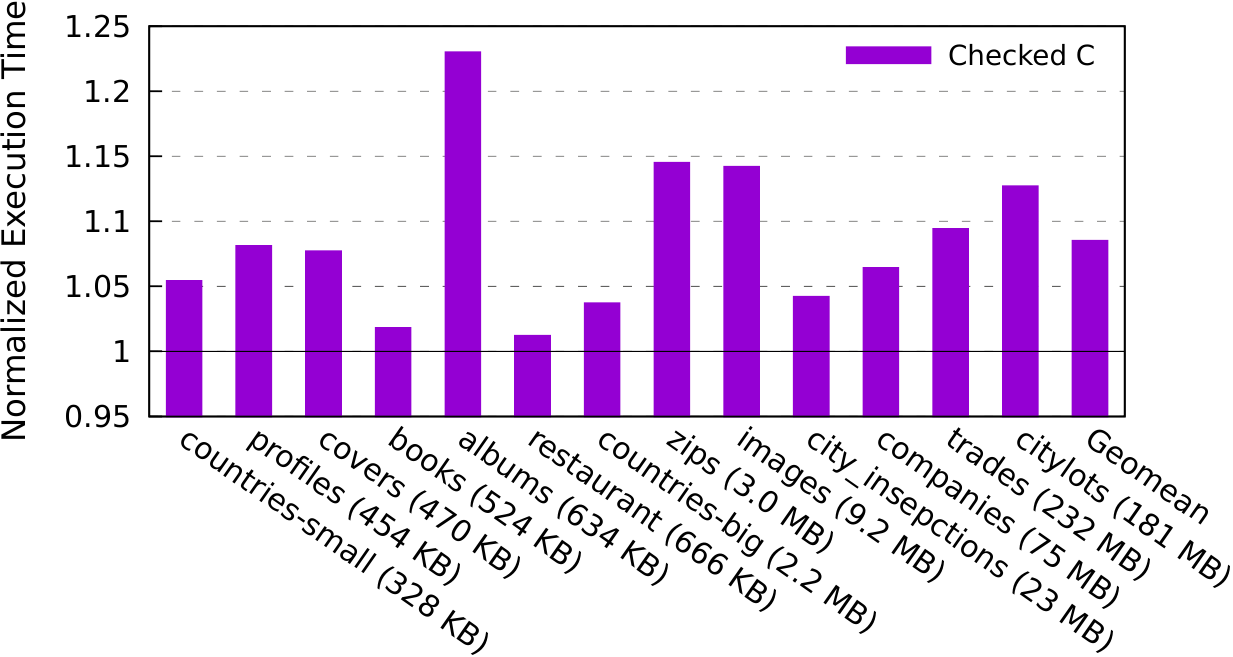}%
        }
    \hfill
    \subfloat[{\CheckedC} Performance on {\tt lzfse} \label{fig:lzfse_perf}]%
        {\includegraphics[width=1.17\microfigwidth]%
            {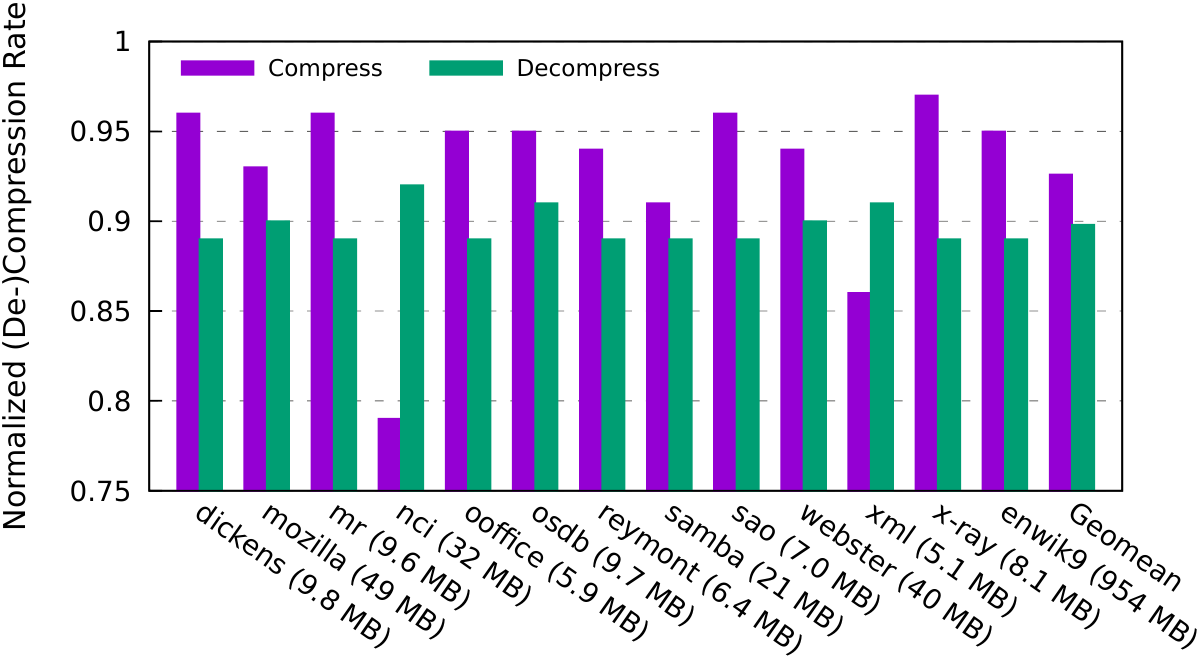}%
        }
    \hfill\strut
    \caption{Checked C Performance Overhead.}
    \label{fig:checkedc_perf}
\end{figure*}

%
%
\begin{figure*}[tb]
    \centering
    \includegraphics[width=1\textwidth]{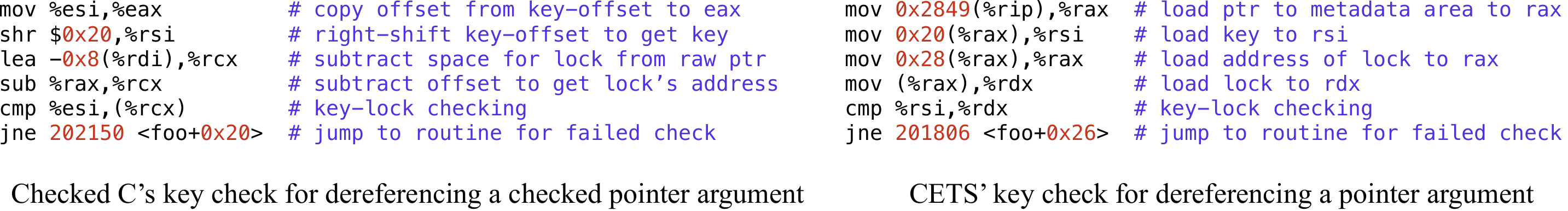}
    \caption{Comparison of {\CheckedC}'s and CETS's Key Check Procedures.
    The code is from a function (compiled with {\tt -O3}) dereferencing a
    pointer argument. For {\CheckedC}, the raw C pointer is in {\tt \$rdi},
    and the key-offset is in {\tt \$rsi}.}
    \label{fig:key_check}
\end{figure*}

\Comment{JTC: JZ, we need to communicate the standard deviation
somehow (either by using error bars in the figures or stating that the
standard deviation was below a certain percent in the figure captions
or main text.}

Figure~\ref{fig:olden_perf} shows the normalized execution time of
our {\CheckedC} and CETS.
Overall, {\CheckedC} incurred a geometric mean of 29.1\% overhead on
the nine benchmarks, and CETS slowed down the six programs by 92.2\%.
For those six programs that can be compiled correctly by CETS,
{\CheckedC}'s overhead is 36.3\%.
\Comment{JTC: JZ, have we run experiments so that we know that this is
why CETS is slower than our approach, or are we only surmising this?
I think we should make clear which we are doing; running some sort of
experiment to verify these conclusions would lead to a stronger argument.}
\Comment{JZ: No. These are educated guesses. I don't think there are easy
ways to verify them, and I believe the reasons are convincing by themselves.}
The vastly improved performance of {\CheckedC} over CETS is mainly
attributed to two factors.
First, {\CheckedC} adds 8 bytes of metadata for each raw C pointer
for temporal memory safety, while CETS adds 16 bytes~\cite{CETS:ISMM10}.
As a result, CETS's pointer propagation is slower.
Second, and more importantly, although using the same key-lock check mechanism,
{\CheckedC} can locate the key and lock significantly faster than CETS.
CETS puts a pointer's key and the address of the lock together and
\emph{dynamically} locates the metadata based on the address of the pointer.
Figure~\ref{fig:key_check} shows the comparison of the assembly code of the
key check procedures of dereferencing a pointer, compiled by {\CheckedC} and
CETS, respectively.
Although having the same number of instructions, {\CheckedC} only uses
one memory instruction (loading the lock), while CETS uses four.
Since memory instructions are usually two orders of magnitude slower than
arithmetic and bitwise instructions~\cite{SysPerf:Gregg},
it is understandable why {\CheckedC}'s in-place metadata is considerably faster
to use than CETS's disjoint metadata.

%
%
{\tt mcf} is a benchmark for single-depot vehicle scheduling.
It is highly pointer intensive: \citet{CETS:ISMM10} report that
CETS incurred the highest overhead (175\%) in all CETS' tested SPEC CPU2006
benchmarks.  In contrast, our {\CheckedC} slowed {\tt mcf} down by 64.2\%.

\subsubsection{thttpd}
\label{section:eval:thttpd_perf}


We used the Apache benchmarking tool {\tt ab}~\cite{ab:Apache} to fetch files
from a {\tt thttpd} server running on the same machine,  and we measured
the average transfer throughput.
We used files of random contents (generated by {\tt Python}'s {\tt random}
library) ranging from 4~KB to 32~MB because
the baseline server reaches full throughput for files larger than 16~MB.
We configured {\tt ab} to run with $10,000$ requests at a concurrency level
of 8. Note that {\tt ab}'s concurrency level option configures how many
requests to send at a time \emph{sequentially} but not in different threads.
We confirmed that {\tt ab} did not hog all our machine's CPU resources so
that the performance measurement for {\tt thttpd} was not affected.
Figure~\ref{fig:thttpd_perf} shows the performance of the baseline and
{\CheckedC} {\tt thttpd}.
When only considering the throughput (i.e., when ignoring noise in the
experiments),
{\CheckedC} incurs 0.4\% overhead on average; when also considering the
standard deviations, {\CheckedC} introduced no measurable overhead
for each file.
\Comment{JTC: JZ, these graphs need error bars, or you need to specify
the maximum standard deviation.}
\Comment{JZ: The figure has error bars. But they are too short to notice.}

\subsubsection{parson}
\label{section:eval:parson_perf}

\Comment{JTC: JZ, we need to provide absolute execution time numbers
for baseline, and this is something that you should be doing for all
experiments (except perhaps Olden in which the text already states
that absolute execution time is 5~seconds or more.  There are two
reasons for this: first, it communicates to the reader that we're
doing the experiment correctly; we're not using execution times that
are so small that the overhead is likely noise from the OS kernel or
other sources.  Second, it puts the overheads into context: 5\% can
either be good or bad depending on the length of the original
computation.}

\Comment{JTC: JZ, again, I think we should cite URLs instead of using
footnotes.  Citations include the date we accessed the URL; they also
take up less space in the main document.}

{\tt parson} is a JSON parsing library~\cite{parson}.
It has a test suite to verify its correctness,
and our {\CheckedC} version of {\tt parson} passed all 339 test cases.
However, the built-in test files are too small for performance evaluation,
so we collected larger JSON files:
one is a dataset for testing MongoDB~\cite{MongoDBJSON};
the other is a large JSON file (189.9~MB) about a city's
districts~\cite{SFCityJSON}.
\Comment{JTC: It sounds like we modified {\tt parson}.  If so, we
should explain where we placed the calls to {\tt clock\_gettime}.}
\Comment{JZ: I think people would assume we put them at the beginning and
the end of the program. Again, I find this level of details a bit distracting.
Plus, we do not have the space for the submission.}
We used {\tt libc}'s {\tt clock\_gettime} timer with the
{\tt CLOCK\_MONOTONIC} flag to measure the execution time. Specifically,
the main computation parses a JSON file into a JSON value (a structure
representing the whole file), serializes the value to a string,
and deserializes the string back to the JSON value.
We excluded the smallest three JSON files from the MongoDB dataset because
they take too little time (around 1~ms) to process,
and thus their coefficient of variation are too high (over 20\%).
We also excluded binary JSON files because {\tt parson}
does not support binary inputs.
Figure~\ref{fig:parson_perf} shows {\CheckedC} {\tt parson}'s normalized
execution time. The geometric mean overhead is 8.5\%, and the
highest overhead is 23\% on {\tt albums}. Only four input files slowed
{\tt parson} down more than 10\%, and only one slowed
{\tt parson} down more than 15\%.



\Comment{JTC: JZ, do we know why albums has so much more overhead than
the other inputs?  When possible, we should determine why we're
getting the results that we see and explain that to the reader.  That
is more insightful than just listing how many programs fall under or
above a certain threshold; it can identify further research problems
for us or others to tackle.}

\Comment{JTC: JZ, I don't see error bars in Figure~\ref{fig:parson_perf}.}

\subsubsection{lzfse}
\label{section:eval:lzfse_perf}

\Comment{JTC: JZ, we should cite sources to back up the claim that
other compresssors use Silesia for benchmarking.}


\Comment{JTC: JZ, why does nci have such a sharp decrease in
performance compared to the other programs?  Also, what about XML?
Why do we have more overhead on decompression than compression?
Summarizing how many benchmarks are above/below a certain threshold is
usseful for ``showing off'' how good our approach is, but as Sandhya taught me,
it doesn't provide any insight as to why the overheads exist and what could be
done to further reduce them in later research.  The paper will be stronger
and more insightful if we can deduce why we're seeing the results that
we see in our experiments.}

\Comment{JTC: JZ, again, absolute numbers (or a summary of them in the text)
will give readers perspective on what the numbers mean.}

{\tt LZFSE} is a lossless data compression algorithm developed by Apple
aiming for a high compression ratio~\cite{lzfse}.
We chose the Silesia compression corpus~\cite{Silesia} for evaluation.
Although small, the Silesia corpus covers a wide range of data types
(plain text, PDF, executables, etc.) and is used by other popular compressors
such as {\tt lz4} and {\tt zstd} for benchmarking.
We also used a large (994~MB) data file {\tt enwik9}~\cite{enwik:compression},
which is also commonly used for benchmarking compressors.
%
Figure~\ref{fig:lzfse_perf} shows the average
compression/decompression rate of {\CheckedC} over
the baseline (a higher rate is better).
{\CheckedC} incurred an average of 7.4\% overhead
for compression, with the minimum and maximum being 3.4\% and 20.9\%.
{\CheckedC} {\tt lzfse} slowed down by less than 5\% for 5 of the 13 data
files and less than 10\% for 11 of them.
For decompression, {\CheckedC}'s average overhead is 10.2\%, and the
performance degradation is reasonably consistent,
ranging from 7.7\% to 11.2\%.

\subsubsection{curl}
\label{section:eval:curl_perf}

{\tt curl} has comprehensive built-in tests. We excluded tests~1014, 1119, 1135,
and 1167. Test~1014 fails with the original {\tt curl} compiled by the
vanilla {\tt clang} compiler. The other three tests do not execute {\tt curl}
but perform sanity checks on the {\tt curl} source code
e.g., checking if file names or function names follow the naming convention.
The names of and the symbols in the header files of our runtime library
(\secref{section:impl:runtime}) break these tests. Our partially-ported
{\tt curl} passed the remaining 1,134 tests. The test suite launches
local servers to/from which {\tt curl} sends and retrieves data.
The test suite reports the execution time of each test and of all test cases
combined.
On average, our checked version took 299.7~s while the
baseline {\tt curl} took 298.7~s in total,
incurring 0.4\% overhead.

\subsection{Memory Overhead}
\label{section:eval:mem}

The key, lock, and necessary memory padding for alignment
(\secref{section:design:lock_heap}) incurs
memory overhead. We measured the memory overhead of Olden and our
applications using {\CheckedC}; we also compared our overheads with
CETS' memory overheads on Olden.
We used the {\tt wss} tool~\cite{WSS:BrendanGregg} to measure the
maximum resident set size (RSS) (total memory consumption at a moment)
and average working set size (WSS) (memory consumption in a period of time).
Table~\ref{table:mem} summarizes the results.

\begin{table}[tb]
    \caption{Memory Overhead. {\bf RSS} and {\bf WSS} are the
    maximum RSS and average WSS of one run of a program, respectively.
    {\bf Min}, {\bf Max}, {\bf Mean} is min, max, and geomean of all
    benchmarks (for Olden), or all input files (for thttpd, parson, and lzfse),
    or all test cases (for curl). mcf only has one input and thus having no
    min and max.}
\label{table:mem}
\centering
{\sffamily
\footnotesize{
\begin{tabular}{@{}llrrrrrrrr@{}}
\toprule
    & & \multicolumn{2}{c}{Olden} & \multirow{2}{*}{mcf} & \multirow{2}{*}{thttpd}
    & \multirow{2}{*}{parson} & \multicolumn{2}{c}{lzfse} & \multirow{2}{*}{curl} \\
    \cline{3-4} \cline{8-9}
    & & {\CheckedC} & CETS & & & & Compress & Decompress & \\
\midrule
    \multirow{3}{*}{\bf RSS} & {\bf Min} & 0  & 70\% & --- & 7.1\% & -0.1\% &
    -1.4\% & -12.6\% & 22.4\% \\
    & {\bf Max} & 150\% & 250\% & --- & 9.9\% & 69.6\% & 0.7\% & 1.2\% & 26.9\% \\
    & {\bf Mean} & 72\% & 202\% & 74.9\% & 9.5\% & 11.2\% & -0.2\% & -1.2\% & 25.1\% \\
\midrule
    \multirow{3}{*}{\bf WSS} & {\bf Min} & 5\% & 250\% & --- & --- & -17.2\% &
    -8.2\% & -9.2\%  & --- \\
    & {\bf Max} & 138\% & 1,977\% & --- & --- & 88.3\% & 2.8\% & 53.1\% & ---\\
    & {\bf Mean} & 44\% & 889\% & 44.7\% & --- & 10.5\% & -3.5\% & 9.6\% & ---  \\
\bottomrule
\end{tabular}
}}
\end{table}

\subsubsection{Olden and mcf}
\label{section:eval:mem:olden}

On average, {\CheckedC} consumed 72\% more memory than the baseline
for RSS and 44\% for WSS, and
CETS' RSS and WSS overhead are 202\% and 889\%, respectively.
There are two reasons why {\CheckedC}'s memory consumption is
significantly less than CETS'.
First, {\CheckedC} adds 8 bytes of metadata for each pointer,
while CETS adds 16 bytes, and CETS uses a trie-based table for
metadata lookup~\cite{CETS:ISMM10}.
Second, in order to reduce the calls to allocate memory for the key and lock
address metadata of each allocation, CETS preallocates two memory pools to
store the metadata for the heap and the stack, which incurs unnecessary
memory overhead (empty space in the pools).
In contrast, our {\CheckedC} does not consume unneeded memory for metadata
except for the padding (8~bytes for each lock on 64-bit systems)
needed to properly align the start address of memory objects.

\citet{CETS:ISMM10} do not report CETS' memory overheads for {\tt mcf}.
Our {\CheckedC}
incurs a RSS overhead of 74.9\% and WSS overhead of 44.7\%.
As Section~\ref{section:eval:perf:olden} mentions, {\tt mcf} is a
highly pointer-intensive program and thus represents an extreme
case for both performance and memory overhead.

\Comment{JTC: JZ, you need to be careful about assuming that CETS
uses more memory because it preallocates memory pools.  If that memory
is allocated but not written, then it may not use any more physical
memory (even though it's consuming virtual address space).  Most
likely, some of the physical memory is allocated at allocation time
and the rest is allocated via demand paging.}


\subsubsection{Applications}
{\CheckedC} incurred a rather consistent RSS overhead of 7.1\% to 9.9\%
on {\tt thttpd}. We did not report {\tt thttpd}'s average WSS because
its WSS stays very low most of the time (when we believe the server is
idle) and peaks periodically (most likely to process new connections and
new data).
For {\tt parson}, the RSS and WSS overheads are 11.2\% and 10.5\%, respectively.
Note that we observed negative memory overhead because {\tt wss} measured the
memory periodically and we used the same time interval for baseline and
the checked version, it is possible that the tool missed the real max RSS
of a checked run and recorded the real max RSS of the baseline.

For {\tt lzfse}, {\CheckedC} introduces negligible
RSS and WSS overhead for compression and at most 1.2\%
RSS overhead for decompression. The geomean of WSS overhead
for decoding is 9.6\%.
Although {\tt lzfse} is a CPU-intensive program and uses pointers frequently,
it does not use data structures that contain many pointers:
the {\tt struct} for maintaining the encoder's
state only has four checked pointers, and the one for decoding has six.
Additionally, most of the buffers allocated by {\tt lzfse} are of single-byte
type ({\tt char} or {uint8\_t}), and they are usually manipulated by
one checked pointer. These reasons largely explain why {\tt lzfse}
has around 9\% performance overhead but very low memory overhead
for most tasks.

Since most of {\tt curl}'s test cases execute for less than one second,
we measured the memory consumption of the nine tests that run longer
than 5~seconds. Like {\tt thttp}, we do not report the WSS as it
remains low most of the time and peaks occasionally.
We observed that {\tt curl}'s RSS stays low and stable during the
execution (5.74~MB--6.04~MB for the baseline), and our checked version incurs
an overhead of 22.4\% to 26.9\%, with a geomean of 25.1\%.
We believe that a major contributor to the overhead is the
{\tt unordered\_set} used to track the safe pointers
(\secref{section:port:challenge}). A pointer itself plus
the metadata needed by the {\tt unordered\_set} to store it can exceed
40~bytes~\cite{CPPSTLMem}.

\section{Porting Experience}
\label{section:port}
{\CheckedC} is a language extension, and the intention is that
programmers write and maintain safe code directly. This gives
them explicit control over the use of safe pointers, offering a
durable approach to safety and one that affords easier long-term
maintenance compared to relying solely on automated
transformations, which have struggled to support all of C reliably.
For example,
\citet{CUP:AsiaCCS18} reported that CETS~\cite{CETS:ISMM10} raised both
false positives and false negatives on the Juliet suite.

Nevertheless, the security and maintenance benefits come at a price for
\emph{legacy} programs: programmers must port them to use our new safe
pointers, as we did for our test programs.
However, we believe that the porting work could be partially automated by
extending 3C~\cite{3C:OOPSLA22},
a tool that assists in porting legacy C to spatially-safe {\CheckedC}.
3C can automatically convert 67.9\% of raw C pointers to be
checked, and follow-on usage modes help programmers iteratively port the
rest~\cite{3C:OOPSLA22}.
As our new safe pointers are a strict extension to {\CheckedC}'s type system,
the core approach of 3C should be easily adaptable to our pointers.

In this section, we describe how we ported the test programs to
use our new safe pointers (\secref{section:port:how}),
the manual porting effort (\secref{section:port:effort}),
and the challenges (\secref{section:port:challenge}) we experienced
in the process. We also compare using Rust versus {\CheckedC} for
porting existing programs and developing new programs from scratch
(\secref{section:port:rust}).



\subsection{How to Port}
\label{section:port:how}
Recall from Section~\ref{section:design} that our new checked pointers follow
the type system of the original {\CheckedC}~\cite{CheckedCTR:Microsoft}.
One of the most important type rules is that a checked pointer and a raw pointer
may not be directly assigned to each other.
We mainly rely on compiler errors from enforcing this rule to guide the
porting process. Specifically, we started porting a program by replacing
its heap memory allocation calls with calls to our {\mmalloc}
(\textsection~\ref{section:impl:runtime}).
The pointer returned by {\mmalloc} is a checked pointer, and
there will be a type mismatch error when it is assigned to a raw C pointer.
The refactored checked pointer will then be propagated to other pointers.
However, we must allow casting of a checked pointer to a raw pointer
(the other way around is strictly prohibited)
in two scenarios: passing a checked pointer to a call to a legacy library
function and to an application function that will not be ported in the
immediate future.
In a full implementation, we should declare a bounds-safe interface
(\textsection~\ref{section:compatibility:type}) for those original C functions
and let the compiler automatically do the casting (mainly stripping off
the metadata).
Our current compiler prototype does not implement this feature yet, and
we used a simple macro to manually do the casting.


\subsection{Porting Effort}
\label{section:port:effort}

We quantitatively measured the porting effort by person-days.
We interspersed our time between porting programs and implementing our compiler
in the early development periods because we often needed to fix compiler bugs
or implement missing language features while porting.
We therefore did not measure the porting effort for
{\tt Olden}, {\tt thttpd}, and {\tt parson}.
For {\tt curl},  we ported its engine which starts the program and does
initialization work such as parsing user inputs.
We omit a large file that prints out {\tt curl}'s manual. The rest of the engine
is around 12~KLoC.
We also ported the main body of HTTP and the major components upon
which it depends.  Additionally, we ported some utility functions shared
by different protocols.  In total, we ported 31~KLoC of {\tt curl}.

Table~\ref{table:apps} shows the ported LoC and person-days we spent.
Note that the LoC is not the lines of code that was changed but is
the size of the source file or function we modified.  For example,
if we modified 5~lines of a 30~line function,
we report 30 ported lines of code.
A ``fully-ported'' function means that all pointers are checked, with
two exceptions. First, pointers to global variables will not be ported
as long as they are not assigned to a checked pointer. Second, pointers
returned from a legacy library or unported application code will
remain untouched.

On average, we ported $\sim$1--2 KLoC per person-day, depending on the
program's complexity. We expect that this rate is a slight
underestimate compared to a real-world production environment, for two reasons.
First, we were unfamiliar with these codebases. Developers familiar
with the code should be able to port faster, e.g., starting from the
least complex components. Second, the reported effort also includes
time diagnosing problems that were caused by
bugs in our compiler or
runtime library. A more robust implementation of our toolchain would eliminate
such debugging time.

Table~\ref{table:apps} also gives the number of checked pointers, calls to
{\mmalloc}, and calls to {\mmfree}.
We counted the number of checked pointers in {\struct} definitions, local
pointer variable declarations, function parameters, and function return types.
We report these numbers because they reflect an estimate on how many inline
changes we need to make manually.

\subsection{Challenges}
\label{section:port:challenge}
As our work is built upon the original {\CheckedC}, we face similar
challenges~\cite{CheckedC:SecDev18} when porting a legacy C program.
Additionally, we experienced new challenges due to the use of in-place
metadata for temporal memory safety. All the challenges discussed in this
section are specific to \emph{partially}-ported application code.
A newly-written or fully-ported {\CheckedC} program are not subject
to these challenges, even directly linked with legacy libraries.

\paragraph{Invalid free}
The first major challenge is that ported code may pass a checked
pointer to legacy code (essentially removing the metadata and passing the
raw pointer) which then attempts to free the pointer. While
we are not aware of common C library functions that free application
pointers, it is possible that an unported function will free an
argument passed from a ported one.
This would cause a runtime error: due to the lock and memory padding
(\secref{section:design:mmptr}), the raw pointer of a checked pointer
does not actually point to the start address returned from a heap allocator,
and freeing such a pointer is illegal.
We propose two solutions. First, the runtime can record the generated checked
pointers by {\mmalloc} in a set, and the compiler instruments all
calls to original {\tt free} to dynamically check if the target pointer
is in the checked pointer set.
Our current prototype uses this solution for {\tt curl} implemented
using C{\tt ++}'s {\tt unordered\_set}.
We regard this solution as a ``debug mode'' as the performance
penalty may get high for allocation-intensive programs.

The second solution is to instrument all calls to the original {\tt malloc}
so that it always allocates the space for a lock and memory
padding as {\mmalloc} does, and to instrument calls to original {\tt free} to
adjust the pointer before freeing.
This solution is more predictable as it takes a constant number
of operations to free a pointer at the cost of allocating extra memory for each
heap object. Note that a fully-ported program does not need to
pay the cost of any of the two solutions.

\paragraph{Functions used by ported and unported code}
The second major challenge is how to port functions with pointer arguments
and/or return values used by both already-ported and unported code
of a program. The simplest solution is to follow the original {\CheckedC}'s
convention by using bounds-safe interface and {\tt itype} for the function
prototype (\secref{section:bg}).
However, it will lose temporal memory safety inside the function
body and/or for the returned pointer. The essential reason is that
the metadata for spatial memory safety, i.e., bounds information,
is \emph{explicit} and controllable by programmers, while the metadata for
temporal memory safety is \emph{implicit} and transparent to programmers.
For example, when porting a function like {\tt strncpy} (Figure~\ref{fig:bsi}),
the spatial bound is provided as an argument. Programmers can therefore port this
function using the bounds information. Both checked and unchecked code
can call it, and the compiler will enforce spatial memory safety.
However, this is infeasible for our safe pointers because the
unchecked part has no corresponding metadata to offer.

Programmers can choose to leave the body of such a function unported,
which loses temporal memory safety, or they can port it and all its callers.
However, the second option can be challenging for large programs.
Our current strategy is hybrid: for commonly used small functions,
we wrote a safe version for them, and we keep other shared functions
unchanged.
We acknowledge that this approach adds extra maintenance burden for
programmers.  However, it provides more memory safety, and the
original function can be removed when all callers are ported.

\paragraph{Variadic functions}
Variadic functions (e.g., {\tt printf}) are inherently
unsafe~\cite{VariadicFn:Sec17}. The
original {\CheckedC}~\cite{CheckedC:SecDev18} disallows them in safe regions
(\secref{section:bg});
we likewise only allow them outside safe regions.
Pointer arguments from a variable argument list should be conservatively
assumed to be raw pointers, even though they may actually be from
checked pointers.


\subsection{Discussion: Comparing to Rewriting or Developing from Scratch in Rust}
\label{section:port:rust}

An alternative to {\CheckedC} for safe systems programming is
Rust~\cite{rustlang}.
We do not have quantitative experimental data about the required manual effort
of porting a C program to {\CheckedC} versus Rust, but we believe porting to
{\CheckedC}
would be much easier mainly because {\CheckedC} is an extension to C rather than
a completely new language with a very different programming model. As
Section~\ref{section:port:how} describes, porting a C program to Checked C
mainly requires changing raw pointers to checked pointers,
changing memory (de)-allocation calls to the
corresponding safe wrappers, and (in a complete implementation) declaring
bounds-safe interfaces for legacy code. In addition, the second paragraph
of Section~\ref{section:port} notes that it should be possible to
adapt 3C~\cite{3C:OOPSLA22} to support our temporally-safe pointers.
In contrast, porting C to Rust is a more drastic change; thus, developing
(semi-)automatic tools is significantly more challenging.

That said, we believe that, in general, for a completely new project,
Rust may be a
better choice due to the advancements it has over {\CheckedC}
(e.g., built-in synchronization primitives~\cite{rustlang}).
However, if there are really strict constraints for performance and/or memory
overhead, {\CheckedC} may still be favorable. It is now a widely held impression
that Rust is on par with C in performance (perhaps partially due to
\cite{PLCmp:SLE17}).  However, a recent study~\cite{RustPerf:ASE22} shows that
when running the exact same functionalities and algorithms, Rust takes 1.75x of
execution time compared to C. One major reason why some projects rewritten in
Rust indeed have competitive performance with their C counterparts is that these
new projects are not a mere line-to-line porting of the original C programs but
completely refactored and optimized ones.  We doubt that if those old C projects
go through the same level of refactoring and optimization that is done on
those Rust programs, they would see a large performance gain as well.

\section{Related Work}
\label{section:related}

\Comment{JTC: JZ, one distinction that isn't coming across in this
section is the choice to modify the langauge versus the choice of
applying dangling pointer detection automatically.  Are there any
advantages to changing the langauge compared to work that requires no
source code changes?  If so, what are they?}

\mwh{Agree with the above. We also discussed the idea that you could
  use this approach instead of GC to save performance. For example,
  you could imagine a version of Go that uses mmptrs.}


In general, our work is related to memory safety and techniques of retrofitting
metadata for raw C/C++ pointers, which covers an extremely rich literature that
we cannot practically discuss in its entirety.  We therefore focus on related
work that tackles the temporal memory safety of C.
There are three main directions of work:
checking the validity of pointer dereferences (\secref{section:related:id}),
invalidating dangling pointers (\secref{section:related:nullptr}),
and safe memory allocation (\secref{section:related:allocator}).
We also briefly discuss related work that demands special architectural
support for temporal memory safety
(\secref{section:related:arch}).

\subsection{Dynamic Key-lock Checks}
\label{section:related:id}


\begin{table}[tb]
    \caption{Comparison of Key-Lock Check Approaches. {\bf FP}: False
    Positive (reporting false bugs); {\bf FN}: False Negative (missing bugs).
    We assume both a program and its libraries are protected with the security
    mechanism in the first column; otherwise, it will suffer false negatives
    as long as it is directly linked with legacy library code.
    ``{\qmark}'' means that we cannot infer the information from
    the paper or publicly available source code. \\
    \textsuperscript{*}CETS provides an API for programmers to manually handle
    the case of casting a pointer to an integer and then casting the integer
    back to a pointer; otherwise, it may incur false positives.}
\label{table:related}
\centering
{\sffamily
\footnotesize{
    \begin{tabular}{@{}lcccrccc@{}}
    \toprule
    \multirow{2}{*}{\bf Key-Lock Check Work} & {\bf No} & {\bf No} &
        {\bf No Stack} & {\bf Key} &
        {\bf No Manual} & {\bf No Arch} & {\bf Supports 32-bit} \\
        & {\bf FP} & {\bf FN} & {\bf UAF} & {\bf Size}  &
        {\bf Intervention} & {\bf Support} & {\bf and 64-bit Sys.} \\
    \midrule
    Safe-C~\cite{SafeC:PLDI94} & {\cmark} & {\xmark} & {\cmark} & 32~bits &
        {\qmark} & {\cmark} & {\cmark} \\
        Guarding~\cite{Guarding:SPE97} & {\cmark} & {\xmark} & {\cmark} &
        32~bits & {\xmark} & {\cmark} & {\cmark} \\
        Xu \& Sekar~\cite{XuMemSafe:FSE04} & {\cmark} & {\xmark} & {\cmark} &
        32~bits & {\qmark} & {\cmark} & {\cmark} \\
        CETS~\cite{CETS:ISMM10} & {\cmark}\textsuperscript{*} & {\cmark} & {\cmark} & 64~bits &
        {\cmark}\textsuperscript{*} & {\cmark} & {\cmark} \\
        CUP~\cite{CUP:AsiaCCS18} & {\cmark} & {\cmark} & {\cmark} & 31~bits &
        {\xmark} & {\cmark} & {\xmark} \\
        Arm MTE~\cite{ArmMTE:2019} & {\cmark} & {\xmark} & {\xmark} & 4~bits &
        {\xmark} & {\xmark} & {\xmark} \\
        PTAuth~\cite{PTAuth:Sec21} & {\xmark} & {\xmark} & {\xmark}  & 16~bits &
        {\xmark} & {\xmark} & {\xmark} \\
        ViK~\cite{ViK:ASPLOS22} & {\cmark} & {\xmark} & {\xmark} & 10~bits &
        {\xmark} & {\cmark} &{\xmark} \\
    \midrule
        {\CheckedC} & {\cmark} & {\cmark} & {\cmark} & 32~bits &
        {\xmark} & {\cmark} & {\cmark} \\
    \bottomrule
\end{tabular}
}}
\end{table}

Our {\CheckedC} work falls into the category of dynamic key-lock
checks.\footnote{A key or lock is also referred to as a
{\it capability}~\cite{SafeC:PLDI94,Guarding:SPE97,XuMemSafe:FSE04,
CUP:AsiaCCS18}, {\it lock-key}~\cite{Guarding:SPE97}, or
{\it ID}~\cite{ViK:ASPLOS22}.
For clarity, we uniformly call the metadata for a pointer
a key and the metadata for a memory object a lock.}
We compare such approaches in this section;
Table~\ref{table:related} summarizes the comparisons.
Particularly, Section~\ref{section:related:lowfat} discusses in detail
one type of approach named \emph{low-fat pointers} and compares our
approach with ViK~\cite{ViK:ASPLOS22} (the latest published work before ours
in this category).

UW-Pascal~\cite{UWPascal:TOSE80} is the first work that proposed
dynamic key-lock checks and applied it to a dialect of Pascal.
Safe-C~\cite{SafeC:PLDI94} adds a newly created lock to a hash table and
removes the lock at memory deallocation.
A pointer dereference invokes a search for the lock.
The search time could be linear when the hash table grows large.
In contrast, our {\CheckedC} key-lock checks always take constant time.
CUP~\cite{CUP:AsiaCCS18} repurposes a 64-bit pointer to embed a 31-bit key
that works as an index into a metadata table to check the validity of the
pointer.
CUP requires transforming all code and will otherwise suffer severe
compatibility problems due to the radical change of pointer representation.
Our {\CheckedC} also changes the representation of pointers
but  maintains good backward compatibility (\secref{section:compatibility}).

Guarding~\cite{Guarding:SPE97} and Xu et al.~\cite{XuMemSafe:FSE04}
pair a pointer with additional {\tt struct}(s) of metadata and
put the locks in disjoint arrays.
A pointer's metadata contains a key and the address of the lock
in the array.  Key checks take constant time, but the
dynamically managed lock array mechanism is slower than storing a lock
together with its memory object, as our {\CheckedC} does.
Similarly, CETS~\cite{CETS:ISMM10} stores locks in disjoint arrays
and takes more memory instructions than {\CheckedC} for
metadata propagation and key checks (\secref{section:eval:perf:olden}).

\subsubsection{Low-fat Pointers}
\label{section:related:lowfat}

On 64-bit systems, the higher order bits of a pointer are usually unused.
For example, systems running on x86-64 and ARM64 processors only use the lower
48~bits of the 64-bit virtual address space~\cite{IntelManual:2021,ARMv8A:2019}.
A set of techniques, dubbed \emph{low-fat pointers} by \citet{LowFatPtr:CCS13},
utilize these unused pointer bits to store metadata.

Several systems use low-fat pointers to detect UAF errors.
ARM MTE~\cite{ArmMTE:2019} is a hardware extension that embeds a 4-bit tag (key)
in a pointer's first byte
and adds a 4-bit tag (lock) to every 16~bytes of memory;  it provides new
native instructions to manipulate the keys and locks.
Two recent compiler-based key-lock approaches, PTAuth~\cite{PTAuth:Sec21}
and ViK~\cite{ViK:ASPLOS22}, also leverage low-fat pointers.
PTAuth computes the key using the unused bits and
ARM's Pointer Authentication Code~\cite{ARMv8A:2019} feature.
ViK~\cite{ViK:ASPLOS22} uses 10~bits for keys. Like our work,
PTAuth and ViK place the lock right before the referent memory object.
PTAuth and ViK focus on heap UAFs while MTE and our solution handles
both heap and stack.

In general, low-fat pointers have two major advantages over our
fat-pointer approach.
First, propagating pointer metadata induces no overhead
although there is a small overhead
of clearing the metadata bits when the pointer is expected to be a raw C/C++
pointer, e.g., before being passed to a legacy library function.
Second, it does not suffer the compatibility problem of passing an array of
pointers to legacy code (\secref{section:compatibility}) unless the
array itself will be modified, e.g., by {\tt qsort}
(\secref{section:compatibility:marshal}).

\Comment{JTC: JZ, why can't low fat pointers use wrapper libraries?
It seems this is a common problem for both low-fat pointers and our
approach, and both approaches can use the same solution.}
\Comment{JZ: Good point. Yes, low-fat pointer approaches can also use
wrapper libraries like ours. Please see the revised text.}

However, low-fat pointers suffer other compatibility issues.
When a low-fat pointer is passed to a legacy library function and returned,
the metadata will be lost because legacy code is unaware of the metadata.
Ideally, the returned raw pointer should be refilled with correct metadata.
Unfortunately, this is extremely challenging to do \emph{automatically} as it
requires a precise pointer analysis on the library code.
One can opt to omit key-lock checks on pointers returned from library
functions, as ViK does~\cite{ViK:ASPLOS22}.
Our work likewise must address this challenge: we solve it with corresponding
library function wrappers that recover the metadata
(\secref{section:impl:runtime}), and we believe low-fat pointers can adopt
our solution.
Additionally, a program itself may be using the higher order
bits of a pointer for special purposes; low-fat pointers break such
programs, but our approach does not.

Low-fat pointers have two other common drawbacks. Because the number of usable
bits in a pointer is limited, not only is the key space significantly
smaller than our
fat-pointer solution (Table~\ref{table:related}), but it may be
difficult to scale to programs that use large memory objects.
To locate the lock, PTAuth~\cite{PTAuth:Sec21} performs a dynamic backward
search starting from the raw pointer until a key-lock check passes
or the search hits a preset maximum distance. PTAuth therefore suffers
false positives when the distance between a pointer and its lock is
greater than the \emph{preset} threshold.
ViK~\cite{ViK:ASPLOS22} uses a portion of the unused bits combined with a
prefixed heap allocation
alignment to locate the lock.  By default, it does not protect objects
larger than 4~KB.  While
4~KB covers 98\% of heap objects in Linux kernels~\cite{ViK:ASPLOS22},
our results in Table~\ref{table:stats} show that it is insufficient for
general programs.
In contrast, our {\CheckedC} enhancement places no restrictions on object size.


\paragraph{Key-check Comparison with ViK}
\label{section:related:vik}

An ideal performance comparison with ViK~\cite{ViK:ASPLOS22} would evaluate
both systems on the same machine running the same benchmarks.
However, this is impossible because ViK is not available.
We therefore built assembly code for ViK's core key-check procedure
and qualitatively compared it to our own.
%
\begin{figure}[tb]
    \centering
    \includegraphics[scale=0.34]{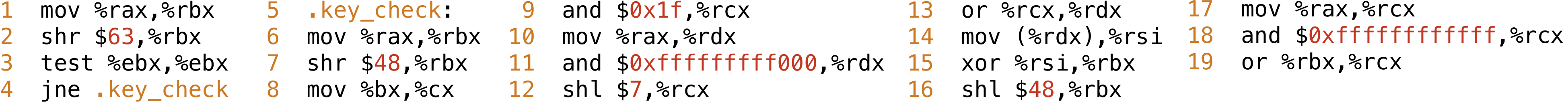}
    \caption{Key-check procedure of ViK for User-space Programs.
    The target pointer is in {\tt \$rax}.}
    \label{fig:vik}
\end{figure}
Figure~\ref{fig:vik} shows the assembly code of ViK's key-check
procedure.\footnote{The ViK paper does not provide concrete assembly code.
The code in Figure~\ref{fig:vik} is based on our understanding of the paper
and direct correspondence with one of ViK's authors.}
ViK targets only a subset of heap objects and it uses the highest-order bit
to indicate whether a pointer points to a protected object,
as Lines~1--4 of Figure~\ref{fig:vik} show.
Lines~6--13 compute the lock's location.
Lines~15--19 does key-lock checking by computing the {\tt xor} of the metadata
in the pointer
and the metadata for the object, putting the result into the pointer's
metadata region. If the {\tt xor} result is $0$, the ``check'' passes
because it generates a valid C pointer; a failed ``check'' results
in an non-canonical pointer which generates a trap if dereferenced.
ViK chose this implementation to avoid a conditional branch~\cite{ViK:ASPLOS22},
and we believe this is to improve performance.

ViK's key-check procedure uses many more instructions than ours
(Figure~\ref{fig:key_check}) because it encodes three pieces of metadata inside
a regular pointer; it thus takes more instructions to extract it.
Conversely, our approach uses a separate register to store two 32-bit pieces
of metadata.
ViK's initial check (lines~1--4) also puts extra pressure on the branch
target buffer.
ViK's major performance advantage over ours is that it does not incur overhead
for propagating pointer metadata. Additionally, two other factors improve ViK's
performance at the cost of weaker security. First, it only protects objects
smaller than a predetermined size.  Second, it omits key checks on pointer
dereferences that are arguably challenging to exploit---e.g., it only
checks the first dereference of a pointer in a function.
In contrast, our solution checks the validity of \emph{every} pointer
dereference.

\subsection{Dangling Pointer Invalidation}
\label{section:related:nullptr}

\Comment{JTC: JZ, we can optinally condense this section.  The key
difference between these systems and ours is that ours doesn't have to
spend time searching through memory to invalidate pointers, giving us
a performance advantage.  We can also claim that we don't miss certain
pointers.  In short, we provide a lot of details that aren't necessary
for comparing to our work; we can therefore probably condense this to
one paragraph if needed.}

\Comment{JTC: JZ, if our approach is faster than these approaches, we
should say so explicitly.}

Another way to prevent use-after-free bugs is to invalidate dangling
pointers so that later dereferences will raise an exception or error.
DANGNULL~\cite{DANGNULL:NDSS15}, FreeSentry~\cite{FreeSentry:NDSS15},
DangSan~\cite{DangSan:EuroSys17}, and pSweeper~\cite{pSweeper:CCS18} are
all compiler-based approaches that instrument programs
to record the points-to relations of a program and invalidate dangling pointers
after their referents are freed. They mainly differ in
the data structures used to maintain the point-to relations.
However, they all maintain metadata in disjoint data structures,
which is an important reason for their high performance overhead
that partially motivated our design choice for in-place metadata.
MemSafe~\cite{MemSafe:SPE13} invalidates a pointer indirectly by
invalidating its bounds information at memory deallocation.
A failed bounds check indicates a UAF bug. This mechanism
can be enhanced with similar techniques described in
Section~\ref{section:design:lock_heap} to catch double free and
invalid free bugs.



One common limitation of dangling pointer invalidation systems is that
a dangling pointer is usually invalidated to a reserved value of which
a later dereference will crash the program or invoke an
exception handling procedure, but any other non-dereference uses,
such as pointer arithmetics or pointer comparison,
are permitted and thus may cause incorrect program execution without
raising attentions. This is not a problem for our {\CheckedC} solution
as it does not modify a pointer's value after its referent is deallocated.
In addition, a more severe limitation of these systems
is that they transform a program's compiler IR and only track pointers
\emph{explicitly} written to memory and ignore pointers in virtual registers.
Consequently, dangling pointers in registers will escape pointer invalidation.
HeapExpo~\cite{HeapExpo:ACSAC20} shows that dangling pointers stored in
virtual registers are common, and previous
works~\cite{DANGNULL:NDSS15,FreeSentry:NDSS15,DangSan:EuroSys17,pSweeper:CCS18}
failed to detect 10 of 19 real-world UAF bugs due to this omission.
HeapExpo solved this problem by also tracking virtual registers albeit
with additional performance and memory overhead.
In contrast, {\CheckedC} does not suffer this limitation because every
checked pointer will be checked regardless of where it is placed when
translated to the compiler's intermediate representation.

\subsection{Safe Memory Allocation and Deallocation}
\label{section:related:allocator}

%
%
Use-after-free bugs become dangerous when freed memory regions get reallocated
and filled with new data. One way to mitigate this
problem is to lower the possibility of reusing memory.
DieHard(er)~\cite{DieHard:PLDI06,DieHarder:CCS10},
FreeGuard~\cite{FreeGuard:CCS17}, and Guarder~\cite{Guarder:Sec18}
allocate heap objects to random locations to provide probabilistic protection
again UAF bugs. SAFECode~\cite{SAFECode:PLDI06}, SVA~\cite{SVA:SOSP07}
and Cling~\cite{Cling:Sec10} only allow reusing memory for
the same type of objects.
Another class of work---Electric Fence~\cite{eFence}, PageHeap~\cite{PageHeap},
Dhurjati et al.~\cite{DAUAF:DSN06}, Oscar~\cite{Oscar:Sec17},
and FFmalloc~\cite{FFmalloc:Sec21}---goes even further to never reuse virtual memory.
Garbage collection techniques have also been explored,
including conservative GC for C~\cite{BoehmGC:PLDI93,BoehmGC:POPL02} and
reference counting~\cite{CRCount:NDSS19}. One state-of-the-art work
in this category, MarkUs~\cite{MarkUs:Oakland20}, achieves low performance
overhead by utilizing extra CPU cores to do live-object traversal.

In general, secure memory allocators often trade memory
for security and performance. Consequently, the memory overhead can be
prohibitive for certain programs. Our {\CheckedC} extension's memory overhead
is mainly proportional to the number of live checked pointers.
There are two other common
limitations of secure allocators. First, they are for the heap
and thus UAF bugs of the stack are possible. Second, some secure
allocators delay freeing memory after a {\tt free} is
called~\cite{CRCount:NDSS19,MarkUs:Oakland20,FFmalloc:Sec21},
which would allow one type of UAF: dereferencing a dangling pointer to
its own stale memory.
This usually does not have security implications but is still
an undefined behavior.
Our {\CheckedC} does not suffer from these two limitations because it
handles the stack and does not delay memory deallocation.

\subsection{Architectural Support}
\label{section:related:arch}
%
%



Several works proposed architectural changes or utilized existing
hardware extensions for temporal memory safety, such as Arm
MTE~\cite{ArmMTE:2019} described in Section~\ref{section:related:id}.
Watchdog~\cite{Watchdog:ISCA12} and WatchdogLite~\cite{WatchdogLite:CGO14}
implemented CETS~\cite{CETS:ISMM10} in hardware by adding new
instructions and cache dedicated for accessing and managing pointer
and object metadata.
Similar to MemSafe~\cite{MemSafe:SPE13}, BOGO~\cite{BOGO:ASPLOS19} also
catches UAF bugs by checking a pointer's bounds information.
It leverages Intel MPX~\cite{IntelManual:2019}
(now deprecated~\cite{IntelManual:2021})
to invalidate the bounds of dangling pointers, and
dereferencing dangling pointers will be caught by MPX as bounds violations.
CHERIvoke~\cite{CHERIvoke:MICRO19} and Cornucopia~\cite{Cornucopia:Oakland20}
periodically scan the address space to revoke capabilities to freed memory,
accelerated by new changes to the CHERI architecture~\cite{CHERI:ISCA14}.
Our solution, in contrast,
requires no specialized hardware components or changes to current
architectures.


\section{Conclusion and Future Work}
\label{section:concl}


This paper presents a new fat-pointer scheme to retrofit temporal memory
safety to C. We demonstrated that when built on a solid
foundation---in our case, {\CheckedC} plus key-lock checking---fat pointers
can provide \emph{full} temporal memory safety \emph{efficiently}.
We showed that, on
a pointer-intensive benchmark suite, use of fat pointers significantly
improves both execution time and memory overhead compared to
using disjoint metadata (29\% vs. 92\% of performance overhead and
72\% vs. 202\% of memory overhead on Olden). Overhead on three full
applications was also low. With findings from analyzing large open-source
C programs and with our hands-on experience of porting real-world
applications, we also showed that our solution does not suffer
serious backward compatibility issues with legacy C code---a formerly major
concern about fat pointers.

There are four main directions for future work.
The first is to integrate {\CheckedC}'s spatial memory safety checks
into our new checked pointers to realize fully memory-safe checked pointers.
Second, we will explore extending the 3C converter~\cite{3C:OOPSLA22} to
support our new checked pointers.
The third is to add multithreading support for the new checked pointers.
Finally, there is room for improving the efficacy of the redundant key check
optimization and for removing unnecessary metadata propagation.

\paragraph{Open Source}
We open-sourced all the artifacts of this paper, including our
{\CheckedC} compiler, the checked versions of the benchmarks
(except {\tt 429.mcf} of SPEC as it is a commercial product),
and the scripts we used for evaluation. They are available at
\url{https://doi.org/10.5281/zenodo.7511299}.

\section*{Acknowledgements}
\label{section:ack}

We thank the anonymous reviewers for their valuable feedback. We also thank
David Tarditi for helping us initiate this project.
In addition, we thank Sreepathi Pai, Deian Stefan, and Aravind Machiry for
their insightful comments and suggestions on an early draft of the paper.
This work was funded by a research gift from Microsoft Research and
NSF award CNS-1955498.

\bibliography{reference}


\begin{thebibliography}{94}


\ifx \showCODEN    \undefined \def \showCODEN     #1{\unskip}     \fi
\ifx \showDOI      \undefined \def \showDOI       #1{#1}\fi
\ifx \showISBNx    \undefined \def \showISBNx     #1{\unskip}     \fi
\ifx \showISBNxiii \undefined \def \showISBNxiii  #1{\unskip}     \fi
\ifx \showISSN     \undefined \def \showISSN      #1{\unskip}     \fi
\ifx \showLCCN     \undefined \def \showLCCN      #1{\unskip}     \fi
\ifx \shownote     \undefined \def \shownote      #1{#1}          \fi
\ifx \showarticletitle \undefined \def \showarticletitle #1{#1}   \fi
\ifx \showURL      \undefined \def \showURL       {\relax}        \fi
\providecommand\bibfield[2]{#2}
\providecommand\bibinfo[2]{#2}
\providecommand\natexlab[1]{#1}
\providecommand\showeprint[2][]{arXiv:#2}

\bibitem[Afek and Sharabani(2007)]%
        {UAF:BlackHatUSA07}
\bibfield{author}{\bibinfo{person}{Jonathan Afek} {and} \bibinfo{person}{Adi
  Sharabani}.} \bibinfo{year}{2007}\natexlab{}.
\newblock \showarticletitle{Dangling pointer: Smashing the Pointer for Fun and
  Profit}.
\newblock  (\bibinfo{year}{2007}).
\newblock
\urldef\tempurl%
\url{https://www.blackhat.com/presentations/bh-usa-07/Afek/Whitepaper/bh-usa-07-afek-WP.pdf}
\showURL{%
\tempurl}


\bibitem[AIDanial(2022)]%
        {cloc}
\bibfield{author}{\bibinfo{person}{AIDanial}.} \bibinfo{year}{2022}\natexlab{}.
\newblock \bibinfo{title}{cloc: Count Lines of Code}.
\newblock
\newblock
\urldef\tempurl%
\url{https://github.com/AlDanial/cloc}
\showURL{%
\tempurl}


\bibitem[Ainsworth and Jones(2020)]%
        {MarkUs:Oakland20}
\bibfield{author}{\bibinfo{person}{Sam Ainsworth} {and}
  \bibinfo{person}{Timothy~M. Jones}.} \bibinfo{year}{2020}\natexlab{}.
\newblock \showarticletitle{MarkUs: Drop-in use-after-free prevention for
  low-level languages}. In \bibinfo{booktitle}{\emph{2020 IEEE Symposium on
  Security and Privacy (SP)}}. \bibinfo{publisher}{IEEE Computer Society},
  \bibinfo{address}{Los Alamitos, CA, USA}, \bibinfo{pages}{578--591}.
\newblock
\urldef\tempurl%
\url{https://doi.org/10.1109/SP40000.2020.00058}
\showDOI{\tempurl}


\bibitem[Akritidis(2010)]%
        {Cling:Sec10}
\bibfield{author}{\bibinfo{person}{Periklis Akritidis}.}
  \bibinfo{year}{2010}\natexlab{}.
\newblock \showarticletitle{Cling: A Memory Allocator to Mitigate Dangling
  Pointers}. In \bibinfo{booktitle}{\emph{Proceedings of the 19th USENIX
  Conference on Security}} (Washington, DC) \emph{(\bibinfo{series}{USENIX
  Security'10})}. \bibinfo{publisher}{USENIX Association},
  \bibinfo{address}{Berkeley, CA, USA}, \bibinfo{pages}{12--12}.
\newblock
\showISBNx{888-7-6666-5555-4}
\urldef\tempurl%
\url{http://dl.acm.org/citation.cfm?id=1929820.1929836}
\showURL{%
\tempurl}


\bibitem[{Apache Software Foundation}(2022)]%
        {ab:Apache}
\bibfield{author}{\bibinfo{person}{{Apache Software Foundation}}.}
  \bibinfo{year}{2022}\natexlab{}.
\newblock \bibinfo{title}{ab - Apache HTTP server benchmarking tool}.
\newblock
\newblock
\urldef\tempurl%
\url{https://httpd.apache.org/docs/2.4/programs/ab.html}
\showURL{%
\tempurl}


\bibitem[{Apple Inc.}(2017)]%
        {lzfse}
\bibfield{author}{\bibinfo{person}{{Apple Inc.}}}
  \bibinfo{year}{2017}\natexlab{}.
\newblock \bibinfo{title}{LZFSE compression library and command line tool}.
\newblock
\newblock
\urldef\tempurl%
\url{https://github.com/lzfse/lzfse}
\showURL{%
\tempurl}


\bibitem[{Arm Ltd.}(2019a)]%
        {ARMv8A:2019}
\bibfield{author}{\bibinfo{person}{{Arm Ltd.}}}
  \bibinfo{year}{2019}\natexlab{a}.
\newblock \bibinfo{booktitle}{\emph{Arm Architecture Reference Manual: Armv8,
  for Armv8-A architecture profile}}.
\newblock
\newblock
\shownote{DDI 0487E.a}.


\bibitem[{Arm Ltd.}(2019b)]%
        {ArmMTE:2019}
\bibfield{author}{\bibinfo{person}{{Arm Ltd.}}}
  \bibinfo{year}{2019}\natexlab{b}.
\newblock \bibinfo{title}{Armv8.5-A Memory Tagging Extension}.
\newblock
\newblock
\urldef\tempurl%
\url{https://developer.arm.com/-/media/Arm%20Developer%20Community/PDF/Arm_Memory_Tagging_Extension_Whitepaper.pdf}
\showURL{%
\tempurl}


\bibitem[Astrauskas et~al\mbox{.}(2020)]%
        {UnsafeRust:OOPSLA20}
\bibfield{author}{\bibinfo{person}{Vytautas Astrauskas},
  \bibinfo{person}{Christoph Matheja}, \bibinfo{person}{Federico Poli},
  \bibinfo{person}{Peter M\"{u}ller}, {and} \bibinfo{person}{Alexander~J.
  Summers}.} \bibinfo{year}{2020}\natexlab{}.
\newblock \showarticletitle{{How Do Programmers Use Unsafe Rust?}}
\newblock \bibinfo{journal}{\emph{Proc. ACM Program. Lang.}}
  \bibinfo{volume}{4}, \bibinfo{number}{OOPSLA}, Article
  \bibinfo{articleno}{136} (\bibinfo{date}{nov} \bibinfo{year}{2020}),
  \bibinfo{numpages}{27}~pages.
\newblock
\urldef\tempurl%
\url{https://doi.org/10.1145/3428204}
\showDOI{\tempurl}


\bibitem[Austin et~al\mbox{.}(1994)]%
        {SafeC:PLDI94}
\bibfield{author}{\bibinfo{person}{Todd~M. Austin}, \bibinfo{person}{Scott~E.
  Breach}, {and} \bibinfo{person}{Gurindar~S. Sohi}.}
  \bibinfo{year}{1994}\natexlab{}.
\newblock \showarticletitle{Efficient Detection of All Pointer and Array Access
  Errors}. In \bibinfo{booktitle}{\emph{Proceedings of the ACM SIGPLAN 1994
  Conference on Programming Language Design and Implementation}} (Orlando,
  Florida, USA) \emph{(\bibinfo{series}{PLDI '94})}. \bibinfo{publisher}{ACM},
  \bibinfo{address}{New York, NY, USA}, \bibinfo{pages}{290--301}.
\newblock
\showISBNx{0-89791-662-X}
\urldef\tempurl%
\url{https://doi.org/10.1145/178243.178446}
\showDOI{\tempurl}


\bibitem[Berger and Zorn(2006)]%
        {DieHard:PLDI06}
\bibfield{author}{\bibinfo{person}{Emery~D. Berger} {and}
  \bibinfo{person}{Benjamin~G. Zorn}.} \bibinfo{year}{2006}\natexlab{}.
\newblock \showarticletitle{{DieHard: Probabilistic Memory Safety for Unsafe
  Languages}}. In \bibinfo{booktitle}{\emph{Proceedings of the 27th ACM SIGPLAN
  Conference on Programming Language Design and Implementation}} (Ottawa,
  Ontario, Canada) \emph{(\bibinfo{series}{PLDI '06})}.
  \bibinfo{publisher}{ACM}, \bibinfo{address}{New York, NY, USA},
  \bibinfo{pages}{158--168}.
\newblock
\showISBNx{1-59593-320-4}
\urldef\tempurl%
\url{https://doi.org/10.1145/1133981.1134000}
\showDOI{\tempurl}


\bibitem[Biswas et~al\mbox{.}(2017)]%
        {VariadicFn:Sec17}
\bibfield{author}{\bibinfo{person}{Priyam Biswas}, \bibinfo{person}{Alessandro
  Di~Federico}, \bibinfo{person}{Scott~A. Carr}, \bibinfo{person}{Prabhu
  Rajasekaran}, \bibinfo{person}{Stijn Volckaert}, \bibinfo{person}{Yeoul Na},
  \bibinfo{person}{Michael Franz}, {and} \bibinfo{person}{Mathias Payer}.}
  \bibinfo{year}{2017}\natexlab{}.
\newblock \showarticletitle{Venerable Variadic Vulnerabilities Vanquished}. In
  \bibinfo{booktitle}{\emph{Proceedings of the 26th USENIX Conference on
  Security Symposium}} (Vancouver, BC, Canada)
  \emph{(\bibinfo{series}{SEC'17})}. \bibinfo{publisher}{USENIX Association},
  \bibinfo{address}{USA}, \bibinfo{pages}{183–198}.
\newblock
\showISBNx{9781931971409}


\bibitem[Boehm(1993)]%
        {BoehmGC:PLDI93}
\bibfield{author}{\bibinfo{person}{Hans-Juergen Boehm}.}
  \bibinfo{year}{1993}\natexlab{}.
\newblock \showarticletitle{Space Efficient Conservative Garbage Collection}.
  In \bibinfo{booktitle}{\emph{Proceedings of the ACM SIGPLAN 1993 Conference
  on Programming Language Design and Implementation}} (Albuquerque, New Mexico,
  USA) \emph{(\bibinfo{series}{PLDI '93})}. \bibinfo{publisher}{Association for
  Computing Machinery}, \bibinfo{address}{New York, NY, USA},
  \bibinfo{pages}{197–206}.
\newblock
\showISBNx{0897915984}
\urldef\tempurl%
\url{https://doi.org/10.1145/155090.155109}
\showDOI{\tempurl}


\bibitem[Boehm(2002)]%
        {BoehmGC:POPL02}
\bibfield{author}{\bibinfo{person}{Hans-J. Boehm}.}
  \bibinfo{year}{2002}\natexlab{}.
\newblock \showarticletitle{Bounding Space Usage of Conservative Garbage
  Collectors}. In \bibinfo{booktitle}{\emph{Proceedings of the 29th ACM
  SIGPLAN-SIGACT Symposium on Principles of Programming Languages}} (Portland,
  Oregon) \emph{(\bibinfo{series}{POPL '02})}. \bibinfo{publisher}{Association
  for Computing Machinery}, \bibinfo{address}{New York, NY, USA},
  \bibinfo{pages}{93–100}.
\newblock
\showISBNx{1581134509}
\urldef\tempurl%
\url{https://doi.org/10.1145/503272.503282}
\showDOI{\tempurl}


\bibitem[Burow et~al\mbox{.}(2018)]%
        {CUP:AsiaCCS18}
\bibfield{author}{\bibinfo{person}{Nathan Burow}, \bibinfo{person}{Derrick
  McKee}, \bibinfo{person}{Scott~A. Carr}, {and} \bibinfo{person}{Mathias
  Payer}.} \bibinfo{year}{2018}\natexlab{}.
\newblock \showarticletitle{CUP: Comprehensive User-Space Protection for
  C/C++}. In \bibinfo{booktitle}{\emph{Proceedings of the 2018 on Asia
  Conference on Computer and Communications Security}} (Incheon, Republic of
  Korea) \emph{(\bibinfo{series}{ASIACCS '18})}.
  \bibinfo{publisher}{Association for Computing Machinery},
  \bibinfo{address}{New York, NY, USA}, \bibinfo{pages}{381–392}.
\newblock
\showISBNx{9781450355766}
\urldef\tempurl%
\url{https://doi.org/10.1145/3196494.3196540}
\showDOI{\tempurl}


\bibitem[Cho et~al\mbox{.}(2022)]%
        {ViK:ASPLOS22}
\bibfield{author}{\bibinfo{person}{Haehyun Cho}, \bibinfo{person}{Jinbum Park},
  \bibinfo{person}{Adam Oest}, \bibinfo{person}{Tiffany Bao},
  \bibinfo{person}{Ruoyu Wang}, \bibinfo{person}{Yan Shoshitaishvili},
  \bibinfo{person}{Adam Doup\'{e}}, {and} \bibinfo{person}{Gail-Joon Ahn}.}
  \bibinfo{year}{2022}\natexlab{}.
\newblock \showarticletitle{ViK: Practical Mitigation of Temporal Memory Safety
  Violations through Object ID Inspection}. In
  \bibinfo{booktitle}{\emph{Proceedings of the 27th ACM International
  Conference on Architectural Support for Programming Languages and Operating
  Systems}} (Lausanne, Switzerland) \emph{(\bibinfo{series}{ASPLOS 2022})}.
  \bibinfo{publisher}{Association for Computing Machinery},
  \bibinfo{address}{New York, NY, USA}, \bibinfo{pages}{271–284}.
\newblock
\showISBNx{9781450392051}
\urldef\tempurl%
\url{https://doi.org/10.1145/3503222.3507780}
\showDOI{\tempurl}


\bibitem[Cimpanu(2020)]%
        {ChromeSecuritySurvey:2020}
\bibfield{author}{\bibinfo{person}{Catalin Cimpanu}.}
  \bibinfo{year}{2020}\natexlab{}.
\newblock \bibinfo{title}{Chrome: 70\% of all security bugs are memory safety
  issues}.
\newblock
\newblock
\urldef\tempurl%
\url{https://www.zdnet.com/article/chrome-70-of-all-security-bugs-are-memory-safety-issues/}
\showURL{%
\tempurl}


\bibitem[Condit et~al\mbox{.}(2007)]%
        {Deputy:ESOP07}
\bibfield{author}{\bibinfo{person}{Jeremy Condit}, \bibinfo{person}{Matthew
  Harren}, \bibinfo{person}{Zachary Anderson}, \bibinfo{person}{David Gay},
  {and} \bibinfo{person}{George~C. Necula}.} \bibinfo{year}{2007}\natexlab{}.
\newblock \showarticletitle{Dependent Types for Low-level Programming}. In
  \bibinfo{booktitle}{\emph{Proceedings of the 16th European Symposium on
  Programming}} (Braga, Portugal) \emph{(\bibinfo{series}{ESOP'07})}.
  \bibinfo{publisher}{Springer-Verlag}, \bibinfo{address}{Berlin, Heidelberg},
  \bibinfo{pages}{520--535}.
\newblock
\showISBNx{978-3-540-71314-2}
\urldef\tempurl%
\url{http://dl.acm.org/citation.cfm?id=1762174.1762221}
\showURL{%
\tempurl}


\bibitem[Criswell et~al\mbox{.}(2007)]%
        {SVA:SOSP07}
\bibfield{author}{\bibinfo{person}{John Criswell}, \bibinfo{person}{Andrew
  Lenharth}, \bibinfo{person}{Dinakar Dhurjati}, {and} \bibinfo{person}{Vikram
  Adve}.} \bibinfo{year}{2007}\natexlab{}.
\newblock \showarticletitle{{Secure Virtual Architecture: A Safe Execution
  Environment for Commodity Operating Systems}}. In
  \bibinfo{booktitle}{\emph{Proceedings of Twenty-first ACM SIGOPS Symposium on
  Operating Systems Principles}} (Stevenson, Washington, USA)
  \emph{(\bibinfo{series}{SOSP '07})}. \bibinfo{publisher}{ACM},
  \bibinfo{address}{New York, NY, USA}, \bibinfo{pages}{351--366}.
\newblock
\showISBNx{978-1-59593-591-5}
\urldef\tempurl%
\url{https://doi.org/10.1145/1294261.1294295}
\showDOI{\tempurl}


\bibitem[curl(2022)]%
        {curl_security}
\bibfield{author}{\bibinfo{person}{curl}.} \bibinfo{year}{2022}\natexlab{}.
\newblock \bibinfo{title}{curl security problems}.
\newblock
\newblock
\urldef\tempurl%
\url{https://curl.se/docs/security.html}
\showURL{%
\tempurl}


\bibitem[Dang et~al\mbox{.}(2017)]%
        {Oscar:Sec17}
\bibfield{author}{\bibinfo{person}{Thurston~H.Y. Dang}, \bibinfo{person}{Petros
  Maniatis}, {and} \bibinfo{person}{David Wagner}.}
  \bibinfo{year}{2017}\natexlab{}.
\newblock \showarticletitle{Oscar: A Practical Page-Permissions-Based Scheme
  for Thwarting Dangling Pointers}. In \bibinfo{booktitle}{\emph{26th {USENIX}
  Security Symposium ({USENIX} Security 17)}}. \bibinfo{publisher}{{USENIX}
  Association}, \bibinfo{address}{Vancouver, BC}, \bibinfo{pages}{815--832}.
\newblock
\showISBNx{978-1-931971-40-9}
\urldef\tempurl%
\url{https://www.usenix.org/conference/usenixsecurity17/technical-sessions/presentation/dang}
\showURL{%
\tempurl}


\bibitem[Deorowicz({[n.\,d.]})]%
        {Silesia}
\bibfield{author}{\bibinfo{person}{Sebastian Deorowicz}.}
  \bibinfo{year}{[n.\,d.]}\natexlab{}.
\newblock \bibinfo{title}{Silesia compression corpus}.
\newblock
\newblock
\urldef\tempurl%
\url{http://sun.aei.polsl.pl/~sdeor/index.php?page=silesia}
\showURL{%
\tempurl}
\newblock
\shownote{{Accessed}: 09-03-2021}.


\bibitem[Dhurjati and Adve(2006)]%
        {DAUAF:DSN06}
\bibfield{author}{\bibinfo{person}{Dinakar Dhurjati} {and}
  \bibinfo{person}{Vikram Adve}.} \bibinfo{year}{2006}\natexlab{}.
\newblock \showarticletitle{Efficiently Detecting All Dangling Pointer Uses in
  Production Servers}. In \bibinfo{booktitle}{\emph{Proceedings of the
  International Conference on Dependable Systems and Networks}}
  \emph{(\bibinfo{series}{DSN '06})}. \bibinfo{address}{Washington, DC, USA},
  \bibinfo{pages}{269--280}.
\newblock
\showISBNx{0-7695-2607-1}
\urldef\tempurl%
\url{https://doi.org/10.1109/DSN.2006.31}
\showDOI{\tempurl}


\bibitem[Dhurjati et~al\mbox{.}(2006)]%
        {SAFECode:PLDI06}
\bibfield{author}{\bibinfo{person}{Dinakar Dhurjati}, \bibinfo{person}{Sumant
  Kowshik}, {and} \bibinfo{person}{Vikram Adve}.}
  \bibinfo{year}{2006}\natexlab{}.
\newblock \showarticletitle{{SAFECode}: Enforcing Alias Analysis for Weakly
  Typed Languages}. In \bibinfo{booktitle}{\emph{Proceedings of the 27th ACM
  SIGPLAN Conference on Programming Language Design and Implementation}}
  (Ottawa, Ontario, Canada) \emph{(\bibinfo{series}{PLDI ’06})}.
  \bibinfo{publisher}{Association for Computing Machinery},
  \bibinfo{address}{New York, NY, USA}, \bibinfo{pages}{144–157}.
\newblock
\showISBNx{1595933204}
\urldef\tempurl%
\url{https://doi.org/10.1145/1133981.1133999}
\showDOI{\tempurl}


\bibitem[{Duan} et~al\mbox{.}(2020)]%
        {FreeBSDCheckedC:SecDev20}
\bibfield{author}{\bibinfo{person}{Junhan {Duan}}, \bibinfo{person}{Yudi
  {Yang}}, \bibinfo{person}{Jie {Zhou}}, {and} \bibinfo{person}{John
  {Criswell}}.} \bibinfo{year}{2020}\natexlab{}.
\newblock \showarticletitle{Refactoring the FreeBSD Kernel with Checked C}. In
  \bibinfo{booktitle}{\emph{2020 IEEE Secure Development (SecDev)}}.
  \bibinfo{pages}{15--22}.
\newblock
\urldef\tempurl%
\url{https://doi.org/10.1109/SecDev45635.2020.00018}
\showDOI{\tempurl}


\bibitem[{Elliott} et~al\mbox{.}(2018)]%
        {CheckedC:SecDev18}
\bibfield{author}{\bibinfo{person}{A.~S. {Elliott}}, \bibinfo{person}{A.
  {Ruef}}, \bibinfo{person}{M. {Hicks}}, {and} \bibinfo{person}{D. {Tarditi}}.}
  \bibinfo{year}{2018}\natexlab{}.
\newblock \showarticletitle{{Checked C: Making C Safe by Extension}}. In
  \bibinfo{booktitle}{\emph{2018 IEEE Cybersecurity Development (SecDev)}}.
  \bibinfo{pages}{53--60}.
\newblock
\urldef\tempurl%
\url{https://doi.org/10.1109/SecDev.2018.00015}
\showDOI{\tempurl}


\bibitem[Enumeration(2020)]%
        {UAF:CWE416}
\bibfield{author}{\bibinfo{person}{Common~Weaknesses Enumeration}.}
  \bibinfo{year}{2020}\natexlab{}.
\newblock \bibinfo{title}{Use After Free}.
\newblock
\newblock
\urldef\tempurl%
\url{https://cwe.mitre.org/data/definitions/416.html}
\showURL{%
\tempurl}


\bibitem[Evans et~al\mbox{.}(2020)]%
        {UnsafeRust:ICSE20}
\bibfield{author}{\bibinfo{person}{Ana~Nora Evans}, \bibinfo{person}{Bradford
  Campbell}, {and} \bibinfo{person}{Mary~Lou Soffa}.}
  \bibinfo{year}{2020}\natexlab{}.
\newblock \showarticletitle{{Is Rust Used Safely by Software Developers?}}. In
  \bibinfo{booktitle}{\emph{Proceedings of the ACM/IEEE 42nd International
  Conference on Software Engineering}} (Seoul, South Korea)
  \emph{(\bibinfo{series}{ICSE '20})}. \bibinfo{publisher}{Association for
  Computing Machinery}, \bibinfo{address}{New York, NY, USA},
  \bibinfo{pages}{246–257}.
\newblock
\showISBNx{9781450371216}
\urldef\tempurl%
\url{https://doi.org/10.1145/3377811.3380413}
\showDOI{\tempurl}


\bibitem[Farkhani et~al\mbox{.}(2021)]%
        {PTAuth:Sec21}
\bibfield{author}{\bibinfo{person}{Reza~Mirzazade Farkhani},
  \bibinfo{person}{Mansour Ahmadi}, {and} \bibinfo{person}{Long Lu}.}
  \bibinfo{year}{2021}\natexlab{}.
\newblock \showarticletitle{PTAuth: Temporal Memory Safety via Robust Points-to
  Authentication}. In \bibinfo{booktitle}{\emph{30th {USENIX} Security
  Symposium ({USENIX} Security 21)}}. \bibinfo{publisher}{{USENIX}
  Association}.
\newblock
\urldef\tempurl%
\url{https://www.usenix.org/conference/usenixsecurity21/presentation/mirzazade}
\showURL{%
\tempurl}


\bibitem[{Fischer} and {LeBlanc}(1980)]%
        {UWPascal:TOSE80}
\bibfield{author}{\bibinfo{person}{Charles~N. {Fischer}} {and}
  \bibinfo{person}{Richard~J. {LeBlanc}}.} \bibinfo{year}{1980}\natexlab{}.
\newblock \showarticletitle{The Implementation of Run-Time Diagnostics in
  Pascal}.
\newblock \bibinfo{journal}{\emph{IEEE Transactions on Software Engineering}}
  \bibinfo{volume}{SE-6}, \bibinfo{number}{4} (\bibinfo{year}{1980}),
  \bibinfo{pages}{313--319}.
\newblock


\bibitem[Fog(2021)]%
        {CPUPerf:AgnerFog}
\bibfield{author}{\bibinfo{person}{Agner Fog}.}
  \bibinfo{year}{2021}\natexlab{}.
\newblock \bibinfo{booktitle}{\emph{4. Instruction tables: Lists of instruction
  latencies, throughputs and micro-operation breakdowns for Intel, AMD, and VIA
  CPUs}}.
\newblock \bibinfo{type}{{T}echnical {R}eport}.
\newblock
\urldef\tempurl%
\url{https://www.agner.org/optimize/instruction_tables.pdf}
\showURL{%
\tempurl}
\newblock
\shownote{{Accessed}: 07-19-2021}.


\bibitem[Gabis(2021)]%
        {parson}
\bibfield{author}{\bibinfo{person}{Krzysztof Gabis}.}
  \bibinfo{year}{2021}\natexlab{}.
\newblock \bibinfo{title}{parson: Lightweight JSON library written in C}.
\newblock
\newblock
\urldef\tempurl%
\url{https://github.com/kgabis/parson}
\showURL{%
\tempurl}


\bibitem[Gregg(2018)]%
        {WSS:BrendanGregg}
\bibfield{author}{\bibinfo{person}{Brendan Gregg}.}
  \bibinfo{year}{2018}\natexlab{}.
\newblock \bibinfo{title}{How To Measure the Working Set Size on Linux}.
\newblock
\newblock
\urldef\tempurl%
\url{https://www.brendangregg.com/blog/2018-01-17/measure-working-set-size.html}
\showURL{%
\tempurl}
\newblock
\shownote{{Accessed}: 10-05-2021}.


\bibitem[Gregg(2020)]%
        {SysPerf:Gregg}
\bibfield{author}{\bibinfo{person}{Brendan Gregg}.}
  \bibinfo{year}{2020}\natexlab{}.
\newblock \bibinfo{booktitle}{\emph{Systems Performance: Enterprise and the
  Cloud, 2nd Edition}}.
\newblock \bibinfo{publisher}{Addison-Wesley}.
\newblock


\bibitem[Gui et~al\mbox{.}(2021)]%
        {UAFSan:ISSTA21}
\bibfield{author}{\bibinfo{person}{Binfa Gui}, \bibinfo{person}{Wei Song},
  {and} \bibinfo{person}{Jeff Huang}.} \bibinfo{year}{2021}\natexlab{}.
\newblock \showarticletitle{UAFSan: An Object-Identifier-Based Dynamic Approach
  for Detecting Use-after-Free Vulnerabilities}. In
  \bibinfo{booktitle}{\emph{Proceedings of the 30th ACM SIGSOFT International
  Symposium on Software Testing and Analysis}} (Virtual, Denmark)
  \emph{(\bibinfo{series}{ISSTA 2021})}. \bibinfo{publisher}{Association for
  Computing Machinery}, \bibinfo{address}{New York, NY, USA},
  \bibinfo{pages}{309–321}.
\newblock
\showISBNx{9781450384599}
\urldef\tempurl%
\url{https://doi.org/10.1145/3460319.3464835}
\showDOI{\tempurl}


\bibitem[Hind(2001)]%
        {IsPASolved:PASTE01}
\bibfield{author}{\bibinfo{person}{Michael Hind}.}
  \bibinfo{year}{2001}\natexlab{}.
\newblock \showarticletitle{Pointer Analysis: Haven't We Solved This Problem
  Yet?}. In \bibinfo{booktitle}{\emph{Proceedings of the 2001 ACM
  SIGPLAN-SIGSOFT Workshop on Program Analysis for Software Tools and
  Engineering}} (Snowbird, Utah, USA) \emph{(\bibinfo{series}{PASTE '01})}.
  \bibinfo{publisher}{Association for Computing Machinery},
  \bibinfo{address}{New York, NY, USA}, \bibinfo{pages}{54–61}.
\newblock
\showISBNx{1581134134}
\urldef\tempurl%
\url{https://doi.org/10.1145/379605.379665}
\showDOI{\tempurl}


\bibitem[Intel Corporation(2019)]%
        {IntelManual:2019}
Intel Corporation \bibinfo{year}{2019}\natexlab{}.
\newblock \bibinfo{booktitle}{\emph{Intel 64 and IA-32 Architectures Software
  Developer’s Manual}}.
\newblock Intel Corporation.
\newblock
\newblock
\shownote{Order Number: 325462-069US}.


\bibitem[Intel Corporation(2021)]%
        {IntelManual:2021}
Intel Corporation \bibinfo{year}{2021}\natexlab{}.
\newblock \bibinfo{booktitle}{\emph{ntel® 64 and IA-32 Architectures Software
  Developer’s Manual}}.
\newblock Intel Corporation.
\newblock
\newblock
\shownote{Order Number: 253665-075US}.


\bibitem[Jim et~al\mbox{.}(2002)]%
        {Cyclone:ATC02}
\bibfield{author}{\bibinfo{person}{Trevor Jim}, \bibinfo{person}{J.~Greg
  Morrisett}, \bibinfo{person}{Dan Grossman}, \bibinfo{person}{Michael~W.
  Hicks}, \bibinfo{person}{James Cheney}, {and} \bibinfo{person}{Yanling
  Wang}.} \bibinfo{year}{2002}\natexlab{}.
\newblock \showarticletitle{{Cyclone: A Safe Dialect of C}}. In
  \bibinfo{booktitle}{\emph{Proceedings of the General Track of the Annual
  Conference on USENIX Annual Technical Conference}}
  \emph{(\bibinfo{series}{ATEC '02})}. \bibinfo{publisher}{USENIX Association},
  \bibinfo{address}{Berkeley, CA, USA}, \bibinfo{pages}{275--288}.
\newblock
\showISBNx{1-880446-00-6}
\urldef\tempurl%
\url{http://dl.acm.org/citation.cfm?id=647057.713871}
\showURL{%
\tempurl}


\bibitem[Kowshik et~al\mbox{.}(2002)]%
        {ControlC:CASES02}
\bibfield{author}{\bibinfo{person}{Sumant Kowshik}, \bibinfo{person}{Dinakar
  Dhurjati}, {and} \bibinfo{person}{Vikram Adve}.}
  \bibinfo{year}{2002}\natexlab{}.
\newblock \showarticletitle{Ensuring Code Safety without Runtime Checks for
  Real-Time Control Systems}. In \bibinfo{booktitle}{\emph{Proceedings of the
  2002 International Conference on Compilers, Architecture, and Synthesis for
  Embedded Systems}} (Grenoble, France) \emph{(\bibinfo{series}{CASES ’02})}.
  \bibinfo{publisher}{Association for Computing Machinery},
  \bibinfo{address}{New York, NY, USA}, \bibinfo{pages}{288–297}.
\newblock
\showISBNx{1581135750}
\urldef\tempurl%
\url{https://doi.org/10.1145/581630.581678}
\showDOI{\tempurl}


\bibitem[Kwon et~al\mbox{.}(2013)]%
        {LowFatPtr:CCS13}
\bibfield{author}{\bibinfo{person}{Albert Kwon}, \bibinfo{person}{Udit Dhawan},
  \bibinfo{person}{Jonathan~M. Smith}, \bibinfo{person}{Thomas~F. Knight},
  {and} \bibinfo{person}{Andre DeHon}.} \bibinfo{year}{2013}\natexlab{}.
\newblock \showarticletitle{Low-Fat Pointers: Compact Encoding and Efficient
  Gate-Level Implementation of Fat Pointers for Spatial Safety and
  Capability-Based Security}. In \bibinfo{booktitle}{\emph{Proceedings of the
  2013 ACM SIGSAC Conference on Computer \& Communications Security}} (Berlin,
  Germany) \emph{(\bibinfo{series}{CCS '13})}. \bibinfo{publisher}{Association
  for Computing Machinery}, \bibinfo{address}{New York, NY, USA},
  \bibinfo{pages}{721–732}.
\newblock
\showISBNx{9781450324779}
\urldef\tempurl%
\url{https://doi.org/10.1145/2508859.2516713}
\showDOI{\tempurl}


\bibitem[Lattner and Adve(2004)]%
        {LLVM:CGO04}
\bibfield{author}{\bibinfo{person}{Chris Lattner} {and} \bibinfo{person}{Vikram
  Adve}.} \bibinfo{year}{2004}\natexlab{}.
\newblock \showarticletitle{{LLVM}: A Compilation Framework for Lifelong
  Program Analysis \& Transformation}. In \bibinfo{booktitle}{\emph{Proceedings
  of the International Symposium on Code Generation and Optimization:
  Feedback-directed and Runtime Optimization}}
  \emph{(\bibinfo{series}{CGO'04})}. \bibinfo{publisher}{IEEE Computer
  Society}, \bibinfo{address}{Palo Alto, CA}, \bibinfo{pages}{75--86}.
\newblock
\showISBNx{0-7695-2102-9}
\urldef\tempurl%
\url{http://dl.acm.org/citation.cfm?id=977395.977673}
\showURL{%
\tempurl}


\bibitem[Lee et~al\mbox{.}(2015)]%
        {DANGNULL:NDSS15}
\bibfield{author}{\bibinfo{person}{Byoungyoung Lee}, \bibinfo{person}{Chengyu
  Song}, \bibinfo{person}{Yeongjin Jang}, \bibinfo{person}{Tielei Wang},
  \bibinfo{person}{Taesoo Kim}, \bibinfo{person}{Long Lu}, {and}
  \bibinfo{person}{Wenke Lee}.} \bibinfo{year}{2015}\natexlab{}.
\newblock \showarticletitle{{Preventing Use-after-free with Dangling Pointers
  Nullification}}. In \bibinfo{booktitle}{\emph{NDSS}}.
\newblock


\bibitem[Lemire(2016)]%
        {CPPSTLMem}
\bibfield{author}{\bibinfo{person}{Daniel Lemire}.}
  \bibinfo{year}{2016}\natexlab{}.
\newblock \bibinfo{title}{The memory usage of STL containers can be
  surprising}.
\newblock
\newblock
\urldef\tempurl%
\url{https://lemire.me/blog/2016/09/15/the-memory-usage-of-stl-containers-can-be-surprising/}
\showURL{%
\tempurl}


\bibitem[Li et~al\mbox{.}(2022)]%
        {CheckedCModel:CSF22}
\bibfield{author}{\bibinfo{person}{Liyi Li}, \bibinfo{person}{Yiyun Liu},
  \bibinfo{person}{Deena~L. Postol}, \bibinfo{person}{Leonidas Lampropoulos},
  \bibinfo{person}{David~Van Horn}, {and} \bibinfo{person}{Michael Hicks}.}
  \bibinfo{year}{2022}\natexlab{}.
\newblock \showarticletitle{A Formal Model of {Checked C}}. In
  \bibinfo{booktitle}{\emph{Proceedings of the Computer Security Foundations
  Symposium (CSF)}}.
\newblock


\bibitem[Liu et~al\mbox{.}(2018)]%
        {pSweeper:CCS18}
\bibfield{author}{\bibinfo{person}{Daiping Liu}, \bibinfo{person}{Mingwei
  Zhang}, {and} \bibinfo{person}{Haining Wang}.}
  \bibinfo{year}{2018}\natexlab{}.
\newblock \showarticletitle{A Robust and Efficient Defense against
  Use-after-Free Exploits via Concurrent Pointer Sweeping}. In
  \bibinfo{booktitle}{\emph{Proceedings of the 2018 ACM SIGSAC Conference on
  Computer and Communications Security}} (Toronto, Canada)
  \emph{(\bibinfo{series}{CCS '18})}. \bibinfo{publisher}{Association for
  Computing Machinery}, \bibinfo{address}{New York, NY, USA},
  \bibinfo{pages}{1635–1648}.
\newblock
\showISBNx{9781450356930}
\urldef\tempurl%
\url{https://doi.org/10.1145/3243734.3243826}
\showDOI{\tempurl}


\bibitem[Liu et~al\mbox{.}(2017)]%
        {PtrSplit:CCS17}
\bibfield{author}{\bibinfo{person}{Shen Liu}, \bibinfo{person}{Gang Tan}, {and}
  \bibinfo{person}{Trent Jaeger}.} \bibinfo{year}{2017}\natexlab{}.
\newblock \showarticletitle{PtrSplit: Supporting General Pointers in Automatic
  Program Partitioning}. In \bibinfo{booktitle}{\emph{Proceedings of the 2017
  ACM SIGSAC Conference on Computer and Communications Security}} (Dallas,
  Texas, USA) \emph{(\bibinfo{series}{CCS '17})}.
  \bibinfo{publisher}{Association for Computing Machinery},
  \bibinfo{address}{New York, NY, USA}, \bibinfo{pages}{2359–2371}.
\newblock
\showISBNx{9781450349468}
\urldef\tempurl%
\url{https://doi.org/10.1145/3133956.3134066}
\showDOI{\tempurl}


\bibitem[{LLVM Developer Group}(2022a)]%
        {LLVMTestSuite}
\bibfield{author}{\bibinfo{person}{{LLVM Developer Group}}.}
  \bibinfo{year}{2022}\natexlab{a}.
\newblock \bibinfo{title}{LLVM Test Suite}.
\newblock
\newblock
\urldef\tempurl%
\url{https://llvm.org/docs/TestSuiteGuide.html}
\showURL{%
\tempurl}


\bibitem[{LLVM Developer Group}(2022b)]%
        {mem2reg:LLVM}
\bibfield{author}{\bibinfo{person}{{LLVM Developer Group}}.}
  \bibinfo{year}{2022}\natexlab{b}.
\newblock \bibinfo{title}{Promote Memory to Register}.
\newblock
\newblock
\urldef\tempurl%
\url{https://llvm.org/docs/Passes.html#mem2reg-promote-memory-to-register}
\showURL{%
\tempurl}


\bibitem[{LLVM Document}(2022)]%
        {LLVMPointer}
\bibfield{author}{\bibinfo{person}{{LLVM Document}}.}
  \bibinfo{year}{2022}\natexlab{}.
\newblock \bibinfo{title}{llvm::PointerType Class Reference}.
\newblock
\newblock
\urldef\tempurl%
\url{https://llvm.org/doxygen/classllvm_1_1PointerType.html}
\showURL{%
\tempurl}


\bibitem[Lu et~al\mbox{.}(2020)]%
        {SystemVAMD64ABI}
\bibfield{author}{\bibinfo{person}{H.J. Lu}, \bibinfo{person}{Michael Matz},
  \bibinfo{person}{Milind Girkar}, \bibinfo{person}{Jan Hubi\^cka},
  \bibinfo{person}{Andreas Jaeger}, {and} \bibinfo{person}{Mark Mitchell}.}
  \bibinfo{year}{2020}\natexlab{}.
\newblock \bibinfo{booktitle}{\emph{System V Application Binary Interface AMD64
  Architecture Processor Supplement}}.
\newblock
\urldef\tempurl%
\url{https://gitlab.com/x86-psABIs/x86-64-ABI}
\showURL{%
\tempurl}
\newblock
\shownote{Version 1.0}.


\bibitem[Luk and Mowry(1996)]%
        {CompilerPrefetch:ASPLOS96}
\bibfield{author}{\bibinfo{person}{Chi-Keung Luk} {and}
  \bibinfo{person}{Todd~C. Mowry}.} \bibinfo{year}{1996}\natexlab{}.
\newblock \showarticletitle{Compiler-Based Prefetching for Recursive Data
  Structures}. In \bibinfo{booktitle}{\emph{Proceedings of the Seventh
  International Conference on Architectural Support for Programming Languages
  an d Operating Systems}} (Cambridge, Massachusetts, USA)
  \emph{(\bibinfo{series}{ASPLOS VII})}. \bibinfo{publisher}{Association for
  Computing Machinery}, \bibinfo{address}{New York, NY, USA},
  \bibinfo{pages}{222–233}.
\newblock
\showISBNx{0897917677}
\urldef\tempurl%
\url{https://doi.org/10.1145/237090.237190}
\showDOI{\tempurl}


\bibitem[Machiry et~al\mbox{.}(2022)]%
        {3C:OOPSLA22}
\bibfield{author}{\bibinfo{person}{Aravind Machiry}, \bibinfo{person}{John
  Kastner}, \bibinfo{person}{Matt McCutchen}, \bibinfo{person}{Aaron Eline},
  \bibinfo{person}{Kyle Headley}, {and} \bibinfo{person}{Michael Hicks}.}
  \bibinfo{year}{2022}\natexlab{}.
\newblock \showarticletitle{{C} to {Checked C} by {3C}}. In
  \bibinfo{booktitle}{\emph{Proceedings of the {ACM} Conference on
  Object-Oriented Programming Languages, Systems, and Applications (OOPSLA)}}.
\newblock
\urldef\tempurl%
\url{https://arxiv.org/abs/2203.13445}
\showURL{%
\tempurl}


\bibitem[Mahoney(2021)]%
        {enwik:compression}
\bibfield{author}{\bibinfo{person}{Matt Mahoney}.}
  \bibinfo{year}{2021}\natexlab{}.
\newblock \bibinfo{title}{Large Text Compression Benchmark}.
\newblock
\newblock
\urldef\tempurl%
\url{http://mattmahoney.net/dc/text.html}
\showURL{%
\tempurl}
\newblock
\shownote{{Accessed}: 09-03-2021}.


\bibitem[{Microsoft Incorporation}({[n.\,d.]})]%
        {PageHeap}
\bibfield{author}{\bibinfo{person}{{Microsoft Incorporation}}.}
  \bibinfo{year}{[n.\,d.]}\natexlab{}.
\newblock \bibinfo{title}{{How to use Pageheap.exe in Windows XP and Windows
  2000.}}
\newblock
\newblock
\newblock
\shownote{\url{https://support.microsoft.com/en-gb/help/286470/how-to-use-pageheap-exe-in-windows-xp-windows-2000-and-windows-server}}.


\bibitem[Miller(2019)]%
        {VulTrend:BlueHatIL19}
\bibfield{author}{\bibinfo{person}{Matt Miller}.}
  \bibinfo{year}{2019}\natexlab{}.
\newblock \bibinfo{title}{Trends, challenge, and shifts in software
  vulnerability mitigation}.
\newblock
\newblock
\urldef\tempurl%
\url{https://github.com/microsoft/MSRC-Security-Research/tree/master/presentations/2019_02_BlueHatIL}
\showURL{%
\tempurl}
\newblock
\shownote{BlueHat IL}.


\bibitem[Mozilla(2023)]%
        {rustlang}
\bibfield{author}{\bibinfo{person}{Mozilla}.} \bibinfo{year}{2023}\natexlab{}.
\newblock \bibinfo{title}{{Rust} {Programming Language}}.
\newblock \bibinfo{howpublished}{\url{https://www.rust-lang.org/}}.
\newblock


\bibitem[Nagaraju et~al\mbox{.}(2013)]%
        {ExploitTrend:MS13}
\bibfield{author}{\bibinfo{person}{Swamy~Shivaganga Nagaraju},
  \bibinfo{person}{Cristian Craioveanu}, \bibinfo{person}{Elia Florio}, {and}
  \bibinfo{person}{Matt Miller}.} \bibinfo{year}{2013}\natexlab{}.
\newblock \bibinfo{title}{Software Vulnerability Exploitation Trends}.
\newblock
\newblock
\newblock
\shownote{Microsoft Technical Report}.


\bibitem[Nagarakatte(2014)]%
        {SoftBoundCETS:Github}
\bibfield{author}{\bibinfo{person}{Santosh Nagarakatte}.}
  \bibinfo{year}{2014}\natexlab{}.
\newblock \bibinfo{title}{SoftBoundCETS for LLVM+Clang version 34}.
\newblock
\newblock
\urldef\tempurl%
\url{https://github.com/santoshn/softboundcets-34}
\showURL{%
\tempurl}
\newblock
\shownote{{Accessed}: 07-25-2021}.


\bibitem[Nagarakatte et~al\mbox{.}(2012)]%
        {Watchdog:ISCA12}
\bibfield{author}{\bibinfo{person}{Santosh Nagarakatte}, \bibinfo{person}{Milo
  M.~K. Martin}, {and} \bibinfo{person}{Steve Zdancewic}.}
  \bibinfo{year}{2012}\natexlab{}.
\newblock \showarticletitle{Watchdog: Hardware for Safe and Secure Manual
  Memory Management and Full Memory Safety}. In
  \bibinfo{booktitle}{\emph{Proceedings of the 39th Annual International
  Symposium on Computer Architecture}} (Portland, Oregon)
  \emph{(\bibinfo{series}{ISCA '12})}. \bibinfo{publisher}{IEEE Computer
  Society}, \bibinfo{address}{USA}, \bibinfo{pages}{189–200}.
\newblock
\showISBNx{9781450316422}


\bibitem[Nagarakatte et~al\mbox{.}(2014)]%
        {WatchdogLite:CGO14}
\bibfield{author}{\bibinfo{person}{Santosh Nagarakatte}, \bibinfo{person}{Milo
  M.~K. Martin}, {and} \bibinfo{person}{Steve Zdancewic}.}
  \bibinfo{year}{2014}\natexlab{}.
\newblock \showarticletitle{WatchdogLite: Hardware-Accelerated Compiler-Based
  Pointer Checking}. In \bibinfo{booktitle}{\emph{Proceedings of Annual
  IEEE/ACM International Symposium on Code Generation and Optimization}}
  (Orlando, FL, USA) \emph{(\bibinfo{series}{CGO '14})}.
  \bibinfo{publisher}{Association for Computing Machinery},
  \bibinfo{address}{New York, NY, USA}, \bibinfo{pages}{175–184}.
\newblock
\showISBNx{9781450326704}
\urldef\tempurl%
\url{https://doi.org/10.1145/2544137.2544147}
\showDOI{\tempurl}


\bibitem[Nagarakatte et~al\mbox{.}(2015)]%
        {SoftBoundCETS:SNAPL15}
\bibfield{author}{\bibinfo{person}{Santosh Nagarakatte}, \bibinfo{person}{Milo
  M.~K. Martin}, {and} \bibinfo{person}{Steve Zdancewic}.}
  \bibinfo{year}{2015}\natexlab{}.
\newblock \showarticletitle{{Everything You Want to Know About Pointer-Based
  Checking}}. In \bibinfo{booktitle}{\emph{1st Summit on Advances in
  Programming Languages (SNAPL 2015)}} \emph{(\bibinfo{series}{Leibniz
  International Proceedings in Informatics (LIPIcs)},
  Vol.~\bibinfo{volume}{32})}, \bibfield{editor}{\bibinfo{person}{Thomas Ball},
  \bibinfo{person}{Rastislav Bodik}, \bibinfo{person}{Shriram Krishnamurthi},
  \bibinfo{person}{Benjamin~S. Lerner}, {and} \bibinfo{person}{Greg Morrisett}}
  (Eds.). \bibinfo{publisher}{Schloss Dagstuhl--Leibniz-Zentrum fuer
  Informatik}, \bibinfo{address}{Dagstuhl, Germany}, \bibinfo{pages}{190--208}.
\newblock
\showISBNx{978-3-939897-80-4}
\showISSN{1868-8969}
\urldef\tempurl%
\url{https://doi.org/10.4230/LIPIcs.SNAPL.2015.190}
\showDOI{\tempurl}


\bibitem[Nagarakatte et~al\mbox{.}(2009)]%
        {SoftBound:PLDI09}
\bibfield{author}{\bibinfo{person}{Santosh Nagarakatte},
  \bibinfo{person}{Jianzhou Zhao}, \bibinfo{person}{Milo~M.K. Martin}, {and}
  \bibinfo{person}{Steve Zdancewic}.} \bibinfo{year}{2009}\natexlab{}.
\newblock \showarticletitle{SoftBound: Highly Compatible and Complete Spatial
  Memory Safety for C}. In \bibinfo{booktitle}{\emph{Proceedings of the 30th
  ACM SIGPLAN Conference on Programming Language Design and Implementation}}
  (Dublin, Ireland) \emph{(\bibinfo{series}{PLDI '09})}.
  \bibinfo{publisher}{ACM}, \bibinfo{address}{New York, NY, USA},
  \bibinfo{pages}{245--258}.
\newblock
\showISBNx{978-1-60558-392-1}
\urldef\tempurl%
\url{https://doi.org/10.1145/1542476.1542504}
\showDOI{\tempurl}


\bibitem[Nagarakatte et~al\mbox{.}(2010)]%
        {CETS:ISMM10}
\bibfield{author}{\bibinfo{person}{Santosh Nagarakatte},
  \bibinfo{person}{Jianzhou Zhao}, \bibinfo{person}{Milo~M.K. Martin}, {and}
  \bibinfo{person}{Steve Zdancewic}.} \bibinfo{year}{2010}\natexlab{}.
\newblock \showarticletitle{{CETS: Compiler-Enforced Temporal Safety for C}}.
  In \bibinfo{booktitle}{\emph{Proceedings of the 2010 International Symposium
  on Memory Management}} (Toronto, Ontario, Canada)
  \emph{(\bibinfo{series}{ISMM '10})}. \bibinfo{publisher}{ACM},
  \bibinfo{pages}{31--40}.
\newblock
\showISBNx{978-1-4503-0054-4}
\urldef\tempurl%
\url{https://doi.org/10.1145/1806651.1806657}
\showDOI{\tempurl}


\bibitem[Necula et~al\mbox{.}(2005)]%
        {CCured:TOPLAS05}
\bibfield{author}{\bibinfo{person}{George~C. Necula}, \bibinfo{person}{Jeremy
  Condit}, \bibinfo{person}{Matthew Harren}, \bibinfo{person}{Scott McPeak},
  {and} \bibinfo{person}{Westley Weimer}.} \bibinfo{year}{2005}\natexlab{}.
\newblock \showarticletitle{CCured: Type-Safe Retrofitting of Legacy Software}.
\newblock \bibinfo{journal}{\emph{ACM Trans. Program. Lang. Syst.}}
  \bibinfo{volume}{27}, \bibinfo{number}{3} (\bibinfo{date}{May}
  \bibinfo{year}{2005}), \bibinfo{pages}{477–526}.
\newblock
\showISSN{0164-0925}
\urldef\tempurl%
\url{https://doi.org/10.1145/1065887.1065892}
\showDOI{\tempurl}


\bibitem[Necula et~al\mbox{.}(2002)]%
        {CCured:POPL02}
\bibfield{author}{\bibinfo{person}{George~C. Necula}, \bibinfo{person}{Scott
  McPeak}, {and} \bibinfo{person}{Westley Weimer}.}
  \bibinfo{year}{2002}\natexlab{}.
\newblock \showarticletitle{CCured: Type-Safe Retrofitting of Legacy Code}. In
  \bibinfo{booktitle}{\emph{Proceedings of the 29th ACM SIGPLAN-SIGACT
  Symposium on Principles of Programming Languages}} (Portland, Oregon)
  \emph{(\bibinfo{series}{POPL '02})}. \bibinfo{publisher}{Association for
  Computing Machinery}, \bibinfo{address}{New York, NY, USA},
  \bibinfo{pages}{128–139}.
\newblock
\showISBNx{1581134509}
\urldef\tempurl%
\url{https://doi.org/10.1145/503272.503286}
\showDOI{\tempurl}


\bibitem[Novark and Berger(2010)]%
        {DieHarder:CCS10}
\bibfield{author}{\bibinfo{person}{Gene Novark} {and} \bibinfo{person}{Emery~D.
  Berger}.} \bibinfo{year}{2010}\natexlab{}.
\newblock \showarticletitle{{DieHarder: Securing the Heap}}. In
  \bibinfo{booktitle}{\emph{Proceedings of the 17th ACM Conference on Computer
  and Communications Security}} (Chicago, Illinois, USA)
  \emph{(\bibinfo{series}{CCS '10})}. \bibinfo{publisher}{ACM},
  \bibinfo{address}{New York, NY, USA}, \bibinfo{pages}{573--584}.
\newblock
\showISBNx{978-1-4503-0245-6}
\urldef\tempurl%
\url{https://doi.org/10.1145/1866307.1866371}
\showDOI{\tempurl}


\bibitem[Patil and Fischer(1997)]%
        {Guarding:SPE97}
\bibfield{author}{\bibinfo{person}{Harish Patil} {and} \bibinfo{person}{Charles
  Fischer}.} \bibinfo{year}{1997}\natexlab{}.
\newblock \showarticletitle{Low-Cost, Concurrent Checking of Pointer and Array
  Accesses in C Programs}.
\newblock \bibinfo{journal}{\emph{Softw. Pract. Exper.}} \bibinfo{volume}{27},
  \bibinfo{number}{1} (\bibinfo{date}{Jan.} \bibinfo{year}{1997}),
  \bibinfo{pages}{87–110}.
\newblock
\showISSN{0038-0644}


\bibitem[Pereira et~al\mbox{.}(2017)]%
        {PLCmp:SLE17}
\bibfield{author}{\bibinfo{person}{Rui Pereira}, \bibinfo{person}{Marco Couto},
  \bibinfo{person}{Francisco Ribeiro}, \bibinfo{person}{Rui Rua},
  \bibinfo{person}{J\'{a}come Cunha}, \bibinfo{person}{Jo\~{a}o~Paulo
  Fernandes}, {and} \bibinfo{person}{Jo\~{a}o Saraiva}.}
  \bibinfo{year}{2017}\natexlab{}.
\newblock \showarticletitle{Energy Efficiency across Programming Languages: How
  Do Energy, Time, and Memory Relate?}. In
  \bibinfo{booktitle}{\emph{Proceedings of the 10th ACM SIGPLAN International
  Conference on Software Language Engineering}} (Vancouver, BC, Canada)
  \emph{(\bibinfo{series}{SLE 2017})}. \bibinfo{publisher}{Association for
  Computing Machinery}, \bibinfo{address}{New York, NY, USA},
  \bibinfo{pages}{256–267}.
\newblock
\showISBNx{9781450355254}
\urldef\tempurl%
\url{https://doi.org/10.1145/3136014.3136031}
\showDOI{\tempurl}


\bibitem[Perens(1993)]%
        {eFence}
\bibfield{author}{\bibinfo{person}{Bruce Perens}.}
  \bibinfo{year}{1993}\natexlab{}.
\newblock \bibinfo{title}{{Electric Fence}}.
\newblock
\newblock
\newblock
\shownote{\url{https://linux.die.net/man/3/efence}}.


\bibitem[Phantasmagoria(2005)]%
        {MallocMaleficarum}
\bibfield{author}{\bibinfo{person}{Phantasmal Phantasmagoria}.}
  \bibinfo{year}{2005}\natexlab{}.
\newblock \bibinfo{title}{The Malloc Maleficarum}.
\newblock
\newblock
\urldef\tempurl%
\url{https://dl.packetstormsecurity.net/papers/attack/MallocMaleficarum.txt}
\showURL{%
\tempurl}


\bibitem[Poskanzer(2018)]%
        {thttpd}
\bibfield{author}{\bibinfo{person}{Jef Poskanzer}.}
  \bibinfo{year}{2018}\natexlab{}.
\newblock \bibinfo{title}{thttpd - tiny/turbo/throttling HTTP server}.
\newblock
\newblock
\urldef\tempurl%
\url{https://acme.com/software/thttpd/}
\showURL{%
\tempurl}


\bibitem[Pratikakis et~al\mbox{.}(2011)]%
        {LOCKSMITH:TOPLS11}
\bibfield{author}{\bibinfo{person}{Polyvios Pratikakis},
  \bibinfo{person}{Jeffrey~S. Foster}, {and} \bibinfo{person}{Michael Hicks}.}
  \bibinfo{year}{2011}\natexlab{}.
\newblock \showarticletitle{LOCKSMITH: Practical Static Race Detection for C}.
\newblock \bibinfo{journal}{\emph{ACM Trans. Program. Lang. Syst.}}
  \bibinfo{volume}{33}, \bibinfo{number}{1}, Article \bibinfo{articleno}{3}
  (\bibinfo{date}{jan} \bibinfo{year}{2011}), \bibinfo{numpages}{55}~pages.
\newblock
\showISSN{0164-0925}
\urldef\tempurl%
\url{https://doi.org/10.1145/1889997.1890000}
\showDOI{\tempurl}


\bibitem[Rogers et~al\mbox{.}(1995)]%
        {Olden:TOPLAS95}
\bibfield{author}{\bibinfo{person}{Anne Rogers}, \bibinfo{person}{Martin~C.
  Carlisle}, \bibinfo{person}{John~H. Reppy}, {and} \bibinfo{person}{Laurie~J.
  Hendren}.} \bibinfo{year}{1995}\natexlab{}.
\newblock \showarticletitle{{Supporting Dynamic Data Structures on
  Distributed-memory Machines}}.
\newblock \bibinfo{journal}{\emph{ACM Trans. Program. Lang. Syst.}}
  \bibinfo{volume}{17}, \bibinfo{number}{2} (\bibinfo{date}{March}
  \bibinfo{year}{1995}), \bibinfo{pages}{233--263}.
\newblock
\showISSN{0164-0925}
\urldef\tempurl%
\url{https://doi.org/10.1145/201059.201065}
\showDOI{\tempurl}


\bibitem[Shen and Dolan-Gavitt(2020)]%
        {HeapExpo:ACSAC20}
\bibfield{author}{\bibinfo{person}{Zekun Shen} {and} \bibinfo{person}{Brendan
  Dolan-Gavitt}.} \bibinfo{year}{2020}\natexlab{}.
\newblock \showarticletitle{HeapExpo: Pinpointing Promoted Pointers to Prevent
  Use-After-Free Vulnerabilities}. In \bibinfo{booktitle}{\emph{Proceedings of
  the 36th Annual Computer Security Applications Conference}}
  \emph{(\bibinfo{series}{ACSAC '20})}. \bibinfo{publisher}{Association for
  Computing Machinery}.
\newblock


\bibitem[Shin et~al\mbox{.}(2019)]%
        {CRCount:NDSS19}
\bibfield{author}{\bibinfo{person}{Jangseop Shin}, \bibinfo{person}{Donghyun
  Kwon}, \bibinfo{person}{Yeongpil~Cho Jiwon~Seo}, {and}
  \bibinfo{person}{Yunheung Paek}.} \bibinfo{year}{2019}\natexlab{}.
\newblock \showarticletitle{CRCount: Pointer Invalidation with Reference
  Counting to Mitigate Use-after-free in Legacy C/C++}. In
  \bibinfo{booktitle}{\emph{NDSS}}.
\newblock


\bibitem[Silvestro et~al\mbox{.}(2017)]%
        {FreeGuard:CCS17}
\bibfield{author}{\bibinfo{person}{Sam Silvestro}, \bibinfo{person}{Hongyu
  Liu}, \bibinfo{person}{Corey Crosser}, \bibinfo{person}{Zhiqiang Lin}, {and}
  \bibinfo{person}{Tongping Liu}.} \bibinfo{year}{2017}\natexlab{}.
\newblock \showarticletitle{FreeGuard: A Faster Secure Heap Allocator}. In
  \bibinfo{booktitle}{\emph{Proceedings of the 2017 ACM SIGSAC Conference on
  Computer and Communications Security}} (Dallas, Texas, USA)
  \emph{(\bibinfo{series}{CCS '17})}. \bibinfo{publisher}{Association for
  Computing Machinery}, \bibinfo{address}{New York, NY, USA},
  \bibinfo{pages}{2389–2403}.
\newblock
\showISBNx{9781450349468}
\urldef\tempurl%
\url{https://doi.org/10.1145/3133956.3133957}
\showDOI{\tempurl}


\bibitem[Silvestro et~al\mbox{.}(2018)]%
        {Guarder:Sec18}
\bibfield{author}{\bibinfo{person}{Sam Silvestro}, \bibinfo{person}{Hongyu
  Liu}, \bibinfo{person}{Tianyi Liu}, \bibinfo{person}{Zhiqiang Lin}, {and}
  \bibinfo{person}{Tongping Liu}.} \bibinfo{year}{2018}\natexlab{}.
\newblock \showarticletitle{Guarder: A Tunable Secure Allocator}. In
  \bibinfo{booktitle}{\emph{Proceedings of the 27th USENIX Conference on
  Security Symposium}} (Baltimore, MD, USA) \emph{(\bibinfo{series}{SEC'18})}.
  \bibinfo{publisher}{USENIX Association}, \bibinfo{address}{USA},
  \bibinfo{pages}{117–133}.
\newblock
\showISBNx{9781931971461}


\bibitem[Simpson and Barua(2013)]%
        {MemSafe:SPE13}
\bibfield{author}{\bibinfo{person}{Matthew~S. Simpson} {and}
  \bibinfo{person}{Rajeev~K. Barua}.} \bibinfo{year}{2013}\natexlab{}.
\newblock \showarticletitle{MemSafe: Ensuring the Spatial and Temporal Memory
  Safety of C at Runtime}.
\newblock \bibinfo{journal}{\emph{Softw. Pract. Exper.}} \bibinfo{volume}{43},
  \bibinfo{number}{1} (\bibinfo{date}{Jan.} \bibinfo{year}{2013}),
  \bibinfo{pages}{93–128}.
\newblock
\showISSN{0038-0644}
\urldef\tempurl%
\url{https://doi.org/10.1002/spe.2105}
\showDOI{\tempurl}


\bibitem[Stenberg(2022)]%
        {curl}
\bibfield{author}{\bibinfo{person}{Daniel Stenberg}.}
  \bibinfo{year}{2022}\natexlab{}.
\newblock \bibinfo{title}{cURL: A command line tool and library for
  transferring data with URLs}.
\newblock
\newblock
\urldef\tempurl%
\url{https://curl.se/}
\showURL{%
\tempurl}


\bibitem[Tarditi(2021)]%
        {CheckedCTR:Microsoft}
\bibfield{author}{\bibinfo{person}{David Tarditi}.}
  \bibinfo{year}{2021}\natexlab{}.
\newblock \bibinfo{booktitle}{\emph{Extending C with Bounds Safety and Improved
  Type Safety}}.
\newblock \bibinfo{type}{{T}echnical {R}eport}.
\newblock
\urldef\tempurl%
\url{https://github.com/microsoft/checkedc/tree/master/spec/bounds_safety}
\showURL{%
\tempurl}
\newblock
\shownote{{Accessed}: 07-14-2021}.


\bibitem[van~der Kouwe et~al\mbox{.}(2017)]%
        {DangSan:EuroSys17}
\bibfield{author}{\bibinfo{person}{Erik van~der Kouwe}, \bibinfo{person}{Vinod
  Nigade}, {and} \bibinfo{person}{Cristiano Giuffrida}.}
  \bibinfo{year}{2017}\natexlab{}.
\newblock \showarticletitle{{DangSan: Scalable Use-after-free Detection}}. In
  \bibinfo{booktitle}{\emph{Proceedings of the Twelfth European Conference on
  Computer Systems}} (Belgrade, Serbia) \emph{(\bibinfo{series}{EuroSys '17})}.
  \bibinfo{publisher}{ACM}, \bibinfo{pages}{405--419}.
\newblock
\showISBNx{978-1-4503-4938-3}
\urldef\tempurl%
\url{https://doi.org/10.1145/3064176.3064211}
\showDOI{\tempurl}


\bibitem[WebAssembly(2021)]%
        {wasm:memory64}
\bibfield{author}{\bibinfo{person}{WebAssembly}.}
  \bibinfo{year}{2021}\natexlab{}.
\newblock \bibinfo{title}{Memory64}.
\newblock
\newblock
\urldef\tempurl%
\url{https://github.com/WebAssembly/memory64/blob/main/proposals/memory64/Overview.md}
\showURL{%
\tempurl}


\bibitem[Wesley~Filardo et~al\mbox{.}(2020)]%
        {Cornucopia:Oakland20}
\bibfield{author}{\bibinfo{person}{Nathaniel Wesley~Filardo},
  \bibinfo{person}{Brett~F. Gutstein}, \bibinfo{person}{Jonathan Woodruff},
  \bibinfo{person}{Sam Ainsworth}, \bibinfo{person}{Lucian Paul-Trifu},
  \bibinfo{person}{Brooks Davis}, \bibinfo{person}{Hongyan Xia},
  \bibinfo{person}{Edward Tomasz~Napierala}, \bibinfo{person}{Alexander
  Richardson}, \bibinfo{person}{John Baldwin}, \bibinfo{person}{David
  Chisnall}, \bibinfo{person}{Jessica Clarke}, \bibinfo{person}{Khilan Gudka},
  \bibinfo{person}{Alexandre Joannou}, \bibinfo{person}{A. Theodore~Markettos},
  \bibinfo{person}{Alfredo Mazzinghi}, \bibinfo{person}{Robert~M. Norton},
  \bibinfo{person}{Michael Roe}, \bibinfo{person}{Peter Sewell},
  \bibinfo{person}{Stacey Son}, \bibinfo{person}{Timothy~M. Jones},
  \bibinfo{person}{Simon~W. Moore}, \bibinfo{person}{Peter~G. Neumann}, {and}
  \bibinfo{person}{Robert N.~M. Watson}.} \bibinfo{year}{2020}\natexlab{}.
\newblock \showarticletitle{Cornucopia: Temporal Safety for CHERI Heaps}. In
  \bibinfo{booktitle}{\emph{2020 IEEE Symposium on Security and Privacy (SP)}}.
  \bibinfo{pages}{608--625}.
\newblock
\urldef\tempurl%
\url{https://doi.org/10.1109/SP40000.2020.00098}
\showDOI{\tempurl}


\bibitem[Wickman et~al\mbox{.}(2021)]%
        {FFmalloc:Sec21}
\bibfield{author}{\bibinfo{person}{Brian Wickman}, \bibinfo{person}{Hong Hu},
  \bibinfo{person}{Insu Yun}, \bibinfo{person}{Daehee~JangJungWon Lim},
  \bibinfo{person}{Sanidhya Kashyap}, {and} \bibinfo{person}{Taesoo Kim}.}
  \bibinfo{year}{2021}\natexlab{}.
\newblock \showarticletitle{Preventing Use-After-Free Attacks with Fast Forward
  Allocation}. In \bibinfo{booktitle}{\emph{30th {USENIX} Security Symposium
  ({USENIX} Security 21)}}. \bibinfo{publisher}{{USENIX} Association},
  \bibinfo{address}{Vancouver, B.C.}
\newblock
\urldef\tempurl%
\url{https://www.usenix.org/conference/usenixsecurity21/presentation/wickman}
\showURL{%
\tempurl}


\bibitem[Woodruff et~al\mbox{.}(2014)]%
        {CHERI:ISCA14}
\bibfield{author}{\bibinfo{person}{Jonathan Woodruff},
  \bibinfo{person}{Robert~N.M. Watson}, \bibinfo{person}{David Chisnall},
  \bibinfo{person}{Simon~W. Moore}, \bibinfo{person}{Jonathan Anderson},
  \bibinfo{person}{Bro~oks Davis}, \bibinfo{person}{Ben Laurie},
  \bibinfo{person}{Peter~G. Neumann}, \bibinfo{person}{Robert Norton}, {and}
  \bibinfo{person}{Michael Roe}.} \bibinfo{year}{2014}\natexlab{}.
\newblock \showarticletitle{{The CHERI Capability Model: Revisiting RISC in an
  Age of Risk}}. In \bibinfo{booktitle}{\emph{Proceeding of the 41st Annual
  International Symposium on Computer Architecture}} (Minneapolis, Minnesota,
  USA) \emph{(\bibinfo{series}{ISCA '14})}. \bibinfo{publisher}{IEEE Press},
  \bibinfo{address}{Piscataway, NJ, USA}, \bibinfo{pages}{457--468}.
\newblock
\showISBNx{978-1-4799-4394-4}
\urldef\tempurl%
\url{http://dl.acm.org/citation.cfm?id=2665671.2665740}
\showURL{%
\tempurl}


\bibitem[Xia et~al\mbox{.}(2019)]%
        {CHERIvoke:MICRO19}
\bibfield{author}{\bibinfo{person}{Hongyan Xia}, \bibinfo{person}{Jonathan
  Woodruf}, \bibinfo{person}{Sam Ainsworth}, \bibinfo{person}{Nathaniel~W.
  Filardo}, \bibinfo{person}{Michael Roe}, \bibinfo{person}{Alexander
  Richardson}, \bibinfo{person}{Peter Rugg}, \bibinfo{person}{Peter~G.
  Neumann}, \bibinfo{person}{Simon~W. Moore}, \bibinfo{person}{Robert N.~M.
  Watson}, {and} \bibinfo{person}{Timothy~M. Jones}.}
  \bibinfo{year}{2019}\natexlab{}.
\newblock \showarticletitle{{CHERIvoke}: Characterising Pointer Revocation
  Using CHERI Capabilities for Temporal Memory Safety}. In
  \bibinfo{booktitle}{\emph{Proceedings of the 52Nd Annual IEEE/ACM
  International Symposium on Microarchitecture}} (Columbus, OH, USA)
  \emph{(\bibinfo{series}{MICRO '52})}. \bibinfo{publisher}{ACM},
  \bibinfo{address}{New York, NY, USA}, \bibinfo{pages}{545--557}.
\newblock
\showISBNx{978-1-4503-6938-1}
\urldef\tempurl%
\url{https://doi.org/10.1145/3352460.3358288}
\showDOI{\tempurl}


\bibitem[Xu et~al\mbox{.}(2004)]%
        {XuMemSafe:FSE04}
\bibfield{author}{\bibinfo{person}{Wei Xu}, \bibinfo{person}{Daniel~C.
  DuVarney}, {and} \bibinfo{person}{R. Sekar}.}
  \bibinfo{year}{2004}\natexlab{}.
\newblock \showarticletitle{An Efficient and Backwards-compatible
  Transformation to Ensure Memory Safety of C Programs}. In
  \bibinfo{booktitle}{\emph{Proceedings of the 12th ACM SIGSOFT Twelfth
  International Symposium on Foundations of Software Engineering}} (Newport
  Beach, CA, USA) \emph{(\bibinfo{series}{SIGSOFT '04/FSE-12})}.
  \bibinfo{publisher}{ACM}, \bibinfo{address}{New York, NY, USA},
  \bibinfo{pages}{117--126}.
\newblock
\showISBNx{1-58113-855-5}
\urldef\tempurl%
\url{https://doi.org/10.1145/1029894.1029913}
\showDOI{\tempurl}


\bibitem[Xu et~al\mbox{.}(2015)]%
        {UAFKernel:CCS15}
\bibfield{author}{\bibinfo{person}{Wen Xu}, \bibinfo{person}{Juanru Li},
  \bibinfo{person}{Junliang Shu}, \bibinfo{person}{Wenbo Yang},
  \bibinfo{person}{Tianyi Xie}, \bibinfo{person}{Yuanyuan Zhang}, {and}
  \bibinfo{person}{Dawu Gu}.} \bibinfo{year}{2015}\natexlab{}.
\newblock \showarticletitle{{From Collision To Exploitation: Unleashing
  Use-After-Free Vulnerabilities in Linux Kernel}}. In
  \bibinfo{booktitle}{\emph{Proceedings of the 22Nd ACM SIGSAC Conference on
  Computer and Communications Security}} (Denver, Colorado, USA)
  \emph{(\bibinfo{series}{CCS '15})}. \bibinfo{publisher}{ACM},
  \bibinfo{address}{New York, NY, USA}, \bibinfo{pages}{414--425}.
\newblock
\showISBNx{978-1-4503-3832-5}
\urldef\tempurl%
\url{https://doi.org/10.1145/2810103.2813637}
\showDOI{\tempurl}


\bibitem[Younan(2015)]%
        {FreeSentry:NDSS15}
\bibfield{author}{\bibinfo{person}{Yves Younan}.}
  \bibinfo{year}{2015}\natexlab{}.
\newblock \showarticletitle{{FreeSentry: Protecting Against Use-After-Free
  Vulnerabilities Due to Dangling Pointers}}. In
  \bibinfo{booktitle}{\emph{NDSS}}.
\newblock


\bibitem[Zeiss(2012)]%
        {SFCityJSON}
\bibfield{author}{\bibinfo{person}{Mirco Zeiss}.}
  \bibinfo{year}{2012}\natexlab{}.
\newblock \bibinfo{title}{Really big json file representing san francisco's
  subdivision parcels}.
\newblock
\newblock
\urldef\tempurl%
\url{https://github.com/zemirco/sf-city-lots-json}
\showURL{%
\tempurl}


\bibitem[Zhang et~al\mbox{.}(2019)]%
        {BOGO:ASPLOS19}
\bibfield{author}{\bibinfo{person}{Tong Zhang}, \bibinfo{person}{Dongyoon Lee},
  {and} \bibinfo{person}{Changhee Jung}.} \bibinfo{year}{2019}\natexlab{}.
\newblock \showarticletitle{{BOGO: Buy Spatial Memory Safety, Get Temporal
  Memory Safety (Almost) Free}}. In \bibinfo{booktitle}{\emph{Proceedings of
  the Twenty-Fourth International Conference on Architectural Support for
  Programming Languages and Operating Systems}} (Providence, RI, USA)
  \emph{(\bibinfo{series}{ASPLOS '19})}. \bibinfo{publisher}{ACM},
  \bibinfo{address}{New York, NY, USA}, \bibinfo{pages}{631--644}.
\newblock
\showISBNx{978-1-4503-6240-5}
\urldef\tempurl%
\url{https://doi.org/10.1145/3297858.3304017}
\showDOI{\tempurl}


\bibitem[Zhang et~al\mbox{.}(2023)]%
        {RustPerf:ASE22}
\bibfield{author}{\bibinfo{person}{Yuchen Zhang}, \bibinfo{person}{Yunhang
  Zhang}, \bibinfo{person}{Georgios Portokalidis}, {and} \bibinfo{person}{Jun
  Xu}.} \bibinfo{year}{2023}\natexlab{}.
\newblock \showarticletitle{Towards Understanding the Runtime Performance of
  Rust}. In \bibinfo{booktitle}{\emph{Proceedings of the 37th IEEE/ACM
  International Conference on Automated Software Engineering}} (Rochester, MI,
  USA) \emph{(\bibinfo{series}{ASE '22})}. \bibinfo{publisher}{Association for
  Computing Machinery}, \bibinfo{address}{New York, NY, USA}, Article
  \bibinfo{articleno}{140}, \bibinfo{numpages}{6}~pages.
\newblock
\showISBNx{9781450394758}
\urldef\tempurl%
\url{https://doi.org/10.1145/3551349.3559494}
\showDOI{\tempurl}


\bibitem[Özler(2019)]%
        {MongoDBJSON}
\bibfield{author}{\bibinfo{person}{Hakan Özler}.}
  \bibinfo{year}{2019}\natexlab{}.
\newblock \bibinfo{title}{A curated list of JSON / BSON datasets from the web
  in order to practice / use in MongoDB}.
\newblock
\newblock
\urldef\tempurl%
\url{https://github.com/ozlerhakan/mongodb-json-files}
\showURL{%
\tempurl}


\end{thebibliography}

\end{document}